\documentclass[11pt,letter]{article}
\usepackage[top=1in,bottom=1in,left=1in,right=1in]{geometry}
\usepackage[noEucal]{main}
\usepackage{changepage}
\def\ci{{\mathscr I}}

\usepackage[bbgreekl]{mathbbol}
\DeclareSymbolFontAlphabet{\mathbbm}{bbold}

\def\d{{\delta}}

\def\D{D}

\def\g{\text{g}}

\def\W{{\text W}}

\def\o{{\omega}}
\def\w{\wedge}
\def\C{{\bf C}}
\def\cc{{\text{c.c.}}}
\def\hc{{\text{c.t.}}}

\def\la{{\langle}}
\def\ra{{\rangle}}
\def\CDt{\D^C}
\def\mfD{\mfd}
\def\Dt{\textrm{D}}
\def\kin{\text{kin}}
\def\mat{\text{mat}}
\def\inter{\text{inter}}
\renewcommand{\avg}[1]{\left\langle #1 \right\rangle}

\def\bi{{\bf i}}
\def\X{{\bf X}}
\def\Y{{\bf Y}}

\def\bdt{{\bf d}}
\def\bk{{\bs\kappa}}
\def\bt{{\bs \t}}
\def\bT{{\bs \Theta}}
\def\bo{{\bs \omega}}
\def\bO{{\bs \Omega}}

\def\soft{{\text{soft}}}
\def\hard{{\text{hard}}}
\def\Th{\text{sign}}
\def\ct{{\text{CT}}}
\def\LT{{\text{LT}}}

\begin{document}
\begin{titlepage}
\unitlength = 1mm
\ \\
\vskip 3cm
\begin{center}
%%%%%%%%%%%%%%%%%%%%---------------------TITLE---------------------%%%%%%%%%%%%%%%%%%%%%%%%

{\LARGE{\textsc{Covariant Phase Space and Soft Factorization in Non-Abelian Gauge Theories}}}

\vspace{0.8cm}
%%%%%%%%%%%%%%%%%%%%---------------------AUTHOR(S)---------------------%%%%%%%%%%%%%%%%%%%%%
Temple He$^\ddagger$, Prahar Mitra$^{\Diamond}$

\vspace{1cm}

{\it  $^\ddagger$Center for Quantum Mathematics and Physics, University of California, Davis, CA 95616, USA}\\
{\it  $^\Diamond$School of Natural Sciences, Institute for Advanced Study, Princeton, NJ 08540, USA}

\vspace{0.8cm}

%%%%%%%%%%%%%%%%%%%%---------------------ABSTRACT---------------------%%%%%%%%%%%%%%%%%%%%%%
\begin{abstract}

We perform a careful study of the infrared sector of massless non-abelian gauge theories in four-dimensional Minkowski spacetime using the covariant phase space formalism, taking into account the boundary contributions arising from the gauge sector of the theory. Upon quantization, we show that the boundary contributions lead to an infinite degeneracy of the vacua. The Hilbert space of the vacuum sector is not only shown to be remarkably simple, but also universal. We derive a Ward identity that relates the $n$-point amplitude between two generic $in$- and $out$-vacuum states to the one computed in standard QFT. In addition, we demonstrate that the familiar single soft gluon theorem and multiple consecutive soft gluon theorem are consequences of the Ward identity.
 \end{abstract}

\vspace{1.0cm}
\end{center}
\end{titlepage}
\pagestyle{empty}
\pagestyle{plain}
\pagenumbering{arabic}
 %%%%%%%%%%%%%%%%---------------------END OF TITLE PAGE AND ABSTRACT---------------------%%%%%%%%%%%%%%%%%%%%%%

\tableofcontents

\section{Introduction}
\label{sec:intro}

Infrared (IR) divergences present in the scattering matrix elements of gauge and gravitational theories have long been known to physicists \cite{Mott_1931,Bloch:1937aa}, and numerous attempts in the 1970s and 1980s have been made to render the scattering matrix elements in such theories IR finite.\footnote{This issue only arises in theories that do not confine and in which there exist massless charged asymptotic states.} A key idea in these approaches is to use modified asymptotic states to define the scattering matrix, wherein the charged external states are dressed with a coherent state of soft (low energy) photons \cite{PhysRev.140.B1110,Greco:1967zza,Kulish:1970ut,Kibble:1968aa,Kibble:1968ab,Kibble:1968ac,Kibble_1968}. The results have since been extended to non-abelian gauge theories in \cite{Greco:1978te,Giavarini:1987ts,DelDuca:1989jt,GRECO1978282,Hannesdottir:2019rqq} and more recently to perturbative gravity in \cite{Ware_2013}.

Although many may argue that there is no need for an IR finite $S$-matrix when the inclusive cross-section is IR finite (this follows from the KLN theorem \cite{Kinoshita_1962,Lee:1964aa}), advances in our understanding of how the soft theorems of quantum field theories (QFTs) are related to asymptotic symmetries have brought newfound appreciation for what the IR divergences in the $S$-matrix elements signify (see \cite{Strominger:2017zoo} for a review, as well as \cite{Kapec:2017tkm,Choi:2017bna,Choi:2017ylo,Carney:2018ygh,Ashtekar:2018lor,Hirai:2019gio,Gonzo:2019fai,Choi:2019rlz,H:2019fvd,Himwich:2020rro,Hirai:2020kzx}). Soft theorems restrict the form of the scattering amplitude in the IR (low energy) sector of any consistent QFT; more precisely, they imply that if $m$ particles in an $(n+m)$-point scattering amplitude are soft (i.e. have parametrically low energy compared to the remaining $n$ particles), the $S$-matrix element necessarily has the form
\begin{align}
    \CA_{n+m} \xrightarrow[]{\text{$m$ soft particles}} \SS_m\CA_{n},
\end{align}
where $\SS_m$ is the soft factor associated to $m$ soft gauge particles.\footnote{$\SS_m$ could be either a $c$-number, as is the case for the leading soft photon and soft graviton theorems; a matrix, as is the case for the leading soft gluon theorem; or a differential operator, as is in the case of the subleading soft photon, gluon, and graviton theorems.} The soft factor is universal in that it depends only on the quantum numbers of the external particles, but not on the detailed interactions in the theory. 

In exploring the connection between soft theorems and asymptotic symmetries, it was discovered that rather than having one unique vacuum state \cite{Strominger:2017zoo}, as is typically assumed in quantum field theories, there is in fact an infinite degeneracy of vacua in such theories parameterized by the soft particles \cite{He:2014laa,He:2014cra,He:2015zea}.\footnote{This vacuum degeneracy is not the degeneracy associated to the $\theta$-vacuum.} Scattering processes that respect the asymptotic symmetries of the theory involve $in$- and $out$-states residing in \emph{different} vacua, and it is the violation of precisely this fact in standard QFT that leads to infrared divergences. As was shown in \cite{Gabai:2016kuf,Kapec:2017tkm}, one can obtain an IR finite $S$-matrix after incorporating this infinite degeneracy and the corresponding $in$- and $out$-states are the coherent states constructed in \cite{PhysRev.140.B1110,Kibble:1968aa,Kibble:1968ab,Kibble:1968ac,Kibble_1968,Kulish:1970ut}.

Although this infinite degeneracy in gauge (and gravitational) theories may seem surprising at first, its existence can be deduced from a careful but straightforward application of the covariant phase space formalism \cite{Crnkovic:1986ex,Lee_1990,Iyer:1994ys,Wald:1999wa,Harlow:2019yfa} to gauge theories. Here, the Hilbert space of the theory is constructed by a careful study of the symplectic form of the theory on asymptotic Cauchy slices of spacetime (on which the $S$-matrix is defined). Of particular import are the boundary terms (i.e. terms localized on the boundary of the Cauchy slice) in the symplectic form, which are responsible for the infinite vacuum degeneracy mentioned earlier. In this paper, we will perform an analysis of the phase space of gauge theories, study the corresponding Hilbert space, including the infinite-dimensional vacuum degeneracy, and derive a factorization formula for the scattering matrix element between \emph{any} two vacuum states in the Hilbert space. 

The outline of our paper is as follows. In Section~\ref{sec:prelims}, we will review the relevant aspects of symplectic geometry and the covariant phase space formalism. In Section~\ref{sec:classicalgauge}, we initiate a careful methodical application of the formalism to generic non-abelian gauge theories. We then focus to the case of four-dimensional gauge theories near $\ci^\pm$ and construct the Dirac brackets associated to the gauge fields. In Section~\ref{sec:hs}, we will canonically quantize the classical theory and construct the Hilbert space. Finally, in Section~\ref{sec:softfactor}, we will explore the vacuum sector of the theory and derive a Ward identity that allows us to relate an $n$-point scattering amplitude involving arbitrary $in$- and $out$-vacuum states to the standard one evaluated in QFT. We then show how the leading soft gluon theorem involving a single soft gluon as well as multiple consecutive soft gluons are consequences of the Ward identity. 

\subsection{Summary of the Paper}

Because some of the derivations are rather tedious, we present in this section a (detailed) summary of the important results in this paper.

Starting in Section \ref{sec:classicalgauge}, we study non-abelian gauge theories with a gauge group $\CG$ and associated Lie algebra $\mfg$ on a globally hyperbolic $d$-dimensional manifold $\CM$. The fields of the theory are a gauge field $A$ and a generic set of matter fields $\Phi^i$ transforming in finite-dimensional irreducible representations $R_i$ of $\CG$ (with $i=1,\ldots,N$). We assume that the theory does not confine so the semi-classical analysis performed here generalizes to the quantum theory as well. After setting up our conventions, in Section \ref{sec:phasespacegeneric} we use the covariant phase space formalism (reviewed in Section \ref{sec:covphasespace}) to construct the phase space $\G$ (reviewed in Section \ref{sec:phasespacereview}) of the theory on a generic Cauchy slice $\S$ of $\CM$. This includes establishing an explicit coordinatization of $\G$ and the construction of the symplectic form $\bO$, which is a closed non-degenerate two-form on $\G$. The main results for this procedure are given in \eqref{omega-explicit} and \eqref{integrate-over-Sigma}. The symplectic form can be inverted to obtain the Poisson brackets on the phase space.  In Section \ref{sec:canonicaltransformgen}, we turn to a study of canonical transformations, which are diffeomorphisms on the phase space that preserve (in the sense of the Lie derivative) the symplectic form. Two types of canonical transformations are studied -- large gauge transformations and isometry transformations. Canonical transformations are generated (in the sense of the Poisson bracket) on the phase space by so-called Hamiltonian charges. The Hamiltonian charge for large gauge and isometry transformations are given in \eqref{largegaugecharge} and \eqref{stress-tensor-explicit} respectively.

Thus far, we have studied generic non-abelian gauge theories in generic spacetimes. In Section \ref{sec:gaugetheories-flatspacetime} and thereafter, we focus our attention to the special case of non-abelian gauge theories with massless scalar matter in four-dimensional Minkowski spacetime. The restriction to scalar matter is only for convenience, and all the central results of this paper generalize with trivial modifications to spinning fields. To simplify all the relevant results derived in Sections \ref{sec:phasespacegeneric} and \ref{sec:canonicaltransformgen} to this special case, we work in flat null coordinates $(u,r,z,\bz)$ where the metric of Minkowski spacetime takes the form
\begin{align}
\dt s^2 = -\dt u\,\dt r + r^2\,\dt z \, \dt\bz.
\end{align}

The particular Cauchy slices on which we construct our phase spaces are taken to be $\ci^+$ and $\ci^-$. These are asymptotic boundaries of the spacetime and are relevant to consider if one is interested in the scattering of massless particles (which is our eventual goal). These surfaces are located at $r\to\pm\infty$ while keeping $(u,z,\bz)$ fixed. Their boundaries are located at $u\to\mp\infty$ and are denoted by $\ci^\pm_\mp$. 

To describe the symplectic form on these surfaces, we define
\begin{equation}
\begin{split}\label{defs}
C_z &= C \p_z C^{-1} = A_z \big|_{\ci^+_-} = A_z \big|_{\ci^-_+}  \\
N_z^\pm &=   C \p_z N^\pm C^{-1} = \int  du\, \p_u A_z^\pm  \\
{\hat A}_z^\pm &= A_z^\pm - C \p_z C^{-1} \\
\phi^{\pm i} &= r \Phi^i \big|_{\ci^\pm} ,
\end{split}
\end{equation}
where $C \in \CG$ and $N^\pm \in \mfg$. The equality and flatness of $A_z |_{\ci^+_-} $ and $A_z |_{\ci^-_+}$ is a natural (though perhaps not necessary) requirement in order for the phase spaces on $\ci^+$ and $\ci^-$ to be isomorphic, and this is explained in the last part of Section \ref{sec:bdycondsec}. With these definitions, the symplectic form on $\ci^\pm$ is then given in \eqref{Og3} to be
\begin{equation}
\begin{split}\label{sympformci2}
	\bO_{\ci^\pm}(\X,\Y) &=  \frac{2}{g^2} \int  d^2 z \, \tr{   \X\big(  \p_z\p_\bz N^\pm  C^{-1}  \big) \Y(C)  - \Y\big(  \p_z\p_\bz N^\pm  C^{-1}  \big) \X(C) } \\
&\qquad +  \frac{2}{g^2}\int d u\, d^2 z\, \tr{ \p_u\X(\hat A_z^\pm)\Y(\hat A_\bz^\pm) - \p_u\Y(\hat A_z^\pm) \X(\hat A_\bz^\pm)  } \\
&\qquad +  \sum_{i=1}^N \int du\,d^2 z\, \Big( \p_u\X(\phi^{\pm i})^\ct \Y(\phi^{\pm i}) - \p_u\Y(\phi^{\pm i})^\ct \X(\phi^{\pm i}) \Big).
\end{split}
\end{equation}

Each line in \eqref{sympformci2} depends on a different set of fields -- the first line on the soft gauge fields $C$ and $N^\pm$, the second on the hard gauge fields ${\hat A}^\pm_z$ and the third on the matter fields $\phi^{\pm i}$. This split in $\O$ implies that the phase space also factorizes into the form
\begin{equation}
\begin{split}\label{phasespacefactor}
	\G = \G^{\pm A,\soft} \times \G^{\pm A,\hard} \times \G^{\pm 1} \times \cdots \times \G^{\pm N},
\end{split}
\end{equation}
where $\G^{\pm i}$ is the phase space of $\Phi^i$. Inverting the symplectic form \eqref{sympformci2}, we determine in \eqref{commutator} the Dirac brackets to be 
\begin{equation}
\begin{split}
\label{commutators-intro}
	\big\{ {\hat A}_z^{\pm a}(u,z,\bz)  , {\hat A}_{\bw}^{\pm b}(u',w,\bw) \big\} &= - \frac{g^2}{4}\d^{ab}\,  \text{sign} (u-u') \d^2(z-w) \\
	\big\{ N^{\pm a}(z,\bz) , C^{bc}(w,\bw) \big\} &= - \frac{g^2}{4\pi} f^{acd} C^{bd}(w,\bw)  \ln|z-w|^2 \\
	\big\{ N^{\pm a} (z,\bz) , N^{\pm b}(w,\bw) \big\} &= -  \frac{g^2}{8\pi^2}  f^{abc}  \int d^2 y  \ln|z-y|^2 \ln|w-y|^2   \p_y\p_\by N^{\pm c} (y,\by) \\
	\big\{ \phi^{\pm i}(u,z,\bz),\phi^{\pm j}(u',w,\bw)^\ct \} &= -\frac{1}{2}\delta^{ij} {\mathbb 1}\,\Th(u-u')\delta^2(z-w)  \\
	\text{all others} &= 0 . 
\end{split}
\end{equation}
The Hamiltonian charges for large gauge and isometry transformations can now also be determined in these variables (see \eqref{adjointcharges} in main text) to be
\begin{align}
\begin{split}\label{adjointcharges-intro}
Q_\ve  &=  \frac{2}{g^2}    \int  d^2 z\, \ve^a C^{ab}  \p_z \p_\bz N^{\pm b}  + \frac{2}{g^2} f^{abc} \int du\, d^2 z \, \ve^a  {\hat A}^{\pm b}_z \p_u {\hat A}_\bz^{ \pm c} \\
&~~~ + \frac{1}{2}\sum_{i=1}^N \int du\, d^2 z\, \ve^a \Big( \p_u(\phi^{\pm i})^\ct T_i^a \phi^{\pm i} - (\phi^{\pm i})^\ct T_i^a \p_u \phi^{\pm i} \Big)   \\
P_f  &=  \int du\, d^2 z\,  f  \bigg(  \frac{2}{g^2} \p_u {\hat A}_z^{\pm a} \p_u {\hat A}_\bz^{\pm a} +  \sum_{i=1}^N \p_u (\phi^{\pm i})^\ct \p_{u} \phi^{\pm i}   \bigg)  \\
J_Y &=    \frac{2}{g^2} \int d^2 z\, Y^z C_z^a C^{ab}  \p_z \p_\bz N^{\pm b}  \\
&~~~ +  \frac{1}{g^2} \int du\, d^2 z\, Y^z \left( \p_z {\hat A}^{\pm a}_\bz \overleftrightarrow{\p_u} {\hat A}^{\pm a}_z  -  u \p_z  \left( \p_u {\hat A}^{\pm a}_{z} \p_u {\hat A}_\bz^{\pm a} \right) \right)   \\
&~~~ + \frac{1}{2} \sum_{i=1}^N \int du\, d^2 z \,Y^z \Big( \p_u (\phi^{\pm i})^\ct \p_z \phi^{\pm i}  + \p_z (\phi^{\pm i})^\ct \p_u \phi^{\pm i}   - u \p_z \left(  \p_u (\phi^{\pm i} )^\ct \p_u \phi ^{\pm i} \right)  \Big) \\
&~~~ +  \cc .
\end{split}
\end{align}
Here, $P_f$ generates translations and $J_Y$ generates Lorentz transformations. In flat null coordinates these are infinitesimally generated by Killing vectors described in \eqref{kv}. We then use the Dirac brackets in \eqref{commutators-intro} to demonstrate that these charges indeed generate the proper transformations on the fields (see \eqref{chargeaction-all}):
\begin{align}
	\{ Q_\ve , \cdot \} = - \d_\ve (\ \cdot\ ) , \qquad  \{ P_f , \cdot \} = - \d_f (\ \cdot\ ),  \qquad \{ J_Y , \cdot \} = - \d_Y (\ \cdot\ ).
\end{align}

Up until this point, all of our results have been strictly classical. In Section \ref{sec:hs}, we canonically quantize the above constructed phase space. In the process, Dirac brackets $\{ \cdot,\cdot\}$ and complex conjugation on the phase space become quantum commutators $-i[\cdot,\cdot]$ and taking the adjoint on the Hilbert space, respectively. Together, \eqref{commutators-intro} and \eqref{adjointcharges-intro} imply that $C$ and $N^\pm$ commute with the translation charges, including the Hamiltonian of the theory. In other words, these operators carry zero energy and thus span an infinite-dimensional vacuum Hilbert space. The space of vacuum states is constructed in Section \ref{sec:hsvac} and we show that it is spanned by a set of basis vectors satisfying
\begin{equation}
\begin{split}
C^{ab} (z,\bz) \ket{U,\pm}  &= U^{ab}   (z,\bz) \ket{U,\pm}  \\
 N^{\pm a}(z,\bz) \ket{U,\pm} &= -  \frac{ig^2}{4\pi}  \int d^2 y\, \ln | z - y |^2 U^{ba}(y,\by) \mfd_{U(y,\by)}^b \ket{U,\pm},
\end{split}
\end{equation}
where the $\pm$ label in the ket state indicates whether it is a state on $\ci^+$ or $\ci^-$, and the $\mfd_{U(y,\by)}$ operator is defined in \eqref{daction} and more throughly explored in Appendix~\ref{app:deriv}. As it turns out, a generic vacuum state is not Lorentz invariant, thereby violating an assumption oftentimes made in standard QFT. However, the $|U=1\ra$ vacuum is Lorentz invariant, and we shall assume throughout this paper that this is the standard perturbative QFT vacuum. For each fixed vacuum, the rest of the hard modes act on it to create a tower of energetic states, thereby creating a Fock space (see Section \ref{sec:hsrad}), whose annihilation operators are given in \eqref{Odef}. Having constructed the isomorphic Hilbert spaces on $\ci^+$ and $\ci^-$, it is natural to consider the overlap of states in the two Hilbert spaces. This quantity is known as the scattering matrix, and is computed via the LSZ reduction formula in QFT. We discuss this construction in Section \ref{sec:Smatrix-def}.

Finally, in Section \ref{sec:softfactor} we arrive at the main result of our paper. We use the large gauge charge $Q_\ve$ and the definition of $C_z$ to derive an elegant factorization formula that relates the $S$-matrix evaluated in any $in$- and $out$-vacuum states to the one evaluated in standard QFT (i.e. in the $U=1$ vacuum). To be precise, in Section \ref{sec:mainresult}, we show
\begin{align}\label{mainresult-intro}
\begin{split}
&\bra{U,+} T \{ \CO_1^{i_1}(p_1) \cdots \CO_n^{i_n}(p_n) \} \ket{ U',- } \\
&\qquad \qquad = \d(U-U') R_1(U(z_1,\bz_1))^{i_1}{}_{j_1} \cdots  R_n(U(z_n,\bz_n))^{i_n}{}_{j_n} \avg{  \CO_1^{j_1}(p_1) \cdots \CO_n^{j_n}(p_n) }_{U=1} .
\end{split}
\end{align}
Here, the left-hand-side denotes an $n$-point scattering amplitude evaluated in a $U'$ and $U$ $in$- and $out$-vacuum respectively, and $i_k$ denotes the flavor indices (with respect to $\CG$) of the particles created/annihilated by the operator $\CO_k$. The last term on the right-hand-side denotes the standard QFT $S$-matrix. Because we know how to compute this using Feynman diagrams, it follows we can determine the $S$-matrix element between any two arbitrary vacua. Indeed, we conclude this paper in Sections \ref{sec:singlesoft1} and \ref{sec:multisoft1} by verifying the above formula in two special cases -- the first in which one gluon is taken to be soft and the second in which two gluons are taken to be soft consecutively. In these cases, the leading soft gluon theorem implies that the scattering amplitude undergoes a soft factorization, and we show that it is a consequence of the factorization formula above.

\section{Preliminaries}
\label{sec:prelims}

\subsection{Symplectic Geometry}\label{sec:phasespacereview}

In this section, we review the relevant aspects of symplectic geometry that will be important in this paper. For a wonderful and more detailed exposition, we refer the reader to Chapter 20 of \cite{Nair:2005iw}. A more recent review can also be found in \cite{Harlow:2019yfa}.

\subsubsection{Conventions}
\label{sec:symp-geo-conventions}

We start by establishing our conventions for differential forms on a symplectic manifold $\G$. The space of functions on $\G$ is denoted by $\CF(\G)$, and a vector field $\X\in T\G$ can be viewed as the map $\X : \CF(\G) \to \CF(\G)$ defined by $\X(f) = \bi_\X \bdt f$, where $\bdt$ and $\bi_\X$ are the the exterior derivative and interior product on $\G$. This is a derivative map so it satisfies the product rule $\X(fg) = \X(f)g+f \X(g)$. The Lie bracket of two vectors is defined as
\begin{equation}
\begin{split}\label{LieBracketDefinition}
[ \X , \Y ](f) \equiv \X ( \Y ( f ) ) - \Y ( \X ( f ) ). 
\end{split}
\end{equation}

A $q$-form is a completely antisymmetric $q$-linear map $\C_q : T\G \otimes \cdots \otimes T\G \to \CF(\G)$ that takes $q$ vectors as inputs, and we denote it by $\C_q(\X_1,\ldots,\X_q)$. The space of $q$-forms on $\G$ is $\O^q(\G)$, and in the special case where $\bO \in \O^2(\G)$ and $\bT\in \O^1(\G)$, the following identities hold:
\begin{align}
\label{psforms1}
\bi_\X \bO ( \Y ) &= \bO  ( \X,\Y)  \\
\label{psforms2}
\bdt \bT ( \X , \Y ) &= \X ( \bT ( \Y ) ) - \Y ( \bT ( \X ) ) - \bT ( [ \X , \Y ] )  \\
\label{psforms3}
\bdt \bO(\X,\Y,{\bs Z}) &= \X ( \bO ( \Y , \bf{Z}) )  - \bO ( [ \X , \Y ] , {\bf Z} )  + \text{(cyclic in $\X,\Y,{\bf Z}$)}  .
\end{align}

Lastly, the Cartan homotopy formula provides a very useful way of determining the Lie derivative of a differential form:
\begin{equation}
\begin{split}
\label{LieDerivativeIdentity}
{\bs\mathsterling}_\X  = \bdt \bi_\X + \bi_\X \bdt . 
\end{split}
\end{equation}
We will use boldface letters throughout this paper to denote forms and vectors on $\G$, in an effort to distinguish them from spacetime forms and vectors.

\subsubsection{Definitions}
\label{sec:phasespacedef}

A phase space or symplectic manifold $(\G,\bO)$ is a smooth manifold $\G$ on which there exists a closed non-degenerate two-form $\bO$ known as the \underline{symplectic form}:
\begin{equation}
\begin{split}\label{Oclosedandinvertible}
	\text{Closed:} \qquad & \bdt\bO = 0 \\
	\text{Non-degenerate:} \qquad & \bi_\X\bO = 0 ~\implies~ \X =0 \quad\forall\quad \X \in T\G.
\end{split}
\end{equation}
Assuming $\CH^2(\G) = 0$,\footnote{If $\CH^2(\G) \neq 0$, then $\bT$ is not globally defined. In such cases, auxilliary variables are required to describe the action of the theory (which is related to the integral of $\bT$). This is the case for the Wess-Zumino terms in the WZW model.} there exists a one-form $\bT$ known as the \underline{symplectic potential} such that
\begin{equation}
\begin{split}
\bO = \bdt \bT   \quad\implies\quad \bO ( \X , \Y   ) = \X ( \bT ( \Y ) ) - \Y ( \bT ( \X ) ) - \bT ( [ \X , \Y ] ) ,
\label{Odtheta}
\end{split}
\end{equation}
where the implication follows from \eqref{psforms2}. The symplectic potential is defined only up to a closed one-form, but if we also assume $\H^1(\G) = 0$ so that all closed one-forms are exact,\footnote{If $\CH^1(\G) \neq 0$, then the holonomies of $\bT$ around the non-contractible curves become relevant in the quantum theory as vacuum angles. An example of this is the $\t$-vacuum angle in non-abelian gauge theories.} it follows that $\bT$ is defined only up to an exact one-form. As we will see in the Section~\ref{sec:canonical}, such shifts in $\bT$ are related to canonical transformations.

We can think of $\bO$ as a map $\bO : T\G \to \O^1(\G)$ defined via
\begin{equation}
\begin{split}\label{Omegamap}
\bO(\X) \equiv - \bi_\X \bO \quad \implies \quad \bO(\X)(\Y) = - \bO(\Y)(\X) =  - \bO(\X,\Y) . 
\end{split}
\end{equation}
Since $\bO$ is non-degenerate, there exists an inverse map $\bO^{-1}: \O^1(\G) \to T\G$ such that $\bO^{-1} ( \bO(\X)) = \X$ and $\bO ( \bO^{-1} ({\bf C}_1) ) = {\bf C}_1$. The inverse map can also be thought of as an antisymmetric bilinear map acting on one-forms defined as
\begin{equation}
\begin{split}
\bO^{-1} ( {\bf C}_1, {\bf C}'_1 ) \equiv {\bf C}_1( \bO^{-1} ( {\bf C}'_1  ) ) =  - {\bf C}'_1 ( \bO^{-1} ({\bf C}_1 ) ) ,
\end{split}
\end{equation}
and we can easily derive the properties
\begin{equation}
\begin{split}\label{Omegaproperty1}
\bO^{-1} ( \bO(\X) , \bO(\Y) ) = - \bO ( \X , \Y )  , \qquad \bO( \bO^{-1} ({\bf C}_1) , \bO^{-1} ({\bf C}_1') ) =   -  \bO^{-1} ({\bf C}_1 , {\bf C}_1'  )  .
\end{split}
\end{equation}

\subsubsection{Canonical Transformations}
\label{sec:canonical}

Given the geometry of the phase space, diffeomorphisms on $\G$ that preserve the symplectic form $\bO$ are special and are known as symplectomorphisms (in the math community) or canonical transformations (in the physics community). Infinitesimally, these are generated by \underline{Hamiltonian vector fields} $\X_f$ satisfying
\begin{equation}
\begin{split}
{\bs\mathsterling}_{\X_f} \bO = 0 .
\end{split}
\end{equation}
Using \eqref{LieDerivativeIdentity} and the fact that $\bO$ is closed (and that $\CH^1(\G) = 0$), we have
\begin{equation}
\begin{split}\label{HVFsolve}
\bi_{\X_f} \bO  = - \bdt f, \qquad f \in \CF(\G). 
\end{split}
\end{equation}
The function $f$ is known as the \underline{Hamiltonian charge} corresponding to $\X_f$, and because $\bO$ is non-degenerate, the above equation defines $f$ uniquely up to an additive constant. This implies that there is then an invertible map between Hamiltonian vector fields and functions on $\G$ modulo constant shifts. Using \eqref{Omegamap}, the map can be described as
\begin{equation}
\begin{split}\label{canonicaltransformdef}
\bO(\X_f) = \bdt f  \quad \Longleftrightarrow \quad \X_f = \bO^{-1} ( \bdt f )  . 
\end{split}
\end{equation}
This leads to the useful sequence of identities
\begin{equation}
\begin{split}\label{sympidentity1}
\X_f(g) = - \X_g ( f ) = \bO ( \X_f , \X_g ) = - \bO^{-1} ( \bdt f , \bdt g ) =  -  \bi_{\X_f} \bi_{\X_g} \bO . 
\end{split}
\end{equation}
Note that while $\bO$ is preserved under canonical transformations, $\bT$ is not. Rather, using \eqref{LieDerivativeIdentity} and \eqref{HVFsolve}, we find
\begin{equation}\label{symppotcan}
\begin{split}
{\bs\mathsterling}_{\X_f} \bT = \bdt ( \bi_{\X_f} \bT - f  ) . 
\end{split}
\end{equation}
Thus, under canonical transformations, $\bT$ transforms as a $U(1)$ gauge potential and $\bO$ is its ``gauge-invariant'' field strength.

\subsubsection{Poisson Bracket}
\label{sec:posson-bracket}

Let $\X_f$ and $\X_g$ be two Hamiltonian vector fields corresponding to functions $f$ and $g$ respectively. Then $[\X_f,\X_g]$ is also a Hamiltonian vector field since 
\begin{align}
	{\bs\mathsterling}_{[\X_f,\X_g]} \bO = [ {\bs\mathsterling}_{\X_f} , {\bs\mathsterling}_{\X_g} ]  \bO = 0.
\end{align}
Consequently, by \eqref{canonicaltransformdef} there exists a function $h$ such that $\bO ( [ \X_f , \X_g ] ) = \bdt h$. To determine $h$, first act on both sides with $\bi_\Y$ to obtain $\Y(h) =  \bO ( \Y , [ \X_f , \X_g ] )$. Then by closedness of $\bO$ and \eqref{psforms3}, we have
\begin{equation}
\begin{split}
	\bdt \bO ( \X_f , \X_g , \Y ) = 0 \quad \implies \quad \bO ( \Y , [ \X_f , \X_g ]  ) = \Y ( \bO ( \X_f , \X_g ) ) = \Y(h)  . 
\end{split}
\end{equation}
Since this is true for an arbitrary vector field $\Y$, it follows that the Hamiltonian charge corresponding to $[ \X_f , \X_g ]$ is (up to an additive constant)
\begin{equation}
\begin{split}
h =  \bO ( \X_f , \X_g )  \equiv - \{ f , g \}  ,  
\end{split}
\end{equation}
where we have defined the Poisson bracket as
\begin{equation}\label{pb}
\begin{split}
\{ f , g \}  \equiv - \bO ( \X_f , \X_g )  = \bO^{-1} ( \bdt f , \bdt g )  .
\end{split}
\end{equation}
Closedness of $\bO$ implies that the Poisson bracket satisfies the Jacobi identity
\begin{equation}
\begin{split}\label{Poisson-Jacobi}
\{  f , \{ g , h \} \}  + \{  h , \{ f , g \} \}  + \{  g , \{ h , f \} \}  = 0 . 
\end{split}
\end{equation}

\subsection{Covariant Phase Space Formalism}
\label{sec:covphasespace}

The dynamics of a system is typically described in terms of a Lagrangian, and the covariant phase space formalism is a recipe that allows us to construct the phase space of a theory \textit{given} the Lagrangian. In this section, we will review the essential and relevant elements of this formalism.

\subsubsection{Conventions}

We start by establishing our conventions for differential forms on spacetime. Let $(\CM,g)$ be a $d$-dimensional globally hyperbolic Lorentzian spacetime described by coordinates $x^\mu$. We then adopt the conventions
\begin{equation}
\begin{split}\label{standardoperations}
( C_q \w C'_{q'})_{\mu_1 \cdots \mu_{q+q'}} &= \frac{ ( q + q' )! }{ q! q'! } ( C_q )_{[\mu_1 \cdots \mu_q} ( C' _{q'} )_{\mu_{q+1} \cdots \mu_{q+q'} ]}  \\
( i_\xi C_q )_{\mu_1 \cdots \mu_{q-1}} &= \xi^\mu (C_q)_{\mu \mu_1 \cdots \mu_{q-1} }  \\
( \dt C_q )_{\mu_1 \cdots \mu_{q+1}} &= ( q + 1 ) \p_{[ \mu_1} (C_q)_{\mu_2 \cdots \mu_{q+1}]}    \\
( \ast C_q )_{\mu_1 \cdots \mu_{d-q} } &= \frac{1}{q!} \e_{\mu_1 \cdots \mu_{d-q} }{}^{ \nu_1 \cdots \nu_q } (C_q)_{\nu_1 \cdots \nu_q} ,
\end{split}
\end{equation}
where $[\,\cdots]$ denotes the weighted antisymmetrization of indices, e.g. $\o^{[\mu\nu]} = \frac{1}{2!} ( \o^{\mu\nu} - \o^{\nu\mu} )$, and $\e$ is the volume form defined via $\e_{0\cdots d-1} = \sqrt{-\det g}$. Vectors and forms in spacetime are \textit{not} in boldface to distinguish them from the vectors and forms on the phase space $\G$. In this paper, we will assume that $\CH^{d-1}(\CM) = 0$ so all closed $(d-1)$-forms are also exact.

A $q$-form can be integrated over a $q$-dimensional submanifold $\S_q$ of $\CM$. Of particular importance in this paper are  the cases $q=d-1$ and $q=d-2$. In this case,
\begin{equation}
\begin{split}\label{formintegral}
\int_{\S_{d-1}} C_{d-1} = - \int_{\S_{d-1}} d\S_\mu\, ( \ast C_{d-1} )^\mu , \qquad \int_{\S_{d-2}} C_{d-2} = - \frac{1}{2} \int_{\S_{d-2}} dS_{\mu\nu} \, ( \ast C_{d-2} )^{\mu\nu} ,
\end{split}
\end{equation}
where $d\S_\mu$ and $dS_{\mu\nu}$ are the area elements on the surfaces $\S_{d-1}$ and $\S_{d-2}$, respectively. 

For general $q$-forms, Stokes' theorem is
\begin{equation}
\begin{split}\label{StokesTheorem}
\int_{\S_q} \dt C_{q-1} = \oint_{\p \S_q} C_{q-1},
\end{split}
\end{equation}
where the orientation of $\p\S_q$ is outward-directed with respect to $\S_q$. For the special case of $q=d$ and $q=d-1$, we can also express Stokes' theorem as
\begin{equation}
\begin{split}\label{StokesTheorem2}
\int_{\S_d} \e \nabla_\mu C^\mu = \oint_{\p \S_d} d\S_\mu \, C^\mu  , \qquad \int_{\S_{d-1} } d\S_\mu \, \nabla_\nu C^{[\mu\nu]} = \frac{1}{2} \oint_{\p \S_{d-1} } dS_{\mu\nu} \,C^{\mu\nu},
\end{split}
\end{equation}
where $\nabla_\mu$ is the covariant derivative with respect to the metric $g$, and has the standard definition when acting on tensors. To define its action more generally, it is convenient to work with the vielbein $e_\mu^A$, which satisfies $g_{\mu\nu} = \eta_{AB} e_\mu^A e_{\nu}^B$. The introduction of the vielbein (which is necessary if there are spinors in the theory) introduces a new symmetry of the theory, namely local Lorentz symmetry. The basic object in a local field theory is a \underline{field} $\varphi_r$, which transforms in some representation of the local Lorentz symmetry, with $r,s,$ etc. being the vector (internal space) indices in this representation. The covariant derivative $\nabla$ is then defined to act via\footnote{It is important to remember that we are assuming that our fields $\varphi$ carry internal space indices only. If they carry additional tensor indices (with respect to $\text{GL}(d,\mrr)$) then we must modify \eqref{covdevdef} to include the (standard) Christoffel symbol terms. Alternatively, tensor indices may be converted to internal space indices using the vielbein, after which \eqref{covdevdef} can be used.}
\begin{equation}
\begin{split}\label{covdevdef}
\nabla_\mu \varphi_r \equiv \p_\mu \varphi_r + \frac{1}{2} \o_{\mu\rho \s }  ( \S^{\rho \s }  )_r{}^s \varphi_s , \qquad \o_{\mu}{}^{\rho}{}_{\s} \equiv \Gamma^\rho_{\mu\s} - e^\rho_A   \p_\mu e_\s^A,
\end{split}
\end{equation}
where $\o_\mu{}^\rho{}_{\s}$ is the spin connection and $\S^{\rho\s}$ are the generators of the Lorentz algebra in the representation under which $\varphi$ transforms. They satisfy the Lorentz algebra\footnote{\label{vector-rep-lorentz}The generators in the vector and spinor representation are $(\S_\text{vec}^{\mu\nu})_\rho{}^\s =  \d^\mu_\rho g^{\nu\s} -   \d^\nu_\rho g^{\mu\s}$ and $\S_\text{spin}^{\mu\nu} = - \frac{1}{4} [ \c^\mu , \c^\nu ]$ respectively ($\c^\mu$ are the Dirac matrices with $\{ \c^\mu , \c^\nu \} = - 2 g^{\mu\nu}$).}
\begin{equation}
\begin{split}\label{Lorentzalgebra}
\big[ \S^{\mu\nu} , \S^{\rho\s} \big] &= -  \big( g^{\mu\rho} \S^{\nu\s}  - g^{\nu\rho} \S^{\mu\s}   - g^{\mu\s} \S^{\nu\rho} + g^{\nu\s} \S^{\mu\rho}   \big) .
\end{split}
\end{equation}
It is useful to note that the commutator of covariant derivatives takes a simple form
\begin{equation}
\begin{split}\label{covdevantisym}
\left[ \nabla_{\mu} ,  \nabla_{\nu} \right]  \varphi_r = \frac{1}{2} \CR_{\mu\nu\rho\s} ( \S^{\rho\s} )_r{}^s \varphi_s ,
\end{split}
\end{equation}
where $\CR$ is the Riemann tensor
\begin{equation}
\begin{split}\label{Riemann-tensor}
\CR^\rho{}_{\s\mu\nu} &\equiv \p_\mu \o_\nu{}^\rho{}_\s - \p_\nu \o_\mu{}^\rho{}_\s + \o_\mu{}^\rho{}_\tau \o_\nu{}^\tau{}_\s - \o_\nu{}^\rho{}_\tau \o_\mu{}^\tau{}_\s . 
\end{split}
\end{equation}
In the rest of this paper, in order to simplify our notation, we will suppress the internal space indices on the fields.

\subsubsection{Solution Space}
\label{sec:solspace}

A field theory living on $\CM$ is described in terms of dynamical fields $\varphi^i$ and background fields $\mathring{\psi}^I$ ($i$ and $I$ are discrete labels). The \underline{configuration space} $\bs{\mathfrak F}$ is the space of all \textit{allowed} field configurations that are defined by imposing boundary conditions on the fields, \eg we can impose Neumann boundary conditions on the fields on all or part of $\p \CM$ that allows for finite energy flux through those boundaries. Note that each \textit{allowed} field configuration is a point in ${\bs{\mathfrak F}}$.

In the next section, we will elevate a subspace of $\bs{\mathfrak F}$ to a phase space, so all quantities on the phase space will be induced from those on $\bs{\mathfrak F}$. For this reason, we will use the same conventions for vectors and forms on $\bs{\mathfrak F}$ as we did for those on the phase space in Section~\ref{sec:phasespacereview}. A vector $\X \in T {\bs{\mathfrak F}}$ is defined as
\begin{equation}
\begin{split}
\X = \sum_i \int_\CM \e  \X^i \left( \nabla_{\mu_1\cdots\mu_n} \varphi^i \, ; \,  g_{\mu\nu}  , \nabla_{\mu_1 \cdots \mu_n} \CR_{\mu\nu\rho\s} ,  \nabla_{\mu_1 \cdots \mu_n} \mathring{\psi}^I \, ; \,  x  \right) \frac{\d}{\d\varphi^i} . 
\end{split}
\end{equation}
where for all $n \geq0$, $\nabla_{\mu_1 \cdots \mu_n} \equiv \nabla_{(\mu_1} \cdots \nabla_{\mu_n)}$ is the symmetric covariant derivative.\footnote{Antisymmetrized covariant derivatives simplify to the Riemann tensor \eqref{covdevantisym}, so without loss of generality all derivatives can be symmetrized.} Note that in general, the vector components $\X^i$ are functions of the dynamical and background fields, their derivatives, the metric, the Riemann tensor and its derivatives, and may also have an explicit dependence on the coordinates. The vector acts on functions via
\begin{equation}
\begin{split}
\X(f) = \sum_i \int_\CM \e  \X^i \left( \nabla_{\mu_1\cdots\mu_n} \varphi^i \, ; \,  g_{\mu\nu}  , \nabla_{\mu_1 \cdots \mu_n} \CR_{\mu\nu\rho\s} ,  \nabla_{\mu_1 \cdots \mu_n} \mathring{\psi}^I \, ; \,  x  \right) \frac{\d f}{\d\varphi^i},
\end{split}
\end{equation}
and we refer to $\X(f)$ as the ``variation of $f$ with respect to $\X$.''

The dynamics of a system can oftentimes be conveniently described by a \underline{Lagrangian form} $L$, which is a $d$-form on $\CM$ and a function on ${\bs{\mathfrak F}}$, i.e.
\begin{equation}
\begin{split}
L  = L \left( \nabla_{\mu_1\cdots\mu_n} \varphi^i \, ; \,  g_{\mu\nu}  , \nabla_{\mu_1 \cdots \mu_n} \CR_{\mu\nu\rho\s} ,  \nabla_{\mu_1 \cdots \mu_n} \mathring{\psi}^I \right) \in \O^d(\CM) \times \CF(\bs{\mathfrak F}) .
\end{split}
\end{equation}
Note that in a local theory, the Lagrangian does not have an explicit dependence on the coordinates. The Lagrangian form is related to the more familiar \underline{Lagrangian density} $\CL$ via
\begin{equation}
\begin{split}\label{Lexp-form}
L = \e \, \CL\left( \nabla_{\mu_1\cdots\mu_n} \varphi^i \, ; \,  g_{\mu\nu}  , \nabla_{\mu_1 \cdots \mu_n} \CR_{\mu\nu\rho\s} ,  \nabla_{\mu_1 \cdots \mu_n} \mathring{\psi}^I \right) . 
\end{split}
\end{equation}
The Lagrangian density is invariant under local Lorentz transformations, which implies 
\begin{equation}
\begin{split}
\label{local-Lorentz-inv}
\sum_i \sum_{n=0}^\infty  \Pi_i^{\mu_1 \cdots \mu_n} \big( \S_i^{\mu\nu} \big)_{\mu_1 \cdots \mu_n}{}^{\nu_1 \cdots \nu_n} \nabla_{\nu_1 \cdots \nu_n} \varphi^i = 0 , \qquad \Pi_i^{\mu_1\cdots\mu_n} \equiv \pd{\CL }{  ( \nabla_{\mu_1 \cdots \mu_n} \varphi^i ) } .
\end{split}
\end{equation}
We now consider the variation of $L$ with respect to a vector $\X\in T{\bs{\mathfrak F}}$. Using the explicit form of the Lagrangian \eqref{Lexp-form} and the fact that $\X$ acts only on dynamical fields, we can write
\begin{equation}
\begin{split}\label{XL}
\X(L) &= \e  \, \X(\CL) = \e \sum_i \sum_{n=0}^\infty \Pi_i^{\mu_1\cdots\mu_n} \nabla_{\mu_1} \cdots \nabla_{\mu_n} \X(\varphi^i) .
\end{split}
\end{equation}
We simplify this further using ``integration by parts''-style manipulations (IBP) to remove all the derivatives from $\X(\varphi^i)$,\footnote{This implies replacing $a (Db) \to D(ab) - (D a) b$ for any derivative operator $D$ and any quantities $a,b$.} and the total derivative terms obtained in the process can then be absorbed into a boundary term. To see this explicitly, note that we can use IBP on the $n\geq 1$ terms in \eqref{XL} to obtain 
\begin{equation}
\begin{split}
	\X(L) &= \e\sum_i \Pi_i\X(\varphi^i) + \e \nabla_{\mu_1}    \sum_i \sum_{n=1}^\infty \Pi_i^{\mu_1\cdots\mu_n} \nabla_{\mu_2} \cdots \nabla_{\mu_n} \X(\varphi^i)  \\
	&\qquad - \e \sum_i \sum_{n=1}^\infty \nabla_{\mu_1}  \Pi_i^{\mu_1\cdots\mu_n} \nabla_{\mu_2} \cdots \nabla_{\mu_n} \X(\varphi^i) .
\end{split}
\end{equation}
Applying IBP again to the third term and noting $\Pi_i^{\mu_1\cdots\mu_n}$ is symmetric in its indices, we get
\begin{equation}
\begin{split}
	\X(L) &= \e\sum_i \Pi_i\X(\varphi^i) +  \e  \sum_i \sum_{n=1}^\infty \nabla_{\mu_1}\nabla_{\mu_2} \Pi_i^{\mu_1\cdots\mu_n}  \nabla_{\mu_3} \cdots \nabla_{\mu_n} \X(\varphi^i) \\
&\qquad + \e \nabla_{\mu_1} \sum_i  \sum_{n=1}^\infty \Big( \Pi_i^{\mu_1\cdots\mu_n} \nabla_{\mu_2} \cdots \nabla_{\mu_n} \X(\varphi^i) - \nabla_{\mu_2}  \Pi_i^{\mu_1\cdots\mu_n} \nabla_{\mu_3} \cdots \nabla_{\mu_n} \X(\varphi^i) \Big) . 
\end{split}
\end{equation}
Continuing in this fashion until the only terms involving a derivative of $\X(\varphi^i)$ are total derivatives, we get
\begin{equation}
\begin{split}
	\X(L) &= \e  \sum_i \sum_{n=0}^\infty (-1)^n \nabla_{\mu_1\cdots\mu_n} \Pi_i^{\mu_1\cdots\mu_n} \X(\varphi^i) \\
	&\qquad - \e \nabla_{\mu_1}  \sum_i \sum_{n=1}^\infty \sum_{k=1}^n (-1)^k \nabla_{\mu_2 \cdots \mu_k} \Pi_i^{\mu_1\cdots\mu_n} \nabla_{\mu_{k+1} \cdots \mu_n} \X(\varphi^i) ,
\end{split}
\end{equation}
where we used $\Pi_i^{\mu_1\cdots\mu_n}$ to symmetrize the derivatives. In the language of differential forms, this result can be written as
\begin{equation}
\begin{split}\label{lagvar}
\X ( L )  =  \sum_i \CE_i \X ( \varphi^i) + \dt  \bt(\X), 
\end{split}
\end{equation}
where
\begin{align}
\label{eom1}\ast\,\CE_i &= - \sum_{n=0}^\infty (-1)^n \nabla_{\mu_1 \cdots \mu_n} \Pi_i^{\mu_1 \cdots \mu_n}  \\
\label{spcd1} [\ast \bt (\X)]^\mu &=  \sum_i \sum_{n=1}^\infty \sum_{k=1}^n (-1)^k \nabla_{\mu_2 \cdots \mu_k} \Pi_i^{\mu \mu_2 \cdots \mu_n} \nabla_{\mu_{k+1} \cdots \mu_n} \X(\varphi^i).
\end{align}
Note that the equations of motion (i.e. the Euler-Lagrange equations) of the theory are
\begin{equation}
\begin{split}\label{Sspacedef}
\CE_i = 0 . 
\end{split}
\end{equation}
The subspace of $\bs{\mathfrak F}$ defined by the equations above is known as the \underline{solution space} ${\bs\CS}$. The tangent bundle $T {\bs \CS}$ consists of vector fields $\X$ satisfying the linearized equations of motion, i.e.
\begin{equation}
\begin{split}\label{TSdef}
\X \in T {\bs \CS} \quad \iff \quad \X ( \CE_i )  = 0. 
\end{split}
\end{equation}
We shall henceforth work exclusively on the solution space, and field configurations that live in ${\bs\CS}$ are said to be \underline{on-shell}.

\subsubsection{Symplectic Form}
\label{sec:sympform}

To elevate the solution space to a phase space, we need to construct the symplectic form. Note that $\X(L)$ is a $d$-form on $\CM$ but a function on $\CS$. Consequently, $\bt$ is a $(d\!-\!1)$-form on $\CM$ and a one-form on $\CS$, i.e.
\begin{align}
	\bt \in \O^{d-1}(\CM) \times \O^1(\bs\CS).
\end{align}
$\bt$ is known as the \underline{symplectic potential current density}. Note that \eqref{lagvar} defines $\bt$ only up to a closed and hence exact form (since $\CH^{d-1}(\CM)=0$) in spacetime, i.e.
\begin{equation}
\begin{split}\label{thetaamb}
\bt \to \bt + \dt  \bk , \qquad \bk \in \O^{d-2} (\CM)  \times  \O^1({\bs\CS}) .
\end{split}
\end{equation}
Next, we define the \underline{symplectic current density} as the exterior derivative of $\bt$ on ${\bs\CS}$, i.e.
\begin{equation}
\begin{split}\label{omegadef}
\bo = \bdt \bt \in \O^{d-1} (\CM ) \times \O^2({\bs\CS}) .
\end{split}
\end{equation}
By construction, $\bo$ is closed in ${\bs\CS}$. However, it is also closed in $\CM$. To see this, note that from \eqref{lagvar} and \eqref{Sspacedef},
\begin{equation}
\begin{split}\label{dthetaid}
\dt  \bt (\X) = \X(L) . 
\end{split}
\end{equation}
By using the definition of exterior derivative \eqref{psforms2}, we find as promised
\begin{equation}
\begin{split}
\label{domegaid}
\dt  \bo(\X,\Y) = \X(\Y(L)) - \Y(\X(L)) - [ \X , \Y ] (L)  = 0 . 
\end{split}
\end{equation}
The \underline{pre-symplectic potential} and \underline{pre-symplectic form} are obtained by integrating $\bt$ and $\bo$, respectively, over a Cauchy slice $\S$ (which is a $(d-1)$-dimensional spacelike submanifold of $\CM$ whose domain of dependence is the entire spacetime $\CM$):
\begin{equation}
\begin{split}\label{presympdef}
\wt \bT_\S (\X) &= \int_\S \bt(\X)  , \qquad 
\wt \bO_\S( \X , \Y ) = \int_\S \bo ( \X , \Y )  , 
\end{split}
\end{equation}
where the orientation of $\S$ is taken so that the normal vector to $\S$ is future-directed. Note that
\begin{equation}
\begin{split}\label{presymp1}
\wt \bT_\S \in \O^1({\bs\CS}) , \qquad \wt \bO_\S = \bdt \wt \bT_\S \in \O^2({\bs\CS}) ,
\end{split}
\end{equation}
so they are candidates for the symplectic potential and symplectic form, respectively. By construction, the pre-symplectic form is closed, but it is not generically non-degenerate. We can remedy this by factoring ${\bs\CS}$ into the degeneracy subspaces of $\wt \bO_\S$ as follows. For each $\X_0 \in \ker \wt \bO_\S$ and a point $\varphi \in \bs\CS$, we construct the curve in $\bs\CS$ to which $\X_0$ is tangent. The equivalence relation $\sim$ on $\bs\CS$ is defined by identifying all the points on this curve, and the phase space is then $\G \equiv {\bs\CS}/\!\!\sim$. By construction, the restriction of the pre-symplectic form to $\G$ is both closed (but not necessarily exact) in $\G$ and non-degenerate, so the symplectic potential and form on $\G$ are
\begin{equation}
\begin{split}\label{inducedsymp1211}
\bT_\S = \wt \bT_\S \big|_\G , \qquad  \bO_\S = \wt \bO_\S \big|_\G . 
\end{split}
\end{equation}
A useful way to define $\G$ is to impose a gauge condition of the form
\begin{equation}
\begin{split}
\label{gaugecond-gen}
G[\varphi] = 0,
\end{split}
\end{equation}
which uniquely maps each equivalence class of $\sim$ to a particular representative element. To be precise, the gauge condition must be chosen so that for every $\varphi \in {\bs\CS}$, there exists a \emph{unique} solution ${\overline \varphi}$ such that $\varphi \sim {\overline \varphi}$ and $G[{\overline\varphi}] = 0$. In this way, we can define $\G$ as a subspace of ${\bs\CS}$. Often, a convenient choice of the gauge condition can dramatically simplify calculations, and we will make such a convenient choice when we study gauge theories in flat spacetime in Section~\ref{sec:gaugetheories-flatspacetime}. 

This completes the construction of the phase space on a Cauchy slice $\S$. Having constructed the phase space and symplectic form, we can now use the ideas developed in Section~\ref{sec:phasespacereview} to discuss canonical transformations and the Poisson bracket. We recall here two formulae regarding canonical transformations that will be useful in the remainder of this paper:
\begin{equation}
\begin{split}\label{mainformulae}
\bO_\S (\Y,\X_f) = \Y(f[\S]) , \qquad \big\{ f[\S] , g[\S] \big\}_\S = - \bO_\S(\X_f, \X_g) ,
\end{split}
\end{equation}
where $f[\S]$ is the Hamiltonian charge generating $\X_f$ on a given $\S$, and $\{\cdot,\cdot\}_\S$ is the associated Poisson bracket on $\S$.

\subsubsection{Isometries}
\label{sec:isometries}

In a local field theory, there is a special class of transformations on the phase space known as isometries. These act on the fields of the theory via the Lie derivative:
\begin{equation}
\begin{split}\label{isometrydef}
\X_\xi = \int_\CM \e  \, \pounds_\xi \varphi^i \frac{\d}{\d\varphi^i}, 
\end{split}
\end{equation}
where
\begin{align}\label{killing}
	(\pounds_\xi g)_{\mu\nu} = 2 \nabla_{(\mu} \xi_{\nu)} = 0  , \qquad \pounds_\xi \mathring{\psi}^I = 0 .
\end{align}
The action of the Lie derivative on the dynamical fields is 
\begin{equation}
\begin{split}\label{Liederdef}
\pounds_\xi \varphi^i \equiv \xi^\mu \nabla_\mu \varphi^i + \frac{1}{2} \nabla_{[\mu} \xi_{\nu]} \S_i^{\mu\nu} \varphi^i ,
\end{split}
\end{equation}
where we recall that $\S_i^{\mu\nu}$ is the Lorentz generator in the representation under which $\varphi^i$ transforms. A similar formula holds for the background fields as well. 

Vector fields $\xi$ satisfying the first of the two equations in \eqref{killing} are known as \underline{Killing vector fields}, and they generate isometry transformations. The second equation then imposes a further restriction and only selects those Killing vectors that preserve all the boundary fields. Generically, vector fields that satisfy \eqref{isometrydef} cannot depend on the dynamical fields, so $\Y(\xi) = 0$ for all $\Y \in T \G$. It follows that $[\Y,\X_\xi] = 0$.\footnote{If $f$ is a local function of the fields, then $\X_\xi(f) = \pounds_\xi f$, implying that $[\Y,\X_\xi] (f) = \pounds_{\Y(\xi)} f = 0$. The result can then be trivially extended to non-local functions that are integrals of local functions.} From this, we have
\begin{equation}
\begin{split}
\bo(\Y,\X_\xi) &= \Y ( \bt (\X_\xi) )  - \X_\xi ( \bt ( \Y ) ) = \Y ( \bt (\X_\xi) )  - \dt i_\xi  \bt ( \Y ) - i_\xi \dt \bt (\Y) ,
\end{split}
\end{equation}
where we used $\X_\xi(\bt(\Y)) = \pounds_\xi \bt(\Y)$ since $\bt(\Y)$ is a local function of the fields (see \eqref{spcd1}). Using \eqref{lagvar} with the on-shell condition \eqref{Sspacedef}, we have
\begin{equation}
\begin{split}
\bo(\Y,\X_\xi) &= \Y ( \bt (\X_\xi) - i_\xi L  )  - \dt i_\xi  \bt ( \Y ) ,
\end{split}
\end{equation}
and integrating this over $\S$, we get
\begin{equation}
\begin{split}\label{isometry-can}
\bO_\S(\Y,\X_\xi) = \Y ( H_\xi [\S] )  - \oint_{\p\S} i_\xi \bt(\Y) , \qquad H_\xi[\S] = \int_\S ( \bt (\X_\xi) - i_\xi L ) .
\end{split}
\end{equation}
Thus, we see by \eqref{mainformulae} that up to an extra term boundary, isometry transformations are canonical transformations with Hamiltonian charge $H_\xi[\S]$, i.e. the \underline{isometry charge}. This means that if we wish to have a phase space on which isometry transformations are canonical, additional restrictions that eliminate the boundary term must be imposed on the fields.

To finish this section, we will determine the explicit form of the isometry charge. Using \eqref{formintegral}, the isometry charge can be written as
\begin{equation}
\begin{split}\label{isom-ch-1}
H_\xi[\S] &= - \int_\S d\S_\mu\, ( \ast \bt (\X_\xi) - \ast  i_\xi L )^\mu .
\end{split}
\end{equation}
Using \eqref{spcd1} and \eqref{Liederdef}, the integrand can be written as
\begin{equation}
\begin{split}\label{asttheta-isometry}
( \ast \bt (\X_\xi) - \ast i_\xi L )^\mu &= \CA^{\mu\nu} \xi_\nu + \CB^{\mu\nu\rho} \nabla_{[\nu} \xi_{\rho]} ,
\end{split}
\end{equation}
where
\begin{equation}
\begin{split}\label{ABdef}
\CA^{\mu\nu} &\equiv  g^{\mu\nu} \CL +  \sum_i  \sum_{n=1}^\infty \sum_{k=1}^n (-1)^k \nabla_{\mu_2 \cdots \mu_k} \Pi_i^{\mu \mu_2 \cdots \mu_n}  \nabla^\nu \nabla_{\mu_{k+1} \cdots \mu_n} \varphi^i  \\
\CB^{\mu\nu\rho} &\equiv \frac{1}{2}  \sum_i \sum_{n=1}^\infty \sum_{k=1}^n (-1)^k \nabla_{\mu_2 \cdots \mu_k} \Pi_i^{\mu \mu_2 \cdots \mu_n} \big( \S_i^{\nu\rho} \big)_{\mu_{k+1} \cdots \mu_n}{}^{\nu_{k+1} \cdots \nu_n}\nabla_{\nu_{k+1} \cdots \nu_n} \varphi^i .
\end{split}
\end{equation}
Note that $\CB^{\mu\nu\rho}$ is antisymmetric in its last two indices. As we show in Appendix \ref{app:id12proof}, $\CA^{\mu\nu}$ and $\CB^{\mu\nu\rho}$ satisfy on-shell the identities
\begin{equation}
\begin{split}
\label{id1} 
\nabla_\mu \CA^{\mu\nu} +  \CR^\nu{}_{\mu\rho\s} \CB^{\mu\rho\s} = 0, \qquad  \CA^{[\mu\nu]}  + \nabla_\rho \CB^{\rho\mu\nu} = 0  .
\end{split}
\end{equation}
We now define the following quantities
\begin{equation}
\begin{split}\label{st-def}
T^{\mu\nu} &\equiv \CA^{(\mu\nu)} - 2  \nabla_\rho \CB^{(\mu|\rho|\nu)} , \qquad ( \ast \SH_\xi )^{\mu\nu} \equiv ( 2 \CB^{[\mu\nu]\rho} - \CB^{\rho\mu\nu}  )  \xi_\rho .
\end{split}
\end{equation}
$T^{\mu\nu}$ is symmetric by construction, and we define $T^{\mu\nu}$ to be the \underline{stress tensor} of the theory. Indeed, to show that it is conserved, note that by \eqref{id1} we have
\begin{equation}
\begin{split}
\nabla_\mu T^{\mu\nu} &= \nabla_\mu \CA^{(\mu\nu)} - 2  \nabla_\mu \nabla_\rho B^{(\mu|\rho|\nu)}  \\
	&= -  \CR^\nu{}_{\mu\rho\s} \CB^{\mu\rho\s} + \left[ \nabla_\mu , \nabla_\rho \right] \CB^{\rho\mu\nu} + \frac{1}{2} \left[ \nabla_\mu ,  \nabla_\rho \right] B^{\nu\mu\rho} .
\end{split}
\end{equation}
Using \eqref{covdevantisym} to write the covariant derivative commutators in terms of the Riemann tensor and then utilizing the symmetries of the Riemann tensor, the above expression vanishes.

Using the identities \eqref{id1}, we can now express \eqref{asttheta-isometry} as
\begin{equation}
\begin{split}\label{eq264}
( \ast \bt (\X_\xi) - \ast i_\xi L )^\mu &= T^{\mu\nu} \xi_\nu  + \nabla_\nu  ( \ast \SH_\xi )^{\mu\nu}  .
\end{split}
\end{equation}
Substituting this into \eqref{isom-ch-1}, we obtain
\begin{equation}
\begin{split}\label{isometrycharge}
H_\xi[\S] &= - \int_\S d\S_\mu\, T^{\mu\nu} \xi_\nu  + \oint_{\p\S} \SH_\xi . 
\end{split}
\end{equation}
The first term is the well-known form of the isometry charge, but the calculation here shows that there is an additional boundary contribution to the charge. Of course, it is important to remember that the charge generates the appropriate transformations if and only if the boundary term in \eqref{isometry-can} vanishes on the phase space.

\subsubsection{Boundary Ambiguities}
\label{sec:sympambiguities}

We noted previously in \eqref{thetaamb} that $\bt$ is not uniquely fixed by the Lagrangian and is ambiguous up to an exact form. This implies a corresponding ambiguity in the symplectic potential and form:
\begin{equation}
\begin{split}
\bT_\S(\X) &\to \bT_\S(\X) + \oint_{\p\S} \bk(\X) , \qquad \bO_\S(\X,\Y) \to \bO_\S(\X,\Y) + \oint_{\p\S} \bdt \bk(\X,\Y) . 
\end{split}
\end{equation}
This ambiguity only modifies the symplectic structure by a boundary term. Strictly speaking, as $\bk$ is not fixed by the Lagrangian, we need extra information to determine it and define the phase space uniquely. Without such additional input, it is natural to consider the most general $\bk$ allowed by locality. However, while such a generalization is interesting, it is outside the scope of this paper, and we leave this for future work.

\subsubsection{Dependence on \texorpdfstring{$\Sigma$}{}}
\label{sec:Sigmadep}

Thus far, the phase space we constructed depends on the choice of Cauchy surface $\S$. To study the dependence of the symplectic potential and symplectic form on $\S$, let $\S$ and $\S'$ be two different Cauchy slices such that they, together with a time-like surface $\CB$, form the boundary of a region $\CV$, as shown in the figure.
\begin{center}
\includegraphics[scale=0.35]{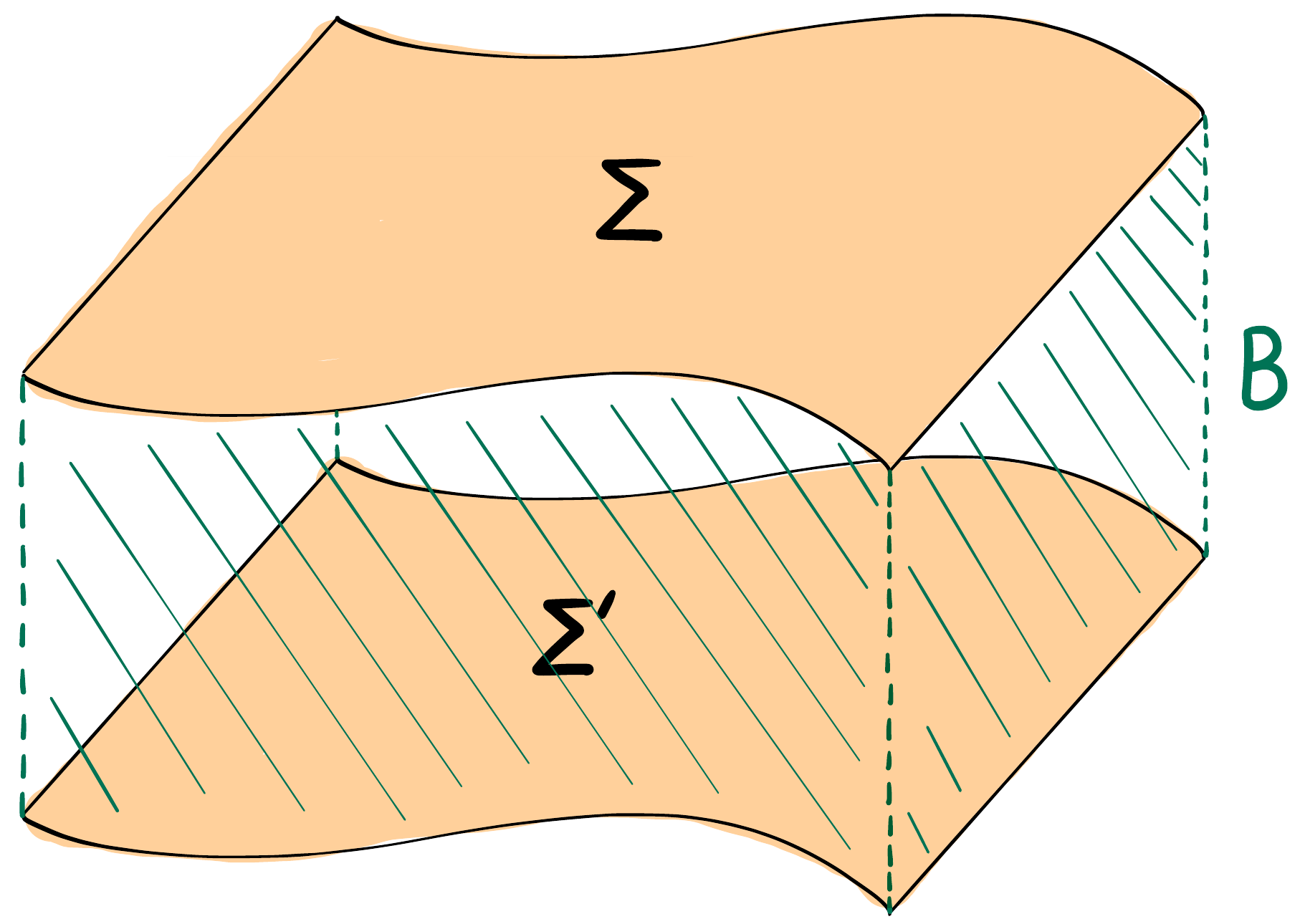}
\end{center}
Integrating \eqref{dthetaid} and \eqref{domegaid} over the region $\CV$ and using Stokes' theorem, we find
\begin{align}
\label{Sdep1}\bT_\S(\X) - \bT_{\S'} (\X)  &= - \bT_\CB(\X) +   \X \left( \int_\CV L \right)   \\
\label{Sdep2}\bO_\S (\X,\Y) - \bO_{\S'} (\X,\Y) &= -  \bO_\CB(\X,\Y) ,
\end{align}
where the sign for $\bT_{\S'}$ and $\bO_{\S'}$ differs from the others since its outward-directed normal vector with respect to $\CV$ is past-directed rather than future-directed. Due to the contribution from the boundary $\CB$ in the equation above, the symplectic form on $\S$ and $\S'$ are in general not equal, which means the deformation $\S \to \S'$ is not a canonical transformation. Since the Hamiltonian charges for canonical transformations are constructed using the symplectic form, this implies that generically Hamiltonian charges on $\S$ and $\S'$ are not equal, i.e. they are not conserved.

\section{Classical Gauge Theories at Null Infinity} \label{sec:classicalgauge}

In this section, we will utilize the covariant phase space formalism to construct the phase space of gauge theories on null infinity. We can then use the ideas developed in Section~\ref{sec:phasespacereview} to construct the Poisson bracket and canonical transformations.

Let us begin by introducing some Lie algebra notations. We are interested in non-abelian gauge theories with compact semi-simple gauge group $\CG$ associated to a Lie algebra $\mfg$. It is possible to choose a basis of generators $X^a$ on $\mfg$ such that
\begin{equation}
\begin{split}
[ X^a , X^b ] = f^{abc} X^c,
\end{split}
\end{equation}
where $f^{abc} \in \mrr$ are known as \underline{structure constants} and satisfy the Jacobi identity
\begin{align}\label{jacobi}
	 f^{d[ab} f^{c]de} = 0.
\end{align}
Note that the sum over repeated indices is implied (because indices are raised and lowered with $\delta^{ab}$, we do not distinguish between raised and lowered indices). A general element of the Lie algebra can be expanded in this basis as $\ve = \ve^a X^a \in \mfg$, and elements in the identity component of the Lie group $\CG_0$ are obtained by exponentiating Lie algebra elements, i.e. $\g = \exp \ve \in \CG_0$.

Finite-dimensional unitary irreducible representations of $\CG$ (and consequently $\mfg$) are denoted by $R_i : \CG \to V^*_i$ ($i$ labels the representation), where $V_i$ is a vector space with elements $\Phi^i$. The generators in a representation $R_i$ are denoted by $T^a_i = R_i(X^a)$, and they satisfy
\begin{equation}\label{basisprop}
\begin{split}
(T^a_i)^\ct = - T^a_i , \qquad \text{tr}_{V_i}\big[ T^a_i T^b_i \big] = - \CT_i \d^{ab} , 
\end{split}
\end{equation}
where $\CT_i > 0$ is known as the index of the representation, and the superscript ``$\ct$'' is used instead of $\dagger$ to denote the conjugate transpose since the latter is reserved later for the adjoint of a quantum operator.\footnote{Assuming the Lie algebra is real, i.e. $\ve^a \in \mrr$, \eqref{basisprop} implies $R_i(g)^\ct = R_i(g)^{-1}$, which means our representation is indeed unitary.} The trace in the definition above depends on the index of the representation, but we can define a representation-independent trace on $\mfg$ via
\begin{equation}\label{rep-indep-trace}
\begin{split}
\text{tr}\big[ X^a X^b \big] = - \frac{1}{\CT_i} \text{tr}_{V_i}\big[ T^a_i T^b_i \big]  = \d^{ab} . 
\end{split}
\end{equation}

A particularly important representation is the adjoint representation, in which the matrix elements of the generators are
\begin{equation}
\begin{split}
	(t^a)^{bc} \equiv \big[\adj(X^a)\big]^{bc} = - f^{abc}  . 
\end{split}
\end{equation}
Note that these generators satisfy $(t^a)^T = - t^a$, as is required by \eqref{basisprop} and the fact that $f^{abc} \in \mrr$. Furthermore, normalizing these generators so that $\CT_\adj = 1$, we have
\begin{equation}\label{adjprop}
\begin{split}
f^{acd} f^{bcd}  = \d^{ab} .
\end{split}
\end{equation}
Lastly, we observe that for all $\g \in \CG$, we have
\begin{equation}
\begin{split}
\label{usefulprop}
\adj(\g)^T = \adj(\g)^{-1} , \qquad R_i(\g^{-1}) T^a_i R_i(\g)  =   [\adj(\g)]^{ab} T_i^b.
\end{split}
\end{equation}

\subsection{The Phase Space}
\label{sec:phasespacegeneric}

The configuration space $\bs{\mathfrak F}$ of a gauge theory is labeled by a Lie algebra-valued one-form gauge field $A = A^a_\mu\, \dt x^\mu \otimes X^a$ and a set of matter fields $\Phi^i \in V_i$, $i=1,\ldots,N$. The field strength $F$ is defined as
\begin{equation}
\begin{split}
F = \dt  A + A\w A = \frac{1}{2} \left[ \p_\mu A_\nu^a - \p_\nu A_\mu^a + f^{abc} A_\mu^b A_\nu^c \right] \dt x^\mu \w \dt x^\nu \otimes X^a . 
\end{split}
\end{equation}
It is convenient to define a \underline{gauge covariant derivative} $\text D$ that acts on adjoint valued $q$-forms $C_q$ and matter fields as
\begin{equation}\label{covD1}
\begin{split}
 \text D C_q \equiv \dt C_q + A \w C_q - (-1)^q C_q \w A , \qquad \D_\mu \Phi^i \equiv \nabla_\mu \Phi^i + R_i(A_\mu) \Phi^i  .
\end{split}
\end{equation}
Note that $\text D$ has the properties 
\begin{equation}
\begin{split}
\text D^2 C_q &= F \w C_q  -  C_q \w F  , \qquad [ \D_\mu , \D_\nu ] \Phi^i = \frac{1}{2} \CR_{\mu\nu\rho\s} \S_i^{\rho\s} \Phi^i + R_i (F_{\mu\nu}) \Phi^i . 
\end{split}
\end{equation}
We consider theories in which the Lagrangian density $\CL$ is a polynomial function of the field strength, matter fields, and their symmetrized gauge covariant derivatives, so that the spacetime $d$-form Lagrangian $L$ is\footnote{This forbids Chern-Simons type terms in the Lagrangian, so they must be considered separately.}
\begin{equation}
\begin{split}
\label{genformLag}
L = \e \, \CL \left( \D_{\a_1\cdots\a_n} F_{\mu\nu} , \D_{\a_1\cdots\a_n} \Phi^i  , \D_{\a_1\cdots\a_n} (\Phi^i)^\ct  \right) ,
\end{split}
\end{equation}
where for all $n \geq 0$, $\D_{\a_1 \a_2 \cdots \a_n}  \equiv \D_{(\a_1} \cdots \D_{\a_n)}$ denotes the symmetrized gauge covariant derivatives. Note that $L$ only depends implicitly on the gauge field $A$ through the field strength and covariant derivative, and we assume for simplicity that $L$ is independent of background fields (but there is implicit dependence on the metric). 

The Lagrangian is invariant under the gauge transformations 
\begin{equation}
\begin{split}\label{finitegaugetransform}
A  \to   \g A \g^{-1} +\g \dt \g^{-1}  , \qquad \Phi^i \to R_i(\g) \Phi^i  , \qquad \g \in \CG. 
\end{split}
\end{equation}
Infinitesimal gauge transformations are generated by the vector field
\begin{equation}
\begin{split}\label{Xepdef}
\X_\ve = \int_\CM \e \left[ - \D_\mu \ve^a \frac{\d}{\d A_\mu^a} + \sum_{i=1}^N \left(  - \ve^a (\Phi^i)^\ct T^a_i  \frac{\d}{\d (\Phi^i)^\ct}  + \hc  \right) \right] \in T \bs{\mathfrak{F}},
\end{split}
\end{equation}
where $\hc$ denotes the conjugate transposed terms. Thus, the variation with respect to $\X_\ve$ is
\begin{equation}
\begin{split}\label{infgaugetransform}
	\X_\ve(A) = - \text D \ve  ,  \qquad  \X_\ve (  \Phi^i )  = R_i(\ve) \Phi^i ,
\end{split}
\end{equation}
and the invariance of the Lagrangian under infinitesimal gauge transformations implies $\X_\ve ( L ) = 0$. 

Now, the variation of the Lagrangian with respect to $\X \in T \bs{\mathfrak F}$ in general takes the form
\begin{equation}
\begin{split}\label{varaction}
\X( L ) =   \tr{  \X ( A )  \w   \CE^A  } + \sum_{i=1}^N \Big( ( \CE_i^\Phi )^\ct \X ( \Phi^i ) + \hc  \Big) + \dt \bt(\X) .
\end{split}
\end{equation}
As we derive in Appendix~\ref{app:explicitcalculations1}, if we define
\begin{equation}
\begin{split}\label{Pi-def}
\Pi^{\a_1 \cdots \a_n ; \mu\nu} \equiv \frac{\p \CL}{ \p ( \D_{\a_1 \cdots \a_n} F_{\mu\nu} ) } , \qquad \Pi_i^{\a_1 \cdots \a_n} \equiv \frac{\p \CL}{ \p ( \D_{\a_1 \cdots \a_n} \Phi^i ) }  ,
\end{split}
\end{equation}
then the equations of motion are
\begin{equation}
\begin{split}\label{eom-explicit}
(\ast \CE^A)^\mu &= - 2 \sum_{n=0}^\infty (-1)^n \D_\nu \D_{\a_1\cdots \a_n} \Pi^{\a_1 \cdots \a_n ; \mu\nu}  \\
&\qquad \qquad  - \sum_{n=1}^\infty  \sum_{k=1}^n (-1)^k  \left[  \D_{\a_2\cdots \a_k}  \Pi^{\mu \a_2 \cdots \a_n ; \a\b},   \D_{\a_{k+1} \cdots \a_n} F_{\a\b} \right] \\
& \qquad \qquad + \sum_{i=1}^N \sum_{n=1}^\infty \sum_{k=1}^n (-1)^k  \left( \D_{\a_n \cdots \a_k}  \Pi^{\mu \a_2 \cdots \a_n}_i T^a_i  \D_{\a_{k+1} \cdots \a_n}\Phi^i + \hc \right) X^a  \\
\ast ( \CE^{\Phi}_i )^\ct &= - \sum_{n=0}^\infty (-1)^n  \D_{\a_1\cdots\a_n}  \Pi^{\a_1 \cdots \a_n}_i ,
\end{split}
\end{equation}
and the symplectic potential current density is
\begin{equation}
\begin{split}\label{spcd-explicit}
[\ast \bt (\X)]^\mu &= - 2 \sum_{n=0}^\infty (-1)^n \tr{  \D_{\a_1\cdots \a_n} \Pi^{\a_1 \cdots \a_n ; \mu\nu} \X(A_\nu)  }  \\
&\qquad \qquad  +  \sum_{n=1}^\infty \sum_{k=1}^n (-1)^k  \tr{  \D_{\a_2 \cdots \a_k }  \Pi^{\mu \a_2 \cdots \a_n ; \a\b} \X ( \D_{\a_{k+1} \cdots \a_n} F_{\a\b} ) } \\
&\qquad \qquad + \sum_{i=1}^N \sum_{n=1}^\infty \sum_{k=1}^n (-1)^k \left( \D_{\a_2 \cdots \a_k}  \Pi^{\mu \a_2 \cdots \a_n}_i  \X(\D_{\a_{k+1} \cdots \a_n}\Phi^i) + \hc \right) .
\end{split}
\end{equation}
The equations of motion $\CE^A = \CE_i^\Phi = 0$ define the solution space ${\bs\CS}$, and vectors in the tangent bundle $T{\bs\CS}$ satisfy the equations $\X(\CE^A) = \X(\CE_i^\Phi) = 0$.

Using \eqref{omegadef} and \eqref{psforms2}, it follows that the symplectic current density is
\begin{equation}
\begin{split}
\label{omega-explicit}
[\ast \bo(\X,\Y)]^\mu  &=  - 2 \sum_{n=0}^\infty (-1)^n \tr{ \X  ( \D_{\a_1 \cdots \a_n} \Pi^{\a_1 \cdots \a_n ; \mu\nu}   )  \Y (A_\nu ) }  \\
&\qquad \qquad + \sum_{n=1}^\infty \sum_{k=1}^n (-1)^k \tr{ \X  ( \D_{\a_2\cdots \a_k} \Pi^{\mu \a_2 \cdots \a_n ; \a\b}    )  \Y ( \D_{\a_{k+1} \cdots \a_n } F_{\a\b} ) }  \\
&\qquad \qquad + \sum_{i=1}^N  \sum_{n=1}^\infty  \sum_{k=1}^n (-1)^k \left[ \X  ( \D_{\a_2 \cdots \a_k } \Pi_i^{\mu \a_2 \cdots \a_n}  )  \Y (  \D_{\a_{k+1} \cdots \a_n } \Phi^i ) + \hc \right] \\
&\qquad \qquad - (\X \leftrightarrow \Y) . 
\end{split}
\end{equation}
Integrating $\bt$ and $\bo$ over a Cauchy slice $\S$ of $\CM$ yields
\begin{equation}
\begin{split}\label{integrate-over-Sigma}
\wt \bT_\S(\X) = \int_\S \bt(\X)  , \qquad \wt \bO_\S(\X,\Y) = \int_\S \bo(\X,\Y) . 
\end{split}
\end{equation}
The next step is to determine the kernel of the pre-symplectic form. In general, this depends on the details of the Lagrangian \eqref{genformLag}. However, gauge invariance implies the existence of at least one class of vectors in the kernel. Letting $\X = \X_\ve$ be the generator of infinitesimal gauge transformations in \eqref{Xepdef}, we derive explicitly in Appendix~\ref{app:explicitcalculations2} that
\begin{equation}
\begin{split}\label{gauge-transform-symp}
\wt \bT_\S(\X_\ve) = \oint_{\p\S} \tr{ \ve \CQ  } , \qquad \wt \bO_\S(\Y,\X_\ve) = \oint_{\p\S} \tr{ \ve \Y (  \CQ )  }   , 
\end{split}
\end{equation}
where
\begin{equation}
\begin{split}\label{Qdef}
\left( \ast \CQ \right)^{\mu\nu} = 2 \sum_{n=0}^\infty  (-1)^n \D_{\a_1 \cdots \a_n} \Pi^{\a_1 \cdots \a_n ; \mu\nu} .
\end{split}
\end{equation}
Importantly, note that the pre-symplectic form is written as an integral over the boundary $\p\S$, so if $\ve|_{\p\S} = 0$ then $\X_\ve \in \ker \wt \bO_\S$. This in turn means that we must identify solutions that differ by such gauge transformations. Exponentiating this, we find that on the phase space we must identity
\begin{equation}
\begin{split}\label{equivalence-rel-small}
( A , \Phi^i )  \sim \left( \g A \g^{-1} + \g \dt \g^{-1}  ,  R_i(\g) \Phi^i  \right) , \qquad \g \big|_{\p \S} = 0 . 
\end{split}
\end{equation}
We refer to such gauge transformations as \underline{small gauge transformations}, and solutions that differ by small gauge transformations are identified on the phase space.

If additional degeneracies exist, then we must identify fields related via these additional degeneracies as well. Letting $\sim$ denote all such equivalences, the phase space for gauge theories is then $\G = {\bs\CS}/\!\!\sim$, and the symplectic potential and symplectic form on the phase space are $\bT_\S = \wt \bT_\S |_\G$ and $\bO_\S = \wt \bO_\S |_\G$, respectively.

\subsection{Canonical Transformations}
\label{sec:canonicaltransformgen}

Having constructed the symplectic form on $\S$ in the previous subsection, we now proceed with a discussion on canonical transformations. In this paper, we will consider two such classes of canonical transformations.

\subsection*{Large Gauge Transformations (LGTs)}

These are gauge transformations generated by a vector field $\X_\ve$ from \eqref{Xepdef}, with $\ve$ satisfying the conditions $\Y(\ve) = 0$ for all $\Y \in T \G$, i.e. $\ve$ is field independent, and $\ve |_{\p\S} \neq 0$. Then \eqref{gauge-transform-symp} implies
\begin{equation}
\begin{split}
	\bO_\S(\Y,\X_\ve) = \Y \left( \oint_{\p\S} \tr{ \ve  \CQ   } \right)  = \Y ( \bT_\S ( \X_\ve ) ), \qquad {\bs\mathsterling}_{\X_\ve} \bT_\S (\Y) = 0 .
\end{split}
\end{equation}
It follows from \eqref{sympidentity1} that LGTs are canonical, and the associated Hamiltonian charge is by \eqref{symppotcan}
\begin{equation}
\begin{split}\label{largegaugecharge}
Q_\ve[\S] = \bT_\S ( \X_\ve ) = \oint_{\p\S} \tr{ \ve  \CQ }  . 
\end{split}
\end{equation}
Using \eqref{sympidentity1}, \eqref{pb}, and \eqref{infgaugetransform}, this charge generates LGTs on the phase space as
\begin{equation}\label{fieldbrackets}
\begin{split}
\big\{ Q_\ve[\S] , A \big\}_\S = \text D \ve , \qquad \big\{ Q_\ve[\S] , \Phi^i \big\}_\S = - R_i(\ve) \Phi^i .
\end{split}
\end{equation}

\subsection*{Isometry Transformations}

The Hamiltonian charge for isometry transformations is given in \eqref{isometry-can}. This can be evaluated using the methods described in Section~\ref{sec:isometries}. The calculation is almost identical, but given the special form of the gauge theory Lagrangian \eqref{genformLag}, the boundary term in the isometry charge takes a special form. We refer the reader to Appendix~\ref{app:explicitcalculations3} for the detailed computation, and will simply claim here that the isometry charge is
\begin{equation}
\begin{split}
\label{stress-tensor-explicit}
H_\xi[\S] = - \int_\S d\S_\mu \, T^{\mu\nu} \xi_\nu  + \oint_{\p\S} \SH_\xi - Q_{i_\xi A}[\S] ,
\end{split}
\end{equation}
where $T^{\mu\nu}$ and $\SH_\xi$ are defined as in \eqref{st-def} with the definitions for the tensors $\CA^{\mu\nu}$ and $\CB^{\mu\nu\rho}$ being
\begin{equation}
\begin{split}\label{ABdef11}
\CA^{\mu\nu} &= g^{\mu\nu} \CL - 2  \sum_{n=0}^\infty (-1)^n \tr{  \D_{\a_1\cdots \a_n} \Pi^{\a_1 \cdots \a_n ; \mu\rho}  F^\nu{}_{\rho}    } \\
&\qquad +   \sum_{n=1}^\infty \sum_{k=1}^n (-1)^k  \tr{  \D_{\a_2 \cdots \a_k }  \Pi^{\mu \a_2 \cdots \a_n ; \a\b} \D^\nu \D_{\a_{k+1} \cdots \a_n} F_{\a\b}  } \\
&\qquad +  \sum_{i=1}^N \sum_{n=1}^\infty \sum_{k=1}^n (-1)^k \left[ \D_{\a_2 \cdots \a_k}  \Pi^{\mu \a_2 \cdots \a_n}_i  \D^\nu \D_{\a_{k+1} \cdots \a_n}\Phi^i + \hc \right] \\
\CB^{\mu\nu\rho} &= \frac{1}{2}  \sum_{n=1}^\infty \sum_{k=1}^n (-1)^k  \tr{  \D_{\a_2 \cdots \a_k }  \Pi^{\mu \a_2 \cdots \a_n ; \a\b} (\S^{\nu\rho})_{\a_{k+1} \cdots \a_n;\a\b}{}^{\a'_{k+1} \cdots \a'_n;\a'\b'} \D_{\a'_{k+1} \cdots \a'_n} F_{\a'\b'}  } \\
&\qquad + \frac{1}{2}   \sum_{i=1}^N \sum_{n=1}^\infty \sum_{k=1}^n (-1)^k \left[ \D_{\a_2 \cdots \a_k}  \Pi^{\mu \a_2 \cdots \a_n}_i  (\S_i^{\nu\rho})_{\a_{k+1} \cdots \a_n}{}^{\a'_{k+1} \cdots \a'_n}  \D_{\a_{k+1} \cdots \a_n}\Phi^i + \hc \right].
\end{split}
\end{equation}
Note that these tensors (and therefore the first two terms in \eqref{stress-tensor-explicit}) are gauge invariant. Furthermore, the last term in \eqref{stress-tensor-explicit} is a boundary term, so while it is invariant under small gauge transformations, it is \textit{not} invariant under large gauge transformations (see \eqref{chargealgebra}). Therefore, in addition to the usual ``bulk'' stress tensor term and a familiar boundary term (c.f. \eqref{isometrycharge}), the isometry charge \eqref{stress-tensor-explicit} also contains the large gauge charge \eqref{largegaugecharge} derived in the previous subsection! Using \eqref{sympidentity1}, \eqref{pb}, and \eqref{infgaugetransform}, this charge generates isometry transformations on the phase space (assuming that the last term in \eqref{isometry-can} vanishes) as
\begin{equation}\label{isobrackets}
\begin{split}
\big\{ H_\xi [\S] , A \big\}_\S &= - \pounds_\xi A , \qquad \big\{ H_\xi [\S] , \Phi^i \big\}_\S = - \pounds_\xi \Phi^i .
\end{split}
\end{equation}

\subsection*{Charge Algebra}

Recall from our discussion in Section~\ref{sec:canonical} that the Hamiltonian charges corresponding to the canonical transformations are defined only up to an additive constant. Therefore, the large gauge and isometry charges given in \eqref{largegaugecharge} and \eqref{stress-tensor-explicit} can in fact be shifted by an arbitrary constant. To fix this constant, note that given our current charge definitions, we can use the Poisson brackets \eqref{fieldbrackets} and \eqref{isobrackets} to determine that the charges satisfy the algebra
\begin{equation}
\begin{split}\label{chargealgebra}
\big\{ Q_\ve [\S] , Q_{\ve'} [\S] \big\}_\S &= Q_{[\ve,\ve']} [\S] \\
\big\{ H_\xi [\S] , H_{\xi'} [\S]  \big\}_\S &= H_{[\xi,\xi']} [\S] \\
\big\{ H_\xi [\S] , Q_\ve [\S]  \big\}_\S &=  Q_{\xi(\ve)} [\S] .\\
\end{split}
\end{equation}
As changing the charges by an additive constant necessarily changes this algebra, we will fix the additive constant so that \eqref{chargealgebra} is satisfied. This completely fixes the large gauge and isometry charges to be \eqref{largegaugecharge} and \eqref{stress-tensor-explicit}.

\subsection{Gauge Theories in Flat Spacetime}
\label{sec:gaugetheories-flatspacetime}

The central quantity of interest in high energy physics is the \underline{scattering amplitude} (or \emph{$S$-matrix}) in our universe, which for many purposes can be approximated as a four-dimensional Minkowski spacetime $\CM = \mrr^{1,3}$. In a quantum theory, the $S$-matrix is a unitary map between the $in$- and $out$-Hilbert spaces that are naturally defined on $\S^- = \ci^- \cup i^-$ and $\S^+ = \ci^+ \cup i^+$, respectively. The two Hilbert spaces are isometric (i.e. they are isomorphic and have identical norms), and classically this implies the existence of a symplectomorphism $\CS$ (i.e. an isomorphism preserving the symplectic form) between the phase spaces on $\S^-$ and $\S^+$.

For the rest of the paper, we will be exploring the structure and properties of this $S$-matrix. Motivated by the reasons given above, we shall restrict ourselves to studying theories obeying the assumptions
\begin{enumerate}

\item We study gauge theories on $\CM = \mrr^{1,3}$ and construct the phase space on the Cauchy surfaces $\S^+ = \ci^+ \cup i^+$ and $\S^- = \ci^- \cup i^-$.\label{rest1}

\item The phase spaces on $\S^+$ and $\S^-$ are symplectomorphic, i.e. $\bO_{\S^+} = \bO_{\S^-}$.

\label{rest3}
\end{enumerate}
The assumptions described above are very general and apply to many theories of interest. However, in order to keep our discussion more focused, we shall also make the assumption
\begin{enumerate}[resume]
\item 
All asymptotic states/particles are massless.
\label{rest4}
\end{enumerate}
This allows us to disregard $i^\pm$ in our discussion and focus exclusively on null infinity $\ci^\pm$, since only massive particles enter and exit the spacetime from $i^\pm$. The absence of massive particles in the far past and future implies that there is no energy flux/excitations through these boundaries, so the fields (or more precisely, the field strengths) are all frozen on $i^\pm$, i.e. $\X(\varphi)|_{i^\pm} = 0$. Thus, we will for the rest of the paper refer to our Cauchy slices as $\ci^\pm$ for convenience, but it is important to remember that we always implicitly mean $\S^\pm$. Most notably, the only boundaries of $\S^\pm$ are $\ci^\pm_\mp$, and does \emph{not} include $\ci^\pm_\pm$.

\subsubsection{Coordinates} 

\begin{figure}[ht!]
\centering
\includegraphics[scale=0.8]{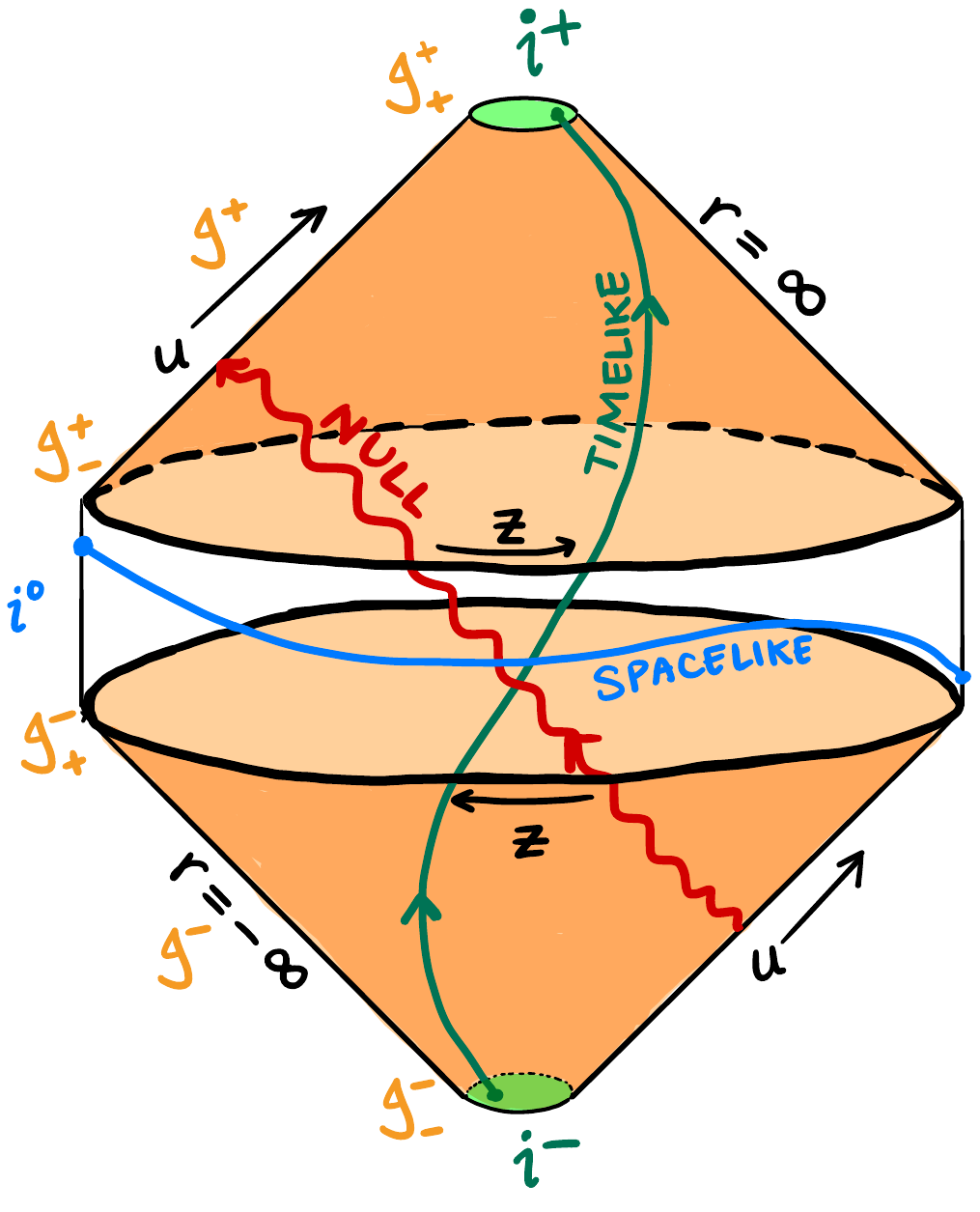}
\caption{Penrose Diagram of Minkowski spacetime}
\label{fig:MinkowskiPenrose}
\end{figure}

To proceed further, we need to establish a coordinate system on $\mrr^{1,3}$. We shall work in flat null coordinates $( u , r , z , \bz )$, which are related to standard Cartesian coordinates by
\begin{equation}
\begin{split}\label{flatnull}
x^\mu(u,r,z,\bz) = \frac{r}{2} \left( 1 + |z|^2 + \frac{u}{r} , z + \bz , - i ( z - \bz ) , 1 - |z|^2 - \frac{u}{r} \right) ,
\end{split}
\end{equation}
so that the metric is
\begin{equation}
\begin{split}\label{flatmetric}
\dt s^2 &= \eta_{\mu\nu}\, \dt x^\mu\, \dt x^\nu = - \dt u \,\dt r + r^2 \,\dt z\, \dt \bz  .
\end{split}
\end{equation}
The Penrose diagram of Minkowski spacetime is shown in Figure \ref{fig:MinkowskiPenrose}. Timelike curves begin and end on the spacelike surfaces $i^-$ and $i^+$, respectively, null curves begin and end on the null surfaces $\ci^-$ and $\ci^+$, respectively, and spacelike curves end on the timelike boundary $i^0$ (spatial infinity).

We now want to focus on the null boundaries $\ci^\pm$,\footnote{The remaining boundaries ($i^\pm$ and $i^0$) in these coordinates are described in Appendix~A.1.1 of \cite{He:2019jjk}, but they will not play a role in this paper.} which are located at $r = \pm\infty$ while keeping $(u,z,\bz)$ fixed. These hypersurfaces have the topology $\mrr  \times \mss^2$, and the future-directed area element is
\begin{equation}
\begin{split}\label{ci-area}
d\S^\mu \big|_{\ci^\pm} = - \frac{ r^2 }{2}\d^\mu_u \, d u\, d^2 z .
\end{split}
\end{equation}
The null generator along $\mrr$ is parameterized by $u$ whereas the $\mss^2$ is parameterized by the stereographic coordinates $(z,\bz)$. A useful feature of this coordinate system is that the point on the celestial $\mss^2$ labeled by $(z,\bz)$ on $\ci^+$ is antipodal to point with the same label on $\ci^-$. The boundaries of $\ci^\pm$ are located at $u=\infty$ ($\ci^\pm_+$) and $u=-\infty$ ($\ci^\pm_-$). These all have the topology of $\mss^2$, and the outward-directed area element on $\ci^\pm_\mp$ (with respect to $\ci^\pm$) is
\begin{equation}
\begin{split}\label{cibdy-area}
d S^{\mu\nu} \big|_{\ci^\pm_\mp}  = \mp 2 r^2\d^{[\mu}_{u} \d^{\nu]}_r  \, d^2 z .
\end{split}
\end{equation}

The isometries of Minkowski spacetime are translations and Lorentz transformations, and their action on Cartesian coordinates is defined as $x^\mu \to \L^\mu{}_\nu x^\nu + a^\mu$ with $\L^T \eta \L = \eta$. We find it convenient to parameterize the translation parameter as $a^\mu = x^\mu(u_0,r_0,z_0,\bz_0)$. Similarly, the Lorentz matrix that generates proper orthochronous Lorentz transformations will be parameterized as
\begin{equation}
\begin{split}
\L^\mu{}_\nu(P) = - \frac{1}{2} \tr{ P^\ct \s^\mu P {\bar \s}_\nu } , \qquad P = \left( \!\!\begin{array}{cc}
d & c \\ b & a 
\end{array} \!\! \right)  \in \text{SL}(2,\mcc)/\mzz_2,
\end{split}
\end{equation}
where $\s^\mu = ({\mathbbm 1},\s^i)$ and ${\bar \s}^\mu = ({\mathbbm 1},-\s^i)$. It can easily be verified that $\L(P)$ satisfies the defining property of a proper orthochronous Lorentz matrix, as well as the identity
\begin{equation}
\begin{split}
\L(P)\L(P') = \L(PP') . 
\end{split}
\end{equation}
The parameterization chosen here highlights the homomorphism between the four-dimensional proper orthochronous Lorentz group $\text{SO}^+(1,3)$ and the two-dimensional global conformal group $\text{SL}(2,\mcc)/\mzz_2$. Using the parameterizations described above, we can determine Poincar\'e transformations of the flat null coordinates to be
\begin{equation}
\begin{split}\label{finitetransformcoord}
(u,r,z) &\xrightarrow{\text{tr}} \left( u + u_0 + \frac{rr_0}{r+r_0} |z-z_0|^2 , r + r_0 , \frac{ r z + r_0 z_0 }{ r + r_0 }  \right)  \\
(u,r,z) &\xrightarrow{\text{LT}} \left( u \frac{|\tau'(z)|}{1 + \frac{u}{4r} \frac{ | \tau''(z) |^2 }{ | \tau'(z)|^2 }  } , \frac{r}{|\tau'(z)|}  \left( 1 + \frac{u}{4r} \frac{ | \tau''(z) |^2 }{ | \tau'(z)|^2 }  \right)  ,  \tau(z) - \frac{\tau'(z)^2}{\tau''(z)}  \frac{ \frac{u}{2r} \frac{ | \tau''(z) |^2 }{ |\tau'(z)|^2  }   }{  1 + \frac{u}{4r} \frac{ | \tau''(z) |^2 }{ | \tau'(z)|^2 }    } \right) ,
\end{split}
\end{equation}
where 
\begin{equation}
\begin{split}\label{mobius}
\tau(z) = \frac{ a z + b }{ c z + d } .
\end{split}
\end{equation}
When we restrict ourselves onto $\ci^\pm$, these transformations drastically simplify to
\begin{equation}
\begin{split}
(u,z) &\xrightarrow{\text{tr}} \left( u + u_0 + r_0 |z-z_0|^2 , z  \right) , \qquad (u,z) \xrightarrow{\text{LT}} \left( u  |\tau'(z)|  ,  \tau(z)  \right) .  
\end{split}
\end{equation}
Thus, four-dimensional Lorentz transformations act as Mobi\"us transformations (equivalently, global conformal transformations) on the coordinates $(z,\bz)$ on $\ci^\pm$. This is why these coordinates are very useful for studying holography in flat spacetime, where the goal is to recast four-dimensional scattering amplitudes (which are Lorentz covariant) as two-dimensional correlators in a putative conformal theory living on (a transverse cut of) $\ci^\pm$.

Infinitesimally, translations and Lorentz transformations are respectively generated by the Killing vectors $\xi_f^\text{tr}$ and $\xi^\text{LT}_Y$, which are
\begin{equation}
\begin{split}\label{kv}
\xi_f^\text{tr} &= f \p_u +   \p_z \p_\bz f \p_r    - \frac{1}{r} ( \p_\bz f \p_z  + \p_z f \p_\bz ) \\
\xi^\text{LT}_Y &=  \frac{1}{2}  \p_z Y^z ( u  \p_u  - r \p_r )  + Y^z  \p_z - \frac{u}{2r} \p_z^2 Y^z  \p_\bz  + \cc , 
\end{split}
\end{equation}
where
\begin{equation}
\begin{split}
f(z,\bz) = \chi^0(1+|z|^2)+\chi^1(z+\bz) - i \chi^2(z-\bz) + \chi^3(1-|z|^2) , \qquad Y^z = a + b z + c z^2,
\end{split}
\end{equation}
with $\chi^\mu\in\mrr$ and $a,b,c\in\mcc$. The Poincar\'e algebra then takes the form
\begin{equation}
\begin{split}\label{kvalgebra}
\big[ \xi^\text{tr}_f , \xi^\text{tr}_{f'} \big] = 0 , \qquad \big[ \xi^\text{LT}_Y , \xi^\text{tr}_f \big] = \xi^\text{tr}_{ \left( Y^z \p_z    - \frac{1}{2}  \p_z Y^z  + \cc \right) f }  , \qquad \big[  \xi^\text{LT}_Y  ,  \xi^\text{LT}_{Y'}   \big]  =  \xi^\text{LT}_{[Y,Y']} .  
\end{split}
\end{equation}

\subsubsection{Boundary Conditions}
\label{sec:bdycondsec}

Returning to our discussion of gauge theories in Minkowski spacetime, we want to to define the configuration space ${\bs{\mathfrak F}}$. This requires us to choose appropriate boundary conditions for the gauge and matter fields on $\ci^\pm$ and $i^0$.\footnote{One also needs boundary conditions on $i^\pm$ to define ${\bs{\mathfrak F}}$, but these details will not be relevant to us.\label{Fbdy}} We obtain these by imposing finiteness of energy-momentum and angular-momentum flux through $\ci^\pm$, which are the isometry charges \eqref{stress-tensor-explicit} corresponding to translations ($\xi = \xi_f^\text{tr}$) and Lorentz transformations ($\xi = \xi_Y^{\text{LT}}$) respectively. To be precise, we define
\begin{equation}
\begin{split}
P_f[\S] \equiv H_{\xi^\text{tr}_f} [\S]  , \qquad J_Y[\S] \equiv H_{\xi^\text{LT}_Y} [\S] ,
\end{split}
\end{equation}
and require that
\begin{equation}
\begin{split}
P_f[\ci^\pm] < \infty , \qquad  J_Y[\ci^\pm] < \infty .
\end{split}
\end{equation}

\subsubsection*{Example I: Scalar Field}
Consider a free massless complex scalar field $\Phi$, which is described by the Lagrangian density
\begin{equation}
\begin{split}
\CL = - \nabla^\mu \Phi^\ct \nabla_\mu \Phi .
\end{split}
\end{equation}
Using \eqref{ABdef} and \eqref{st-def}, we find that
\begin{equation}
\begin{split}\label{scalar-st}
T^{\mu\nu} =  \nabla^{\mu} \Phi^\ct \nabla^{\nu} \Phi + \nabla^{\nu} \Phi^\ct \nabla^{\mu} \Phi  - g^{\mu\nu} \nabla^\rho \Phi^\ct \nabla_\rho \Phi  , \qquad (\ast \SH_\xi)^{\mu\nu}  = 0 . 
\end{split}
\end{equation}
We can now determine the isometry charge for translations and Lorentz transformations on $\ci^\pm$ using \eqref{isometrycharge} and \eqref{kv}. In particular, the (null) energy flux through $\ci^\pm$ is
\begin{equation}
\begin{split}
P_{f=1} [ \ci^\pm ] = \int du\, d^2 z \lim\limits_{r \to \pm\infty}  r^2  \p_u \Phi^\ct  \p_u \Phi .
\end{split}
\end{equation}
This is finite only if $ \p_u \Phi = O(r^{-1})$ at large $|r|$, leading to the following asymptotic behavior for the scalar field near $\ci^\pm$:\footnote{We adopt the notation where $f(r) = o(g(r))$ means $\lim\limits_{|r|\to\infty} \frac{f(r)}{g(r)} = 0$, while $f(r) = O(g(r))$ means $\lim\limits_{|r|\to\infty} \frac{f(r)}{g(r)} <\infty$.}
\begin{equation}
\begin{split}
\label{scalarbdycond}
\Phi(u,r,z,\bz) = \frac{1}{r} \phi^\pm(u,z,\bz) + o(r^{-1}) \quad \text{near $r=\pm\infty$.}
\end{split}
\end{equation}
It is obvious then that the charge can also be written as
\begin{equation}
\begin{split}
P_{f=1} [ \ci^\pm ] =  \int du\, d^2 z \, \p_u \phi^{\pm\ct} \p_{u} \phi^\pm .
\end{split}
\end{equation}

Similarly, the angular-momentum flux through $\ci^\pm$ can be computed by substituting \eqref{scalar-st} and $\xi^\LT_Y$ from \eqref{kv} into \eqref{isometrycharge}. In particular, we have
\begin{align}
\begin{split}
	J_{Y^z=1}[\ci^\pm] &= \frac{1}{2}\int du\,d^2z \,\big( \p_u\phi^{\pm\ct}\p_z \phi^{\pm} + \p_z\phi^{\pm\ct}\p_u\phi^\pm \big),
\end{split}
\end{align}
where we used \eqref{scalarbdycond}. Finiteness of this charge requires that the integral over $u$ be finite, which is satisfied assuming\footnote{Strictly speaking, finiteness of the charge imposes the slightly weaker condition $\phi^\pm = O(1)$ at large $|u|$. However, the constant piece of $\phi^\pm$ is associated to soft scalar modes, which is beyond the scope of this paper.}
\begin{equation}
\begin{split}\label{scalarlargeu}
\phi^\pm(u,z,\bz) = o(1) \quad \text{near $u=\pm\infty$.} 
\end{split}
\end{equation}
One can then verify that with these boundary conditions \emph{all} the isometry charges are finite. In addition, any phase space defined with these boundary conditions also has isometry transformations as canonical transformations since \eqref{scalarlargeu} implies that the last term in \eqref{isometry-can} vanishes, as required!

This completely describes all the relevant boundary conditions for a free scalar field. In fact, these boundary conditions generalize to interacting massless scalar fields as well. More specifically, if all the interactions in the Lagrangian are irrelevant, then the above fall-offs continue to hold because irrelevant interaction terms do not affect the infrared, or long-distance, physics. Relevant deformations of a free Lagrangian typically renormalize the mass of the field and will thus end up violating Assumption~\eqref{rest4}. For this reason, we shall assume that all interactions are irrelevant so that \eqref{scalarbdycond} and \eqref{scalarlargeu} are valid for all scalar fields.

\subsubsection*{Example II: Gauge Field}
Consider a non-abelian gauge field described by the Yang-Mills Lagrangian
\begin{equation}
\begin{split}
\CL = - \frac{1}{4g^2} \tr{ F_{\mu\nu} F^{\mu\nu} } . 
\end{split}
\end{equation}
In this case, we use \eqref{ABdef11}, \eqref{st-def}, and \eqref{Qdef} to determine
\begin{equation}
\begin{split}\label{gauge-st}
T^{\mu\nu} = \frac{1}{g^2} \tr{ F^{\mu\rho} F^\nu{}_\rho - \frac{1}{4} g^{\mu\nu} F_{\rho\s} F^{\rho\s} }  , \qquad (\ast \SH_\xi)^{\mu\nu} = 0 , \qquad  (*\CQ)^{\mu\nu} = -\frac{1}{g^2} F^{\mu\nu}. 
\end{split}
\end{equation}
Using \eqref{stress-tensor-explicit}, \eqref{largegaugecharge}, and \eqref{kv}, the energy flux through $\ci^\pm$ is derived to be
\begin{equation}
\begin{split}\label{freegauge-st}
P_{f=1}[\ci^\pm]  = \frac{2}{g^2}  \int d u \, d^2 z \lim_{r \to \pm \infty} \tr{ F_{uz}  F_{u\bz} }  \pm \frac{1}{g^2} \int d^2 z \lim_{u\to\mp\infty} \lim_{r\to\pm\infty} r^2 \,\tr{ A_u F_{ur} } .
\end{split}
\end{equation}
The first term is finite only if $F_{uz} = O(1)$ at large $|r|$, which naturally suggests
\begin{equation}
\begin{split}\label{gaugebdycond'}
A_z(u,r,z,\bz) = A_z^\pm(u,z,\bz) + o(1) \quad \text{near $r=\pm\infty$.}
\end{split}
\end{equation}

The boundary conditions for the remaining components of the gauge field, as well as the large $|u|$ fall-offs of the gauge field, can be determined by examining the angular-momentum flux.  For instance, using \eqref{stress-tensor-explicit}, \eqref{largegaugecharge}, and \eqref{kv} again, we have 
\begin{equation}
\begin{split}
\label{freegauge-st1}
J_{Y^z=1}[\ci^\pm] &= \frac{1}{g^2} \int du \, d^2 z\lim_{r\to\pm\infty} \tr{  F_{uz} \big( r^2  F_{ur} + F_{z\bz} \big) } \\
&\qquad \qquad \pm \frac{1}{g^2} \int d^2 z \lim_{u\to\mp\infty} \lim_{r\to\pm\infty} r^2\, \tr{ A_z F_{ur} }  .
\end{split}
\end{equation}
The first term is finite at large $|r|$ only if $F_{ur} = O(r^{-2 })$, so it is natural to impose $A_r = O(r^{-2})$ and $A_u = O(r^{-1})$ at large $|r|$. Since we already determined that $A_z = O(1)$ at large $|r|$ above, this also implies that the second term in \eqref{freegauge-st} is finite. Furthermore, as in the case of the scalar field, we also require the integral over $u$ to be finite. The first term immediately implies we should have
\begin{equation}
\begin{split}
	A_z^\pm(u,z,\bz) = O(1)\quad \text{near $u=\pm \infty$.}
\end{split}
\end{equation}
Then, finiteness of the last term in \eqref{freegauge-st1} requires that the coefficient of $r^{-2}$ in the large $r$ expansion of $F_{ur}$ should be finite at large $|u|$. As with the scalar field, these boundary conditions also ensure that the boundary term in \eqref{isometry-can} vanishes so that isometry transformations are indeed canonical!

To summarize, the large $|r|$ fall-offs for the components of the gauge field are
\begin{equation}
\begin{split}\label{gaugebdycond}
A_u(u,r,z,\bz) &= O(r^{-1})   \\
A_r(u,r,z,\bz) &= O(r^{-2}) \\
A_z(u,r,z,\bz) &= A_z^\pm(u,z,\bz) + o(1) . 
\end{split}
\end{equation}
As with the scalar field, these fall-offs generally hold for interacting theories as well. Indeed, gauge invariance prohibits the presence of any relevant interaction terms. Following the same procedure as above, we can determine the boundary fall-offs for all the fields in the theory near $\ci^\pm$. Although our results are completely general, we shall assume for simplicity that all matter fields $\Phi^i$ are scalars so the preceding discussion above will suffice.

\subsubsection*{Gauge Condition}

As described in the paragraph below \eqref{inducedsymp1211}, we need to impose a gauge condition to describe the phase space as a subspace of the solution space ${\bs\CS}$. The gauge choice we adopt is
\begin{equation}
\begin{split}\label{gaugecond}
A_u = 0 , \qquad A_r   \big|_{u=0} = 0 . 
\end{split}
\end{equation}
Before proceeding, we need to verify that this is indeed a \emph{good} gauge condition. To be precise, we need verify that every equivalence class of \eqref{equivalence-rel-small} contains a \emph{unique} solution that satisfies the gauge condition. This means we need to show that for every solution $({\bar A},{\bar \Phi}^i) \in {\bs\CS}$, there exists a \emph{unique} $\g \in \CG$ such that 
\begin{equation}
\begin{split}
\g {\bar A}_u \g^{-1} + \g \p_u \g^{-1} = 0 , \qquad ( \g {\bar A}_r \g^{-1} + \g \p_r \g^{-1} ) \big|_{u=0} = 0 , \qquad \g \big|_{\ci^\pm_\mp} = 0 . 
\end{split}
\end{equation}
It is clear that this system of first order differential equations has a unique solution. The first equation can be solved up to an integration constant $c_1(r,z,\bz)$. Substituting this solution into the second equation yields a first order differential equation for $c_1(r,z,\bz)$, which in turn is solvable up to an integration constant $c_2(z,\bz)$. The final condition is then used to uniquely solve for $c_2(z,\bz)$, thus completing the proof. Upon imposing this gauge, the allowed LGTs on $\G$ are now generated by $\g\equiv \g(z,\bz) \in \CG$.

\subsubsection*{Boundary Condition on $i^0$}
\label{sec:bdyconditionspatial}

Finally, to complete our discussion of the configuration space, we need to describe the boundary conditions for the fields near spatial infinity. In particular, we need to choose boundary conditions so that Assumption~\eqref{rest3} holds. From \eqref{Sdep2} with $\S= \S^+, \S' = \S^-,$ and $\CB = i^0$, we see that $\S^+$ is symplectomorphic to $\S^-$ only if $\bO_{i^0} = 0$. The simplest way to achieve this is to require that the gauge and matter fields induced on $i^0$ vanish, i.e.\footnote{Note that \eqref{bcspatial1} fixes three of the four components of the gauge field to zero, while the remaining (normal) component is fixed by Gauss' law.}
\begin{equation}
\begin{split}\label{bcspatial1}
A \big|_{i^0} \stackrel{?}{=} 0 , \qquad \Phi^i \big|_{i^0} \stackrel{?}{=} 0 . 
\end{split}
\end{equation}
These are the usual boundary conditions assumed in field theories, but they are actually too strong for our purposes. In fact, \eqref{bcspatial1} is not preserved under LGTs, so they preclude the existence of LGTs in the phase space. The non-existence of LGTs would be problematic, as the presence of infrared divergences in QFTs has been shown to be intimately related to LGTs \cite{Gabai:2016kuf,Kapec:2017tkm,Choi:2017bna,Choi:2017ylo,Carney:2018ygh,Ashtekar:2018lor,Hirai:2019gio,Gonzo:2019fai,Choi:2019rlz,H:2019fvd,Himwich:2020rro,Hirai:2020kzx}.

Instead, we propose to include all boundary conditions that are large gauge equivalent to \eqref{bcspatial1} at $i^0$, so that we have
\begin{align}\label{bcspatial2}
	A \big|_{i^0} = C\dt C^{-1} \quad\implies\quad F \big|_{i^0} = 0, \quad \Phi^i \big|_{i^0} = 0, \quad C \equiv C(z,\bz) \in \CG.
\end{align}
Here, $C$ is restricted to depend only on $(z,\bz)$ since we are working in the gauge \eqref{gaugecond}, where the allowed LGTs are generated by $\g(z,\bz)\in\CG$. This extended phase space then allows for LGTs by construction. Furthermore, this extended phase space is obtained from the previous phase space by an LGT, and since LGTs are canonical transformations, the vanishing of the symplectic form $\bO_{i_0} = 0$ is preserved. This implies by \eqref{Sdep2} that the extended phase space remains symplectomorphic, so Assumption~\eqref{rest3} is satisfied.

It is important to remember that \eqref{bcspatial2} is a constraint on the phase space. In the theory of symplectic geometry (equivalently, see Dirac's formulation of constrained phase spaces \cite{Dirac_1950}), it is generically not possible to impose constraints on a phase space while simultaneously preserving the invertibility of the symplectic form. Oftentimes, additional ``gauge conditions'' (i.e. second class constraints) are required. These gauge conditions are determined in the manner described in the paragraph below \eqref{presymp1}, and as we shall see, we will have to explicitly impose them in Section~\ref{sec:constraints}.

\subsubsection{Symplectic Structure on \texorpdfstring{$\ci^\pm$}{}} \label{sec:sympstructure}

The fact that the fields fall off on $\ci^\pm$, as is evidenced by \eqref{scalarbdycond} and \eqref{gaugebdycond}, greatly simplifies the symplectic structure there. This is because derivatives and products of fields fall off faster than the fields themselves, so terms in the Lagrangian involving too many derivatives or fields do not contribute on $\ci^\pm$. Indeed, as we will now demonstrate, only the quadratic term contributes.

First, it is useful to separate the terms in the Lagrangian that contribute on $\ci^\pm$ from those that do not by decomposing the Lagrangian as
\begin{equation}
\begin{split}\label{Lag-decomposition}
L =  L^\ym + L^\mat , \qquad L^\ym = - \frac{1}{2g^2} \tr{ F \w \ast F }  , \qquad L^\mat = \sum_i L^\kin_i   + L^\inter, 
\end{split}
\end{equation}
where $L^\ym$ is the pure Yang-Mills Lagrangian, $g$ the gauge field coupling constant, and $L^\mat$ the part of the Lagrangian that includes the matter kinetic terms $L^\kin_i$ and the interaction terms $L^\inter$. The precise structure of the matter kinetic terms depends on the Lorentz spin of the field, but since we are assuming for simplicity that all matter fields are scalars, we have\footnote{The matter kinetic terms in \eqref{kinex} contain some interactions as well, but we keep them to preserve manifest gauge invariance.}
\begin{equation}\label{kinex}
\begin{split}
L^\text{kin}_i &= \e \CL^\text{kin}_i , \qquad\quad \CL^\kin_i = - ( \D^\mu \Phi^i )^\ct ( \D_\mu \Phi^i ).
\end{split}
\end{equation}
Since we are planning to demonstrate that only quadratic terms contribute, and $L^\inter$ only contains terms that are cubic order or higher, we will not need to worry about it in the limit $\S \to \ci^\pm$.

Using the decomposition \eqref{Lag-decomposition}, the equations of motion for the gauge field is by \eqref{eom-explicit} 
\begin{equation}
\begin{split}\label{gauge-eom}
\text D \ast F = g^2 \ast J^\mat ,
\end{split}
\end{equation}
where the one-form current $J^\mat$ is covariantly conserved, i.e. $\text D \ast J^\mat = 0$, and is explicitly given by
\begin{equation}\label{J-explicit}
\begin{split}
( J^\mat )^\mu &= 2 \sum_{n=0}^\infty (-1)^n \D_\nu \D_{\a_1}\cdots \D_{\a_n} (\Pi^\mat)^{\a_1 \cdots \a_n ; \nu\mu}  \\ 
&\qquad - \sum_{n=1}^\infty  \sum_{k=1}^n (-1)^k \left[ \D_{\a_2} \cdots \D_{\a_k} (\Pi^\mat)^{\mu \a_2 \cdots \a_n ; \a\b}  , \D_{\a_{k+1} } \cdots \D_{\a_n} F_{\a\b} \right]   \\
&\qquad + \sum_{i=1}^N \sum_{n=1}^\infty \sum_{k=1}^n (-1)^k \bigg[ \D_{\a_2} \cdots \D_{\a_k} (\Pi^\mat_i)^{\mu \a_2 \cdots \a_n }   T^a_i \D_{\a_{k+1}}  \cdots \D_{\a_n} \Phi^i + \hc \bigg] X^a ,
\end{split}
\end{equation}
where $\Pi^\mat$ is defined as in \eqref{Pi-def} with the replacement $\CL \to \CL^\mat$. Given the structure of the matter Lagrangian \eqref{Lag-decomposition} and \eqref{kinex}, the matter current takes the form
\begin{equation}
\begin{split}\label{mattercurrent}
( J^\mat)^\mu = \sum_{i=1}^N \left( \left( D^\mu \Phi^i \right)^\ct T^a_i \Phi^i - ( \Phi^i )^\ct T^a_i D^\mu \Phi^i \right) + ( J^\text{inter} )^\mu . 
\end{split}
\end{equation}
where $J^\text{inter}$ is the contribution from $L^\text{inter}$. 

Likewise, we can determine from \eqref{spcd-explicit} and \eqref{integrate-over-Sigma} that the symplectic potential on $\ci^\pm$ is
\begin{equation}
\begin{split}
	\bT_{\ci^\pm}(\X) &=  - \frac{1}{g^2} \int_{\ci^\pm} \tr{ \X( A ) \w \ast F  } + \sum_{i=1}^N \big(\bT_i^{\kin}\big)_{\ci^\pm} ( \X ) + \bT_{\ci^\pm}^\inter ( \X ) ,
\end{split}
\end{equation}
where terms with the superscripts ``$\kin$'' and ``$\inter$'' are determined using \eqref{spcd-explicit} and \eqref{integrate-over-Sigma} with the replacement $\CL \to \CL^{\kin},\CL^\inter$. 

The contribution from the matter kinetic terms can be determined using the explicit form of $\CL^{\text{kin}}$. Since we are assuming that $\Phi^i$ is a scalar field, applying \eqref{kinex} to \eqref{spcd-explicit} and then integrating yields
\begin{equation}
\begin{split}\label{thetakini}
	\big(\bT_{i}^\kin\big)_{\ci^\pm} ( \X ) &= \int_{\ci^\pm} \left( \X (\Phi^i)^\ct  \ast \Dt \Phi^i  + \ast \Dt (\Phi^i)^\ct  \X ( \Phi^i ) \right) \\
	&=  \frac{1}{2}\int du\, d^2 z  \left[ \X ( \phi^{\pm i} )^\ct \p_u \phi^{\pm i}  + \p_u (\phi^{\pm i})^\ct \X ( \phi^{\pm i} ) \right] ,
\end{split}
\end{equation}
where we used \eqref{scalarbdycond} in the last equality.\footnote{If $\X \in T{\bs\CS}$, then it is true that $\X(\Phi^i)$ has the same fall-off near $\ci^\pm$ as $\Phi^i$ itself. However, for certain types of transformations, e.g. symmetry transformations that are spontaneously broken, it is interesting to consider vectors for which $\X(\Phi^i)$ and $\Phi^i$ do not have the same fall-off. Such vectors play a role in the subleading soft theorems and have been studied in \cite{Campiglia:2016efb,Campiglia:2016jdj,Campiglia:2016hvg,Laddha:2017ygw,Laddha:2017vfh}, but we will not consider such cases in this paper.} The contribution of $L^\text{inter}$, on the other hand, vanishes on $\ci^\pm$. To see this, we recall that all terms in $L^\inter$ contain Lorentz invariant products of three or more fields, and therefore so does $\bt^\text{inter}(\X)$. This implies that each term in the integrand $\bt^\text{inter}(\X)$ falls off at least as fast as $O(r^{-3})$. However, the integration measure grows as $O(r^2)$, so upon integrating each term falls off at least as fast as $O(r^{-1})$, which means $\bT_{\ci^\pm}^\inter ( \X )$ vanishes. 

Thus, the full symplectic potential on $\ci^\pm$ is
\begin{align}\label{symppotci}
	\bT_{\ci^\pm}(\X) = -\frac{1}{g^2}\int_{\ci^\pm}\tr{\X(A) \wedge *F } + \sum_{i=1}^N \big(\bT_i^{\kin}\big)_{\ci^\pm}(\X).
\end{align}
Using \eqref{psforms2}, it immediately follows that the symplectic form is $\bO = \bO^A_{\ci^\pm} + \bO^\mat_{\ci^\pm}$, where
\begin{align}
\label{sympformci}
\begin{split}
	\bO^A_{\ci^\pm}(\X,\Y) &= \frac{1}{g^2}\int_{\ci^\pm} \tr{ \X(A) \wedge * \Dt\Y(A) -  (\X \leftrightarrow\Y)}  \\
	\bO^\mat_{\ci^\pm}(\X,\Y) &=  \sum_{i=1}^N \int du \, d^2 z \left[ \p_u \X(\phi^{\pm i})^\ct \Y(\phi^{\pm i})  - (\X \leftrightarrow \Y) \right] .
\end{split}
\end{align}
Due to Assumption~\eqref{rest3}, the full symplectic form $\bO$ on $\ci^+$ and $\ci^-$ are equal, which is why it does not require a $\pm$ label. Furthermore, because the symplectic form is in block diagonal form, the phase space on $\ci^\pm$ factorizes into many components, one for each field, so that\footnote{Note that while the full phase space $\G$ is the same on $\ci^+$ and $\ci^-$, its factorization into components is not.}
\begin{align}\label{Hfactor1}
\G = \G^{\pm A} \times \G^{\pm 1} \times \cdots \times \G^{\pm N}.
\end{align}
For the rest of the section, we will focus on $\G^{\pm A}$, the gauge field component of the phase space.

\subsubsection{Constraints} \label{sec:constraints}

Thus far in Section~\ref{sec:sympstructure}, we have derived the symplectic form \eqref{sympformci} without imposing any constraints. We now want to impose the boundary condition \eqref{bcspatial2} on the gauge field. Since spatial infinity meets $\ci^+$ and $\ci^-$ at the boundaries $\ci^+_-$ and $\ci^-_+$ respectively, we can write the constraints as
\begin{equation}
\begin{split}\label{cons1}
A \big|_{\ci^+_-} = A \big|_{\ci^-_+} = C \dt C^{-1} , \qquad C \equiv C(z,\bz) \in \CG.
\end{split}
\end{equation} 
Once we impose this constraint, we claim that the gauge field symplectic form $\bO^A_{\ci^\pm}$ from \eqref{sympformci} is no longer invertible in the constrained phase space. To see why, first write out the symplectic form in flat null coordinates so that 
\begin{equation}
\begin{split}\label{presympform}
	\bO^A_{\ci^\pm}  ( \X , \Y ) =  \frac{1}{g^2} \int du\, d^2 z \,  \tr{ \p_u \X ( A_z^\pm ) \Y (  A_\bz^\pm )  + \p_u\X(A_\bz^\pm)\Y(A_z^\pm) - (\X \leftrightarrow \Y)   },
\end{split}
\end{equation}
where we have used \eqref{gaugebdycond} to compute the large $r$ limit. Next, it is convenient to introduce the boundary fields
\begin{equation}
\begin{split}\label{softconstraint1}
C_z \equiv A_z^\pm \big|_{\ci^\pm_\mp} = C \p_z C^{-1}  , \qquad N_z^\pm \equiv \int du\, \p_u A_z^\pm  , \qquad {\hat A}_z^\pm \equiv A_z^\pm - C_z,
\end{split}
\end{equation}
where $C_z$ does not have a $\pm$ superscript since it is independent of $\ci^\pm$ by \eqref{cons1}. For reasons that will become clear shortly, we refer to $C_z$ and $N_z$ as soft gauge modes and ${\hat A}_z$ as hard gauge modes. In terms of these soft and hard modes, the symplectic form becomes 
\begin{equation}
\begin{split}\label{presympform1}
	\bO^A_{\ci^\pm}  ( \X , \Y ) &=  \frac{2}{g^2} \int du\, d^2 z\,   \tr{ \p_u \X ( {\hat A}_z^\pm ) \Y (  {\hat A}_\bz^\pm ) - \p_u\Y(\hat A_{z}^\pm) \X(\hat A_\bz^\pm) }  \\
	&\qquad \qquad  +  \frac{1}{g^2} \int  d^2 z\,   \tr{  \X ( N_z^\pm ) \Y (  C_\bz ) + \X(N_\bz^\pm)\Y(C_z) - (\X \leftrightarrow \Y)    } .
\end{split}
\end{equation}
Thus, we see that the symplectic form breaks up into two pieces, one involving the soft modes (the second term) and one involving the hard modes (the first term), indicating that the gauge field phase space further factorizes to
\begin{equation}\label{Hfactor2}
\begin{split}
	\G^{\pm A} = \G^{\pm A,\text{soft}} \times \G^{\pm A,\text{hard} }  . 
\end{split}
\end{equation}
We now want to impose the constraint \eqref{softconstraint1} and write the symplectic form in terms of $C$ and $N_z^\pm$. Substituting the constraint into the soft part of \eqref{presympform1}, we get after some algebra 
\begin{equation}
\begin{split}\label{Og1}
\bO^{A,\text{soft}}_{\ci^\pm} ( \X , \Y )  &= \frac{1}{g^2}   \int d^2 z\,  \tr{ \X \Big( C^{-1} \big( \CDt_z N^\pm_\bz + \CDt_\bz N^\pm_z \big)  \Big)  \Y (  C )  - (\X \leftrightarrow \Y)  } ,
\end{split}
\end{equation}
where we have defined the gauge covariant derivative with respect to $C$ so that for any $M$ in the adjoint representation we have
\begin{equation}
\begin{split}\label{Dzdef}
	\CDt_z  M  \equiv \p_z M + [ C_z ,  M  ]  = C \p_z ( C^{-1} M C ) C^{-1} . 
\end{split}
\end{equation}

We can now finally demonstrate why \eqref{Og1} is not invertible. Consider the vector
\begin{equation}\label{Nequiv1}
\begin{split}
\X_v^{\pm}  = \int d^2 z \, \tr{ i \CDt_z v(z,\bz) \frac{\d}{\d N_z^\pm(z,\bz) } - i \CDt_\bz v (z,\bz)  \frac{\d}{\d N_\bz^\pm(z,\bz) }  } ,
\end{split}
\end{equation}
where $v (z,\bz) \in \mfg$. It can easily be verified that for all $\Y \in T \G^{\pm A,\text{soft}}$, we have
\begin{equation}
\begin{split}
\bO^{A,\text{soft}}_{\ci^\pm} (\Y,\X_v^\pm) = 0  \quad \implies \quad \X_v^\pm \in \ker \bO^{A,\text{soft}}_{\ci^\pm}.
\end{split}
\end{equation}
This means that the symplectic form (which is actually the pre-symplectic form in the constrained system) is non-invertible, and we need to follow the procedure outlined in the paragraph below \eqref{presymp1}. Introducing the equivalence $\X \sim \X + \X_v^\pm$ on the tangent space, we can exponentiate this to determine the equivalence on the phase space to be
\begin{equation}\label{Nequiv}
\begin{split}
	 N_z^{\pm}  \sim N_z^\pm +  i  \CDt_z v  . 
\end{split}
\end{equation}
Analogous to \eqref{gaugecond}, we can define the soft phase space uniquely by imposing a gauge condition that maps every element on $ \G^{\pm A,\text{soft}}$ to its equivalence class. From \eqref{Nequiv}, $N_z^\pm$ is defined up to an arbitrary $D_z^C v(z,\bz)$ function, so we can fix our gauge by choosing
\begin{equation}
\begin{split}\label{secondclassconstraint}
\CDt_z N_\bz^\pm - \CDt_\bz N_z^\pm = 0 . 
\end{split}
\end{equation}
Using \eqref{Dzdef}, we can rewrite this condition as
\begin{equation}
\begin{split}
\p_z \big( C^{-1} N_\bz^\pm C \big)  - \p_\bz \big( C^{-1} N_z^\pm C \big)  = 0,
\end{split}
\end{equation}
and on $\mss^2$ (which is topologically trivial), this equation has the unique solution
\begin{equation}
\begin{split}\label{Nzsol}
N_z^\pm = C \p_z N^\pm C^{-1}  = \CDt_z \big( C N^\pm C^{-1} \big), \qquad N^{\pm a} = ( N^{\pm a} )^*.
\end{split}
\end{equation}
Substituting this back into \eqref{Og1}, the soft part of our gauge field symplectic form becomes in terms of the fields $C$ and $N^\pm$
\begin{equation}
\begin{split}\label{Og2}
\bO^{A,\text{soft}}_{\ci^\pm} ( \X , \Y )  &= \frac{2}{g^2} \int d^2 z \, \tr{ \X  \big(\p_z \p_\bz N^\pm C^{-1}  \big) \Y ( C ) - (\X \leftrightarrow \Y) } . 
\end{split}
\end{equation}

Unfortunately, even after imposing the equivalence relation \eqref{Nequiv}, the symplectic form $\bO^{A,\soft}_{\ci^\pm}$ is still not invertible. To demonstrate this, we will need to construct another vector in $T\G^{\pm A,\soft}$ belonging to $\ker\bO^{A,\soft}_{\ci^\pm}$. To this end, we first introduce the derivative operator $\mfd^a_{C(z,\bz)}$, which is defined so that its action on $C$ is given by
\begin{align}\label{daction}
	\mfd^a_{C(w,\bw)} C(z,\bz) = -X^a C(z,\bz)\delta^2(z-w).
\end{align}
The properties of this operator (and why it is a derivative operator) are more fully explored in Appendix~\ref{app:deriv}, but for our purposes here it suffices to know that such an operator exists. Using \eqref{daction}, we can define the vector
\begin{align}
\begin{split}
	&\X_{\ve,\eta}^\pm = \int d^2z \, \bigg[ -\big(C(z,\bz)\ve C^{-1}(z,\bz)\big)^a \mfD^a_{C(z,\bz)} + \big([N^\pm(z,\bz),\ve] + \eta \big)\frac{\delta}{\delta N^\pm(z,\bz)} \bigg] ,
\end{split}
\end{align}
where $\ve,\eta \in \mfg$ are independent of $(z,\bz)$.\footnote{Actually, we only require $\eta$ to obey $\p_z\p_\bz \eta = 0$, but we will not consider this more general possibility.}
It is straightforward to check that $\bO^{A,\soft}_{\ci^\pm}(\X_{\ve,\eta}^\pm,\Y) = 0$ for any vector $\Y$, which means that $\X_{\ve,\eta}^\pm \in \ker \bO^{A,\soft}_{\ci^\pm}$. We must therefore introduce another equivalence relation $\X \sim \X + \X^\pm_{\ve,\eta}$, and upon exponentiating we obtain the phase space equivalence relation
\begin{align}
\begin{split}\label{gidentification}
( C , N^\pm )  \sim \big( C \g ,    \g^{-1} N^\pm \g + \eta \big) ,
\end{split}
\end{align}
where $\g \in \CG$ and $\eta \in \mfg$ are spacetime constants. Strictly speaking, to define the phase space, we must quotient out by this equivalence relation by imposing a gauge condition. However, it is more convenient for our purposes to not gauge fix but rather implicitly identify the fields related via the equivalence \eqref{gidentification}, which means our results must be invariant under \eqref{gidentification}.

\subsubsection{Charges}

Having constructed the phase space, we can turn to constructing the large gauge and isometry charges. 

\paragraph{Large Gauge Transformations} The LGT charge was derived in \eqref{largegaugecharge} with $\CQ$ being defined in \eqref{Qdef}. Using the explicit form of the Lagrangian \eqref{Lag-decomposition}, we compute 
\begin{align}
	\CQ = \frac{1}{g^2} \ast F + \CQ^\inter,
\end{align}
where $\CQ^\inter$ is the contribution from $L^\inter$. For the same reasons as those described below \eqref{thetakini}, $\CQ^\text{inter}$ does not contribute on $\ci^\pm$, which means the charge generating LGTs on $\ci^\pm$ is simply
\begin{equation}\label{chargeformdef}
\begin{split}
Q_\ve[\ci^\pm] &= \frac{1}{g^2} \int_{\ci^\pm_\mp} \tr{ \ve \ast F }  = \mp \frac{1}{g^2} \int d^2 z \lim_{u\to\mp\infty} \lim_{r\to\pm\infty} r^2 \tr{ \ve F_{ur} } .
\end{split}
\end{equation}
Note that due to Assumption~\eqref{rest3}, the charges on $\ci^+$ and $\ci^-$ are equal, i.e. $Q_\ve[\ci^+] = Q_\ve[\ci^-]$, so we shall simply denote these by $Q_\ve$. Recall from \eqref{chargealgebra} that these charges satisfy the charge algebra 
\begin{align}\label{QQcom}
\begin{split}
	\big\{ Q_\ve, Q_{\ve'} \big\} = Q_{[\ve,\ve']},
\end{split}
\end{align}
where as with the charges, we have dropped the subscript $\ci^+$ or $\ci^-$ from the Poisson bracket due to Assumption \eqref{rest3}. 

It is illuminating to rewrite the above charge so that it consists of a soft part and a hard part. Using Stokes' theorem, we can rewrite the charge as
\begin{equation}
\begin{split}\label{lgc1}
Q_\ve &= \frac{1}{g^2} \int_{\ci^\pm} \dt\big( \tr{ \ve \ast F } \big)  = \frac{1}{g^2} \int_{\ci^\pm} \tr{ \text D \ve  \w \ast F }  + \int_{\ci^\pm}  \tr{ \ve \ast J^\mat  }  .
\end{split}
\end{equation}
Writing this explicitly in flat null coordinates and using the constraints \eqref{softconstraint1} and \eqref{Nzsol}, we get 
\begin{equation}
\begin{split}\label{lgc2}
Q_\ve  &= \frac{2}{g^2}    \int  d^2 z\, \tr{  C^{-1} \ve C \p_z \p_\bz N^\pm } + \frac{1}{2} \int du\, d^2 z\, \tr{ \ve  \left( \frac{4}{g^2} \ \big[ {\hat A}^\pm_z ,  \p_u {\hat A}_\bz^\pm  \big]  + J_u^{\pm\mat} \right)   }  ,
\end{split}
\end{equation}
where using \eqref{scalarbdycond}, \eqref{gaugebdycond} and \eqref{mattercurrent}, we get
\begin{align}
\begin{split}
 J_u^{\pm\mat} &=  \lim_{r\to\pm\infty} r^2 J^\mat_u = \sum_{i=1}^N  \Big( \p_u(\phi^{\pm i})^\ct T_i^a \phi^{\pm i} - (\phi^{\pm i})^\ct T_i^a \p_u \phi^{\pm i} \Big) X^a .
\end{split}
\end{align}
Note that the contribution from $J^\text{inter}$ vanishes on $\ci^\pm$ for the same reasons as those outlined in the paragraph below \eqref{thetakini}.

As promised, we see that just like the phase space, the LGT charge decomposes into a soft part (the first term in \eqref{lgc2}) and a hard part (the second term in \eqref{lgc2}), and we denote the soft (hard) part by $Q_\ve^{\pm \soft}$ ($Q_\ve^{\pm \hard}$). Again, note that while the total LGT charge does not depend on $\ci^+$ or $\ci^-$, its decomposition into a soft and a hard part does.

\paragraph{Isometries} The isometry charge is given by \eqref{stress-tensor-explicit} and \eqref{ABdef11} with the definitions \eqref{st-def}. Using the explicit form of the Lagrangian \eqref{Lag-decomposition}, we find that $(\ast \SH_\xi)^{\mu\nu} = 0$ and
\begin{equation}\label{nonabelian-T}
\begin{split}
T_{\mu\nu} &= \frac{1}{g^2} \tr{ F_{\mu\rho} F_\nu{}^\rho - \frac{1}{4} g_{\mu\nu} F_{\rho\s} F^{\rho\s} } + T_{\mu\nu}^\mat  ,
\end{split}
\end{equation}
where $T_{\mu\nu}^\mat$ is the contribution from the matter fields. Given the structure of the matter Lagrangian \eqref{Lag-decomposition} and \eqref{kinex}, this takes the form
\begin{align}
\begin{split}
	T_{\mu\nu}^\mat = \sum_{i=1}^N \Big(   ( D_{\mu} \Phi^i )^\ct D_{\nu} \Phi^i  +  ( D_{\nu} \Phi^i )^\ct D_{\mu} \Phi^i   - g_{\mu\nu} (D^\rho \Phi^i)^\ct D_\rho \Phi^i \Big) + T_{\mu\nu}^\text{inter} ,
\end{split}
\end{align}
and $T_{\mu\nu}^{\text{inter}}$ is the contribution from $L^{\text{inter}}$. As was the case with the LGT charge, this does not contribute on $\ci^\pm$. It follows upon substituting the explicit form of the Killing vectors \eqref{kv}, the stress tensor \eqref{nonabelian-T}, and the LGT charge \eqref{lgc2} into  \eqref{stress-tensor-explicit} that the isometry charges for translations and Lorentz transformations on $\ci^\pm$ are
\begin{equation}
\begin{split}\label{isom2}
P_f[\ci^\pm] &=  \int du\, d^2 z\,   f  \bigg(  \frac{2}{g^2} \tr{ \p_u {\hat A}_z^\pm \p_u {\hat A}_\bz^\pm  }  +  \sum_{i=1}^N \p_u (\phi^{\pm i})^\ct \p_{u} \phi^{\pm i}   \bigg)  \\
J_Y[\ci^\pm] &=    \frac{2}{g^2} \int d^2 z\, Y^z \tr{  \p_z C^{-1} C \p_z \p_\bz N^\pm  }  \\
& +  \frac{1}{g^2} \int du\, d^2 z\, Y^z \tr{  \p_z {\hat A}^\pm_\bz \overleftrightarrow{\p_u} {\hat A}^\pm_z  -  u \p_z  \left( \p_u {\hat A}^\pm_{z} \p_u {\hat A}_\bz^\pm \right) }   \\
& + \frac{1}{2} \sum_{i=1}^N \int du\, d^2 z \,Y^z \Big(\!\p_u (\phi^{\pm i})^\ct \p_z \phi^{\pm i}  + \p_z (\phi^{\pm i})^\ct \p_u \phi^{\pm i}   - u \p_z \left(  \p_u (\phi^{\pm i} )^\ct \p_u \phi ^{\pm i} \right) \!\! \Big) \\
& +  \cc ,
\end{split}
\end{equation}
where we have used the constraints \eqref{softconstraint1} and \eqref{Nzsol}, as well as the boundary fall-off conditions \eqref{scalarbdycond} and \eqref{gaugebdycond}, to extract the leading non-vanishing terms. By Assumption \eqref{rest3}, we have $P_f[\ci^+] = P_f[\ci^-]$ and $J_Y[\ci^+]=J_Y[\ci^-]$, so we shall simply denote them as $P_f$ and $J_Y$, respectively. These charges also satisfy the charge algebra \eqref{chargealgebra}. Explicitly, using \eqref{kvalgebra}, we can work out
\begin{equation}
\begin{split}
\big\{ P_f , P_{f'} \big\} = 0 , \qquad \big\{ J_Y , P_f \big\} = P_{ ( Y^z\p_z - \frac{1}{2} \p_z Y^z + \cc ) f }  , \qquad \big\{ J_Y , J_{Y'} \big\} = J_{[Y,Y']} .
\end{split}
\end{equation}
Similarly, we can work out the action of these charges on the LGT charge to be
\begin{equation}
\begin{split}
\big\{ P_f , Q_\ve \big\} = 0 , \qquad \big\{ J_Y , Q_\ve \big\} = Q_{Y(\ve)} . 
\end{split}
\end{equation}

\subsubsection{Dirac Brackets}
\label{sec:diracbrackets}

Recall from \eqref{sympformci}, \eqref{presympform1}, and \eqref{Og2} that the full symplectic form is given by
\begin{align}\label{Og3}
\begin{split}
	\bO_{\ci^\pm}(\X,\Y) &=  \frac{2}{g^2} \int d^2 z \, \tr{   \X\big(  \p_z\p_\bz N^\pm  C^{-1}  \big) \Y(C)  - \Y\big(  \p_z\p_\bz N^\pm  C^{-1}  \big) \X(C) } \\
	&\qquad +  \frac{2}{g^2}\int d u\, d^2 z\, \tr{ \p_u\X(\hat A_z^\pm)\Y(\hat A_\bz^\pm) - \p_u\Y(\hat A_z^\pm) \X(\hat A_\bz^\pm)  } \\
	&\qquad +  \sum_{i=1}^N \int du\,d^2 z\, \Big( \p_u\X(\phi^{\pm i})^\ct \Y(\phi^{\pm i}) - \p_u\Y(\phi^{\pm i})^\ct \X(\phi^{\pm i}) \Big),
\end{split}
\end{align}
where the first line is the contribution from the soft part of the gauge field, the second line that from the hard part of the gauge field, and the last line that from the scalar matter fields. We can now use this to determine the Dirac brackets (i.e. the Poisson brackets in the constrained phase space) between the various fields. As we mentioned previously, the fact that the symplectic form breaks up into the three above pieces means that the phase space factorizes into the soft gauge sector, the hard gauge sector, and the matter sector. Thus, the fields living in different sectors have vanishing Dirac brackets, and we can determine the remaining Dirac brackets by studying each sector separately.

We begin by examining the soft sector of the symplectic form, given by the first line of \eqref{Og2}. To rewrite the trace in terms of explicit coordinates, it is convenient to work in the adjoint representation and express $C$ as a matrix in the adjoint representation so that $C(z,\bz) \equiv C^{ab}(z,\bz)$. The components $C^{ab}$ however are not free and satisfy the constraints \eqref{usefulprop}, which are in explicit coordinates given by
\begin{equation}
\begin{split}\label{Cprop}
(C^{ab})^* = C^{ab} , \qquad C^{ac} C^{bc} = C^{ca} C^{cb} = \d^{ab} , \qquad f^{def} C^{ad} C^{be} C^{cf} =  f^{def} C^{da} C^{eb} C^{fc}   = f^{abc} . 
\end{split}
\end{equation}
Recalling that the representation-independent trace $''\text{tr}''$ in \eqref{Og3} is negative the trace in the adjoint representation (see \eqref{rep-indep-trace} with the fact $\CT_{\adj} = 1$), the first line of \eqref{Og3} becomes
\begin{equation}
\begin{split}\label{Og2-gauge-soft}
	\bO^{A,\text{soft}}_{\ci^\pm} ( \X , \Y )  &= \frac{2}{g^2} f^{bcd}  \int d^2 z\,  \Big( \X  (  C^{ac} \p_z \p_\bz N^{\pm d}   ) \Y ( C^{ab} ) - (\X \leftrightarrow \Y) \Big) .  
\end{split}
\end{equation}
While we are now able to determine the Dirac brackets using the equations from Section~\ref{sec:posson-bracket}, given the special form of the above symplectic form, we instead employ an alternative and quicker method, which we shall now describe. 

In classical mechanics, the standard symplectic form is written in Darboux coordinates and is given by
\begin{align}\label{sympclassical}
	\bO(\X,\Y) = \X ( p_i ) \Y (  x^i )  - \Y ( p_i ) \X( x^i ),
\end{align}
and the associated Poisson brackets are $\{ x^i , p_j \} =  \d^i_j$, $\{ x^i , x^j \} = \{ p_i, p_j \} = 0$. Upon comparison, we see that the symplectic form \eqref{Og2-gauge-soft} is precisely of this form, albeit with more complicated quantities in the place of $p_i$ and $x^i$. It then immediately follows that
\begin{equation}
\begin{split}\label{comm2}
\big\{ C^{ab}(z,\bz) , C^{cd}(w,\bw) \big\} &= 0  \\
\big\{ f^{bcd}  C^{ac}  \p_z \p_\bz N^{\pm d} (z,\bz) , f^{b'c'd'}  C^{a'c'}  \p_{w}\p_{\bw} N^{\pm d'} (w,\bw) \big\} &= 0  \\
\big\{ f^{bcd}  C^{ac} \p_z \p_\bz N^{\pm d} (z,\bz) , C^{a'b'} (w,\bw) \big\} &= - \frac{g^2}{2} \d^{aa'} \d^{bb'}  \d^2(z-w).
\end{split}
\end{equation}
From these equations, we can extract the Dirac brackets between $C$ and $N^\pm$ as well as $N^\pm$ with itself. These are given by 
\begin{align}\label{soft-comm}
\begin{split}
	\big\{ N^{\pm a} (z,\bz) , C^{bc} (w,\bw) \big\}&= - \frac{g^2}{4\pi} f^{acd} C^{bd}(w,\bw) \ln|z-w|^2 \\
	\big\{ N^{\pm a} (z,\bz) , N^{\pm b} (w,\bw) \big\} &= -  \frac{g^2}{8\pi^2}  f^{abc}  \int d^2 y \,\ln|z-y|^2 \ln|w-y|^2  \p_y\p_\by N^{\pm c} (y,\by),
\end{split}
\end{align}
where in deriving them we used \eqref{jacobi}, \eqref{adjprop}, and \eqref{Cprop} repeatedly.

Next, we turn to the hard gauge sector of the gauge field symplectic form, which is given by the second line of \eqref{Og3}. This also has the same form as \eqref{sympclassical}, so it immediately follows that the only non-vanishing Dirac bracket is
\begin{equation}\label{hard-comm}
\begin{split}
\big\{ \p_u A_z^{\pm a} (u,z,\bz) , {\hat A}_{\bw}^{\pm a'} (u',w,\bw) \big\}&= - \frac{g^2}{2} \d^{aa'} \d(u-u') \d^2(z-w).
\end{split}
\end{equation}
Integrating in $u$ and fixing the integration constant using the antisymmetry of the bracket, we get
\begin{equation}
\begin{split}
\big\{ {\hat A}_z^{\pm a}(u,z,\bz)  , {\hat A}_{\bw}^{\pm b}(u',w,\bw) \big\} &= - \frac{g^2}{4}\d^{ab}  \,\text{sign} (u-u') \d^2(z-w) .
\end{split}
\end{equation}
Likewise, we want to determine the Dirac bracket involving the matter fields from the third line of \eqref{Og3}. Note that the matter sector and the hard sector of the symplectic form have exactly the same structure, so essentially repeating the same calculation as above we get 
\begin{align}\label{mat-comm}
\begin{split}
    \big\{ \phi^{\pm i}(u,z,\bz),\phi^{\pm j}(u',w,\bw)^\ct \big\} &= -\frac{1}{2}\delta^{ij}\mathbb 1\, \Th(u-u')\delta^2(z-w) ,
\end{split}
\end{align}
where $\mathbb 1$ is the identity matrix in vector space associated to representation $R_i$. This completes our computation of all the Dirac brackets, and to summarize, we collect here all the Dirac brackets on our constrained phase space determined in \eqref{soft-comm}, \eqref{hard-comm}, and \eqref{mat-comm}:
\begin{equation}
\begin{split}
\label{commutator}
\big\{ {\hat A}_z^{\pm a}(u,z,\bz)  , {\hat A}_{\bw}^{\pm b}(u',w,\bw) \big\} &= - \frac{g^2}{4}\d^{ab}  \,\text{sign} (u-u') \d^2(z-w)  \\
\big\{ N^{\pm a}(z,\bz) , C^{bc}(w,\bw) \big\} &= - \frac{g^2}{4\pi} f^{acd} C^{bd}(w,\bw)  \ln|z-w|^2   \\
\big\{ N^{\pm a} (z,\bz) , N^{\pm b}(w,\bw) \big\} &= -  \frac{g^2}{8\pi^2}  f^{abc}  \int d^2 y \, \ln|z-y|^2 \ln|w-y|^2   \p_y\p_\by N^{\pm c} (y,\by) \\
\big\{ \phi^{\pm i}(u,z,\bz),\phi^{\pm j}(u',w,\bw)^\ct \big\} &= -\frac{1}{2}\delta^{ij}\,\mathbb 1 \, \Th(u-u')\delta^2(z-w) \\
\text{all others} &= 0 . 
\end{split}
\end{equation}
For future use, we also compute the Dirac brackets involving the constrained fields $C_z$ and $N_z^\pm$. Using \eqref{softconstraint1} and \eqref{Nzsol} and substituting them into \eqref{commutator}, we get
\begin{align}\label{conscomm}
\begin{split}
	\big\{ C_z^a, C_w^b \big\} &= \big\{ C_z^a , C_\bw^b \big\} = 0 \\
	\big\{ C_z^a, N_w^{\pm b}\big\} &= -\frac{g^2}{4\pi} \frac{C^{ac}(z,\bz)C^{bc}(w,\bw)}{(z-w)^2} \\
	\big\{ C_\bz^a, N_w^{\pm b} \big\} &= \frac{g^2}{2}\delta^{ab}\delta^2(z-w) \\
	\big\{ N_z^{\pm a}, N_w^{\pm b} \big\} &= 0 \\
	\big\{ N_z^{\pm a}, N_\bw^{\pm b} \big\} &= \frac{g^2}{2}f^{abc}C^{cd}(w,\bw)N^{\pm d}(w,\bw)\delta^2(z-w) \\
	&\qquad - \frac{g^2}{8\pi^2}f^{dec}\int d^2y\,\frac{C^{ad}(z,\bz)C^{be}(w,\bw)}{(\bz-\by)^2(w-y)^2}N^{\pm c}(y,\by).
\end{split}
\end{align}
Using the brackets \eqref{commutator} and \eqref{conscomm}, we can now verify that \eqref{lgc2} and \eqref{isom2} indeed generate the appropriate canonical transformations on the phase space. First, we write the charges out in the adjoint representation as
\begin{align}
\begin{split}\label{adjointcharges}
Q_\ve  &=  \frac{2}{g^2}    \int  d^2 z\, \ve^a C^{ab}  \p_z \p_\bz N^{\pm b}  + \frac{2}{g^2} f^{abc} \int du\, d^2 z \, \ve^a  {\hat A}^{\pm b}_z \p_u {\hat A}_\bz^{ \pm c} \\
&~~~ + \frac{1}{2}\sum_{i=1}^N \int du\, d^2 z\, \ve^a \Big( \p_u(\phi^{\pm i})^\ct T_i^a \phi^{\pm i} - (\phi^{\pm i})^\ct T_i^a \p_u \phi^{\pm i} \Big)  , \\
P_f  &=  \int du\, d^2 z\,  f  \bigg(  \frac{2}{g^2} \p_u {\hat A}_z^{\pm a} \p_u {\hat A}_\bz^{\pm a} +  \sum_{i=1}^N \p_u (\phi^{\pm i})^\ct \p_{u} \phi^{\pm i}   \bigg)  \\
J_Y &=    \frac{2}{g^2} \int d^2 z\, Y^z C_z^a C^{ab}  \p_z \p_\bz N^{\pm b}  \\
&~~~ +  \frac{1}{g^2} \int du\, d^2 z\, Y^z \left( \p_z {\hat A}^{\pm a}_\bz \overleftrightarrow{\p_u} {\hat A}^{\pm a}_z  -  u \p_z  \left( \p_u {\hat A}^{\pm a}_{z} \p_u {\hat A}_\bz^{\pm a} \right) \right)   \\
&~~~ + \frac{1}{2} \sum_{i=1}^N \int du\, d^2 z \,Y^z \Big( \p_u (\phi^{\pm i})^\ct \p_z \phi^{\pm i}  + \p_z (\phi^{\pm i})^\ct \p_u \phi^{\pm i}   - u \p_z \left(  \p_u (\phi^{\pm i} )^\ct \p_u \phi ^{\pm i} \right)  \Big) \\
&~~~ +  \cc .
\end{split}
\end{align}
It then follows from \eqref{commutator} that
\begin{equation}
\begin{split}
	\{ Q_\ve , \cdot \} = - \d_\ve (\ \cdot\ ) , \qquad  \{ P_f , \cdot \} = - \d_f (\ \cdot\ ),  \qquad \{ J_Y , \cdot \} = - \d_Y (\ \cdot\ ) ,
\end{split}
\end{equation}
where
\begin{equation}
\begin{split}\label{chargeaction-all}
	\d_\ve C &= \ve C , \qquad \d_\ve N^\pm = 0 , \qquad \d_\ve {\hat A}_z^\pm = - [ {\hat A}_z^\pm , \ve ] , \qquad \d_\ve \phi^{\pm i} = \ve^a T^a_i \phi^{\pm i} , \\
\d_f C &= \d_f N^\pm = 0 , \qquad \d_f {\hat A}_z^\pm = f \p_u {\hat A}_z^\pm , \qquad \d_f \phi^{\pm i} = f \p_u \phi^{\pm i} , \\
\d_Y C &= ( Y^z \p_z  + Y^\bz \p_\bz ) C    , \qquad \d_Y N^\pm = ( Y^z \p_z  + Y^\bz \p_\bz )  N^\pm , \\
\d_Y {\hat A}_z^\pm &= \left[ Y^z \p_z + Y^\bz \p_\bz + \p_z Y^z + \frac{1}{2} ( \p_z Y^z + \p_\bz Y^\bz ) u \p_u \right] {\hat A}_z^\pm  , \\
\d_Y \phi^{\pm i} &= \left[ Y^z \p_z + Y^\bz \p_\bz + \frac{1}{2} ( \p_z Y^z + \p_\bz Y^\bz ) u \p_u \right] \phi^{\pm i}.
\end{split}
\end{equation}

\section{Canonical Quantization}\label{sec:hs}

In the previous section, we examined in detail the phase space for classical non-abelian gauge theories, and in particular described it separately in terms of local $\ci^+$ and $\ci^-$ variables. We now want to elevate our classical fields to quantum fields and quantize the phase space via canonical quantization. This involves finding an irreducible representation $\CR : \CF ( \G )  \to \CH^*$, where $\CH$ is a Hilbert space, so that functions on the phase space $\G$ (or equivalently, points in $\G$) are mapped to linear operators acting on a Hilbert space $\CH$. Under this map, Dirac brackets $\{\cdot ,\cdot \}$ on the phase space become quantum commutators $-i[\cdot , \cdot ]$ on the Hilbert space, and complex conjugation on the phase space becomes taking the adjoint on $\CH$, i.e.
\begin{equation}
\begin{split}\label{canonicalquantprop}
\big[ \CR (f) , \CR(g) \big] = i \CR( \{ f , g \} ) , \qquad \CR(f^*) = \CR(f)^\dagger.
\end{split}
\end{equation}
As we will work exclusively in the Hilbert space, we will simply denote $\CR(f)$ by $f$.

Recall from \eqref{Hfactor1} and \eqref{Hfactor2} that the phase space $\G$ factorizes into the soft and hard gauge sectors and the matter sectors. This implies that the corresponding Hilbert space also factorizes as
\begin{equation}
\begin{split}
\CH =  \CH^{\pm A,\text{soft}}   \times \CH^{\pm A,\text{hard}} \times \CH^{\pm1} \times \cdots \times \CH^{\pm N}.
\end{split}
\end{equation}
We now proceed to explore each of these sectors below.

\subsection{Radiative Hilbert Space: \texorpdfstring{$\CH^{\pm A,\text{hard}} \times \CH^{\pm1} \times \cdot \cdot \cdot \times \CH^{\pm N}$}{}}\label{sec:hsrad}

We start by describing the space of hard states, which are spanned by the hard operators ${\hat A}^\pm_z$ and the matter fields $\phi^{\pm i}$. Using \eqref{canonicalquantprop}, the quantum commutators for the hard fields are determined from  \eqref{commutator} to be
\begin{equation}
\begin{split}\label{Ahatcomm}
\left[ {\hat A}_z^{\pm a}(u,z,\bz)  , {\hat A}_{\bz'}^{\pm b}(u',z',\bz')  \right]&= - \frac{i g^2}{4} \d^{ab} \, \text{sign}(u-u') \d^2(z-z')  \\
\left[ \phi^{\pm i} ( u , z , \bz ) , \phi^{\pm j} ( u' , z' , \bz' )^{\dag} \right] &= - \frac{i}{2} \d^{ij}  \mathbb 1\, \text{sign}(u-u') \d^2(z-z') ,
\end{split}
\end{equation}
and the fields obey the adjoint property
\begin{equation}\label{realitycond}
\begin{split}
({\hat A}^{\pm a}_z)^\dagger = {\hat A}^{\pm a}_\bz , \qquad ( \phi^{\pm i} )^\dagger = \phi^{\pm i*} . 
\end{split}
\end{equation}
We now attempt to construct the radiative Hilbert space as a Fock space. The first step will be to define creation and annihilation operators. For $\o > 0$, the annihilation operators are defined as\footnote{When we integrate these operators over $\o$, e.g. when taking the Fourier transform, we will adopt the Cauchy prinicpal value method for resolving the $\frac{1}{\o}$ singularity.}
\begin{equation}
\begin{split}\label{Odef}
	\CO^{\pm a}_\+ (p(\o,z,\bz)) &\equiv - \frac{4\sqrt{2}\pi}{g}  \frac{1}{\o} \int du\, e^{ \frac{i}{2} \o u} F_{uz}^{\pm a}  \\
	\CO^{\pm a}_\- (p(\o,z,\bz)) &\equiv - \frac{4\sqrt{2}\pi}{g}  \frac{1}{\o} \int du \, e^{ \frac{i}{2} \o u} F_{u\bz}^{\pm a}  \\
	\CO_{\Phi^i}^\pm ( p  ( \o , z , \bz ) ) &\equiv   -\frac{4\pi}{\o} \int  du \, e^{\frac{i}{2} \o u } \p_u\phi^{\pm i}  \\
	\CO_{\Phi^{i \dagger}}^\pm ( p  ( \o , z , \bz ) ) &\equiv    -\frac{4\pi}{\o} \int du \, e^{\frac{i}{2} \o u } \p_u\phi^{\pm i\dagger} ,
\end{split}
\end{equation}
where the $(\pm)$ subscript denotes the helicity, and the (on-shell) momentum is parameterized in flat null coordinates as
\begin{equation}
\begin{split}\label{mompar}
p^\mu(\o,z,\bz) = \frac{\o}{2} \left( 1+|z|^2  ,  z+\bz ,  -i(z-\bz)  ,  1-|z|^2  \right) . 
\end{split}
\end{equation}
Using \eqref{realitycond}, we note that the creation operators are related to $\eqref{Odef}$ by the sign of $\o$, so that
\begin{align}\label{Odaggerprop}
\begin{split}
	\CO^{\pm a}_{(\pm)} (p)^\dag &= -\CO^{\pm a}_{(\mp)}(-p), \quad \CO^{\pm}_{\Phi^i} (p)^\dag = -\CO^{\pm }_{\Phi^{i\dag}}(-p).	
\end{split}
\end{align}
Thus, we see that the subscript $(\pm)$ indicates the helicity only if $p^0 > 0$, and negative the helicity if $p^0 < 0$. Using \eqref{Ahatcomm}, it is straightforward to verify that the nonzero commutators between creation and annihilation operators are
\begin{equation}
\begin{split}\label{Ocomm}
	\Big[ \CO^{\pm a}_{(h)} (p) , \CO^{\pm b}_{(h')} (p')^\dag \Big] &=    (2\pi)^3 \d_{h,h'} \d^{ab} (2p^0) \d^3( \vec{p} - \vec{p}\,' )  \\
	\Big[ \CO^{\pm}_{\Phi^i} (p) , \CO^{\pm}_{\Phi^j} (p')^\dag \Big] &= \Big[ \CO^{\pm}_{\Phi^{i\dag}} (p) , \CO^{\pm}_{\Phi^{j\dag}} (p')^\dag \Big] =    (2\pi)^3  (2p^0) \delta^{ij}\mathbb 1\, \d^3( \vec{p} - \vec{p}\,' ) ,
\end{split}
\end{equation}
where we used the identity
\begin{equation}
\begin{split}
(2p^0) \d^\3 ( \vec{p} - \vec{p}\,')  = \frac{ 4 }{ \o } \d^2 ( z - z' ) \d ( \o - \o' )  . 
\end{split}
\end{equation}
Because \eqref{Ocomm} is the standard commutation relation for creation and annihilation operators, this verifies that the operators in \eqref{Odef} can indeed be understood as annihilation operators. 

We can now construct the Hilbert space in the usual way, and define vacuum states as those that are annihilated by all annihilation operators, i.e.
\begin{equation}
\begin{split}
\CO^{\pm a}_{(h)} ( p )  \ket{ U , \pm } = \CO_{\Phi^i}^\pm ( p )  \ket{ U, \pm }  = \CO_{\Phi^{i\dag}}^\pm (p)\ket{U, \pm} = 0 . 
\end{split}
\end{equation}
As we will explore in Section~\ref{sec:hsvac}, the vacuum state is not unique, but is instead an infinite-dimensional space spanned by basis states labeled by $U(z,\bz) \in \CG$. The remaining (basis) states in the hard Hilbert space are then constructed by acting on the vacuum state with creation operators. A typical hard state has the form
\begin{equation}
\begin{split}
\CO_{(h_1)}^{\pm a_1} (p_1)^\dag   \cdots \CO_{(h_l)}^{\pm a_l} (p_l)^\dag \CO^{\pm}_{\Phi^{i_1}} (p'_1)^\dag \cdots \CO^{\pm}_{\Phi^{i_m}} (p'_{m} )^\dag   \CO^{\pm}_{\Phi^{j_1\dagger}} (p''_1)^\dag   \cdots \CO^{\pm}_{\Phi^{j_n\dagger}} (p''_{n} )^\dag    \ket{U, \pm } .
\end{split}
\end{equation}

\subsection{Soft Hilbert Space: \texorpdfstring{$\CH^{\pm A,\text{soft}}$}{}}\label{sec:hsvac}

The soft Hilbert space is spanned by the soft operators $C$ and $N^\pm$. These soft operators commute with the translation generators $P_f$ from \eqref{isom2}, and hence in particular with the Hamiltonian of the theory $H = P_{f=1}$.\footnote{This Hamiltonian generates translations in $u$, while the Hamiltonian that generates translations in $x^0$ is $P_{f=1+|z|^2}$.} This means $C$ and $N^\pm$ must carry zero energy and momentum, which immediately implies that the vacuum state is not unique, and that there is an infinite-dimensional space of vacua generated by acting on any vacuum state repeatedly with the soft operators. In this subsection, we shall characterize this space of vacua. 

The algebra of operators in the soft Hilbert space is obtained by applying \eqref{canonicalquantprop} to \eqref{commutator} to yield
\begin{equation}
\begin{split}\label{softcomm1}
\big[ C^{ab}(z,\bz) , C^{cd}(w,\bw) \big] &= 0 \\
\big[ N^{\pm a}(z,\bz) , C^{bc}(w,\bw) \big] &= - \frac{ig^2}{4\pi} f^{acd} C^{bd}(w,\bw)  \ln|z-w|^2   \\
\big[ N^{\pm a} (z,\bz) , N^{\pm b}(w,\bw) \big] &= -  \frac{ig^2}{8\pi^2}  f^{abc}  \int d^2 y \, \ln|z-y|^2 \ln|w-y|^2   \p_y\p_\by N^{\pm c} (y,\by),
\end{split}
\end{equation}
which also implies the commutators by \eqref{conscomm}
\begin{align}\label{conscomm1}
\begin{split}
	\big[ C_z^a, C_w^b \big] &= \big[ C_z^a , C_\bw^b \big] = 0 \\
	\big[ C_z^a, N_w^{\pm b}\big] &= -\frac{ig^2}{4\pi} \frac{C^{ac}(z,\bz)C^{bc}(w,\bw)}{(z-w)^2} \\
	\big[ C_\bz^a, N_w^{\pm b} \big] &= \frac{ig^2}{2}\delta^{ab}\delta^2(z-w) \\
	\big[ N_z^{\pm a}, N_w^{\pm b} \big] &= 0 \\
	\big[ N_z^{\pm a}, N_\bw^{\pm b} \big] &= \frac{ig^2}{2}f^{abc}C^{cd}(w,\bw)N^{\pm d}(w,\bw)\delta^2(z-w) \\
	&\qquad - \frac{ig^2}{8\pi^2}f^{dec}\int d^2y\,\frac{C^{ad}(z,\bz)C^{be}(w,\bw)}{(\bz-\by)^2(w-y)^2}N^{\pm c}(y,\by).
\end{split}
\end{align}
From \eqref{Nzsol}, \eqref{Cprop} and \eqref{canonicalquantprop}, we have
\begin{equation}
\begin{split}
C^{ab}(z,\bz)^\dagger = C^{ab}(z,\bz) , \qquad N^{\pm a}(z,\bz)^\dagger = N^{\pm a}(z,\bz) . 
\end{split}
\end{equation}

We now construct the soft Hilbert space as follows. Since the Hermitian operators $C^{ab}$ commute with each other, there exists an orthogonal basis on $\CH^{\pm A,\soft}$ that diagonalizes these operators. Labeling these basis states for $\CH^{\pm A,\soft}$ by $\ket{U,\pm}$, we have 
\begin{equation}
\begin{split}\label{Ceigenstatedef}
C^{ab} (z ,\bz) \ket{U,\pm} = U^{ab}(z,\bz) \ket{U,\pm}, \qquad U (z,\bz) \in \CG.
\end{split}
\end{equation}
These states can be normalized so that
\begin{equation}
\begin{split}
\braket{ U,\pm }{ U',\pm } = \d ( U - U' ) , \quad \int [\dt U] \, \d(U-U') f(U') = f(U) ,
\end{split}
\end{equation}
where the measure $[dU]$ is taken to be the left-invariant Haar measure on $\CG$ so that
\begin{equation}
\begin{split}\label{measureginv}
[dU] = [d(g U)] , \qquad g \in \CG \quad \implies \quad \d(g U-g U') = \d(U-U') . 
\end{split}
\end{equation}
Therefore, a generic state in $\CH^{\pm A,\soft}$ can be written in the $U$ basis as
\begin{equation}
\begin{split}
\ket{f,\pm} = \int [ \dt U] \ket{U,\pm} f(U) \quad \implies \quad f(U) = \braket{U,\pm}{f,\pm},
\end{split}
\end{equation}
where $f(U)$ is known as the vacuum wave-function.

We now want to determine how $N^\pm$ acts on $|U,\pm\ra$. Inserting the commutator involving $N^\pm$ and $C$ between two vacuum states, we get using \eqref{commutator} 
\begin{align}\label{eqn1}
\begin{split}
	 (U(w,\bw)  - U'(w,\bw))\la U',\pm|N^{\pm a}(z)|U,\pm\ra &=  -\frac{ig^2}{4\pi}U(w,\bw)t^a\ln|z-w|^2 \delta(U-U') .
\end{split}
\end{align}
If we recall the derivative operator $\mfd_{U(y,\by)}^b$ introduced in \eqref{daction} (and explored more fully in Appendix~\ref{app:deriv}), which is defined so that its action on $U$ is
\begin{align}\label{Uaction}
    \mfd_{U(y,\by)}^bU(z,\bz) = -X^b U(y,\by)\delta^2(z-y),
\end{align}
then one can show after some algebra that \eqref{eqn1} is satisfied given that
the action of $N^\pm$ on the basis states is
\begin{equation}
\begin{split}
\label{Naction}
N^{\pm a}(z,\bz) \ket{U,\pm} = -  \frac{ig^2}{4\pi}  \int d^2 y\, \ln | z - y |^2 U^{ba}(y,\by) \mfd_{U(y,\by)}^b \ket{U,\pm}.
\end{split}
\end{equation}

We can now determine how the charge $Q_\ve$ and $J_Y$ act on the vacuum states (recall $P_f$ annihilates the vacuum states). Applying \eqref{Naction} to \eqref{lgc2} and \eqref{isom2} and noting that the hard part of the charges annihilate the vacuum, we obtain
\begin{equation}
\begin{split}\label{vacuumaction}
	Q_\ve \ket{U,\pm} &= -   i \int d^2z \, \ve^a(z,\bz)  \mfd_{U(z,\bz)}^a  \ket{ U,\pm }  , \qquad J_Y \ket{ U ,\pm} = i  \int d^2 z\,  Y^z U_z^a  \,\mfd_{U(z,\bz)}^a \ket{ U,\pm } ,
\end{split}
\end{equation}
where $U_z = U \p_z U^{-1}$, and we had to utilize the third property of \eqref{Dproperties} in deriving this result. The fact that $J_Y^\pm$ acts non-trivially on the vacuum states $|U,\pm\ra$ means that states in the infinite-dimensional space of vacua are generically not Lorentz invariant. This may be surprising, since it is in contradiction to a standard assumption made in perturbative QFT regarding gauge theories -- that the vacuum is unique and Lorentz invariant. As we hope to explore further in future work, it is precisely this dissonance that leads to the presence of infrared divergences. For now, however, we observe that by \eqref{vacuumaction}, the $U=1$ vacuum state is Lorentz invariant. We will assume that this is the standard vacuum from perturbative QFT, and shall refer to this as the ``QFT vacuum.''
 
Finally, we want to determine how $|U,\pm\ra$ transforms under a finite LGT. The charge that generates finite LGTs is obtained by exponentiating \eqref{vacuumaction}, so that 
\begin{equation}
    \begin{split}
        \O_{\g} = \exp [ - i Q_\ve ] , 
    \end{split}
\end{equation}
where $\g = \exp \ve$. Using \eqref{canonicalquantprop} to elevate \eqref{fieldbrackets} and \eqref{QQcom} to quantum commutators, i.e.
\begin{align}
	\big[ Q_\ve , \Phi^i \big] = -iR_i(\ve)\Phi^i,  \qquad \big[ Q_\ve, Q_{\ve'} \big] = iQ_{[\ve,\ve']},
\end{align}
we can derive the properties
\begin{equation}
\begin{split}\label{finitelgt}
	\O_\g \O_{\g'} = \O_{\g\g'} , \qquad \O_\g^{-1} \CO \O_\g = R(\g)\CO ,
\end{split}
\end{equation}
where $\CO$ is any operator living in a representation $R$. Using in particular the second property above, we can show that
\begin{equation}
\begin{split}\label{finitelgt1}
	\O_\g \ket{U, \pm} = \ket{ \g U, \pm} ,
\end{split}
\end{equation}
where the overall normalization of $|\g U, \pm \ra$ is fixed using \eqref{measureginv}. Thus, we see that an LGT parameterized by $\g$ takes an eigenstate of $C^{ab}$ with eigenvalue $U^{ab}$ to another eigenstate of $C^{ab}$ with eigenvalue $(\g U)^{ab}$.

\subsection{The \texorpdfstring{$S$}{}-matrix}
\label{sec:Smatrix-def}

A fundamental quantity of interest in QFTs is the $S$-matrix, or the scattering amplitude, and it captures the overlap between an $(n-m)$-particle $in$-state with an $m$-particle $out$-state. Given an $in$-vacuum $|U',-\ra$  and an $out$-vacuum $|U,+\ra$, the $S$-matrix is given via the LSZ reduction formula (we have suppressed explicit color/flavor indices on the operators to avoid notational clutter)
\begin{equation}
\begin{split}\label{Andef}
	&\CA_n ( U,+ | p_1  , \ldots , p_n  | U',- ) = \bra{ U,+ } T \left\{ \big[\CO_1\big]_{h_1}(p_1) \cdots \big[\CO_n\big]_{h_n}(p_n) \right\} \ket{ U' ,- },
\end{split}
\end{equation}
where $T$ is the time-ordering operator (it moves all $out$-operators to the left and $in$-operators to the right), and
\begin{equation}
\begin{split}\label{OdefLSZ}
\big[ \CO_k \big]_{h} (p) \equiv  i \lim_{p^2 \to0}   p^2  \int d^4 x \,  e^{- i p \cdot x } \ve_{(h)}^{\mu_1 \cdots \mu_{|h|} } ( p ) [\Phi_k]_{\mu_1 \cdots \mu_{|h|} } (x) , 
\end{split}
\end{equation}
where $h$ labels the helicity of particle if the energy $p^0$ is positive (the particle is outgoing), and labels negative the helicity of the particle if the energy $p^0$ is negative (the particle is incoming).\footnote{In general, we adopt the convention where the $\pm$ subscript labeling an operator indicates the helicity of the operator if the corresponding particle is outgoing. This means when we refer to the helicity of a particle, we implicitly assume that it is outgoing unless otherwise specified.} Note that $\Phi^i$ is \emph{any} normalized local operator that creates or annihilates the one-particle state corresponding to $\CO_i$ with polarization tensor $\ve_{(h)}^{\mu_1\cdots\mu_{|h|}}$. To write the polarization tensor explicitly, note that it satisfies the properties
\begin{equation}
\begin{split}
\ve_{(\pm|h|)}^{\mu_1 \cdots \mu_{|h|} } ( p ) = \ve_{(\pm)}^{\mu_1} ( p ) \cdots \ve_{(\pm)}^{\mu_{|h|}} (p) , \qquad \ve_\+(p)^* = \ve_\-(p) , \qquad \ve_{(h)}(p) \cdot \ve_{(h')}(p)^* = \d_{h,h'},
\end{split}
\end{equation}
so in the gauge $A_u = 0$, we have
\begin{equation}
\begin{split}\label{polpar} 
\ve_{(+)}^\mu = \frac{1}{\sqrt{2}} ( \bz , 1 , - i , - \bz )  , \qquad \ve_{(-)}^\mu = \frac{1}{\sqrt{2}} ( z , 1 ,  i , - z ), 
\end{split}
\end{equation}
where we parameterized the momentum using \eqref{mompar}. In particular, note that $\ve^\mu_{(+)}$ labels either an outgoing positive helicity gluon or an incoming negative helicity gluon, whereas $\ve^\mu_{(-)}$ labels either an outgoing negative helicity gluon or an incoming positive helicity gluon.

We now want to apply \eqref{OdefLSZ} to the case of the gauge field $A$, i.e. we want to evaluate
\begin{equation}\label{OALSZ}
\begin{split}
\CO_\pm^a (p) \equiv  \frac{i}{g} \lim_{p^2 \to0}   p^2  \int d^4 x \,  e^{- i p \cdot x } \ve_\pmm^\mu (p) A_\mu^{a}(x) . 
\end{split}
\end{equation}
Parametrizing the off-shell momentum in flat null coordinates as
\begin{equation}
\begin{split}\label{coordpar}
p^\mu &= \frac{\o}{2}  \left( 1 + |z|^2 + \mu ,  z + \bz , -i ( z - \bz ) , 1 - |z|^2 - \mu \right) , 
\end{split}
\end{equation}
so that $p^2 = -\mu\o^2$, we can rewrite the on-shell limit $p^2 \to 0$ as $\mu\to 0$. Evaluating \eqref{OALSZ} for an outgoing positive helicity (or incoming negative helicity) operator in these coordinates, we obtain
\begin{equation}
\begin{split}
	\CO_+^a(p)  &= - \frac{\sqrt{2} i \o^2}{4g}  \lim_{\mu \to 0}   \mu   \int du \, dr \,d^2 w \,    r e^{\frac{i\o u}{2} + \frac{i\o r}{2}( |z-w|^2 + \mu )  } A_w^a(u,r,w,\bw) \\
	&=-\frac{\sqrt 2i\o^2}{4g}\lim_{\mu\to0} \frac{1}{\mu} \int du\,dr\,d^2w\, r e^{\frac{i\o u}{2} + \frac{i\o r}{2\mu}( |z-w|^2+\mu)}\Big(  A_z^a\big(u,r\mu^{-1},z,\bz \big) + O(z-w) \Big)
\end{split}
\end{equation}
where $O(z-w)$ captures all terms proportional to $z-w$ and we rescaled $r \to r/\mu$ in the second line. Dividing the integral into the regions $r>0$ and $r<0$, and noting that the $\mu \to 0$ limit sends $A_z \to A_z^\pm$ in those regions, we obtain
\begin{equation}
\begin{split}
	\CO_{+}^a(p)  &= \frac{\sqrt{2}i}{g} \lim_{\mu \to 0}  \int  du\, d^2 w \, e^{\frac{i\o u}{2} }  \frac{ \mu}{(|z-w|^2 + \mu   )^2 }  \big( A_z^{+a} (u,z,\bz) - A_z^{-a} (u,z,\bz)  + O(z-w) \big) . 
\end{split}
\end{equation}
Next, taking the on-shell limit $\mu \to 0$, and observing the identity
\begin{equation}\label{deltaLim}
\begin{split}
 \lim_{\mu \to 0}   \frac{\mu}{(|z-w|^2 + \mu   )^2 }  = 2\pi \delta^2(z-w),
\end{split}
\end{equation}
we get
\begin{equation}
\begin{split}
	\CO_{+}^a(p)  &=  \frac{2\pi\sqrt{2}i}{g} \int du \, e^{\frac{i\o u}{2} } \big( {\hat A}_z^{+a}(u,z,\bz) - {\hat A}_z^{-a}(u,z,\bz) \big) ,
\end{split}
\end{equation}
where we have decomposed the gauge field into the soft and hard modes via \eqref{softconstraint1}; notice that the delta function from \eqref{deltaLim} has eliminated all the $O(z-w)$ terms. Finally, noting that $F_{uz}^{\pm a} = \p_u \hat A_z^{\pm a}$ and using \eqref{Odef}, we get
\begin{equation}\label{gluonO}
\begin{split}
\CO_{+}^a(p)  &=  \CO_\+^{+a}(p) -  \CO_\+^{-a}(p)  .
\end{split}
\end{equation}

When all the particles are \emph{hard}, the $S$-matrix evaluated via the LSZ reduction formula is simply an overlap between the \emph{in}- and \emph{out}-states. To see this, we note that when $\o > 0$, the time-ordering operator in \eqref{Andef} moves $\CO^{-a}_\+(p)$ all the way to the right to annihilates the ket vacuum state, so only the first term contributes. If $\o < 0$, the operators are creation operators according to \eqref{Odaggerprop}, and the time-ordering operator moves $\CO^{+a}_\+(p)$ all the way to the left to annihilates the bra vacuum state, so only the second term contributes. Therefore, when inserting $\CO_{+}^a$ into an $S$-matrix element with $\o \not= 0$,
\begin{equation}
\begin{split}
\CO_{+}^a(p) &=  \begin{cases}
\CO^{+a}_\+(p)  & \o > 0  \\
\CO^{-a}_\-(-p)^\dag  & \o < 0 , 
\end{cases}
\end{split}
\end{equation}
where we used \eqref{Odaggerprop} to write the $\o < 0$ case explicitly as a creation operator.

On the other hand, consider the operator insertion of \eqref{gluonO} in the soft ($\o \to 0$) limit. Expanding the operator insertion near $\o = 0$, we get by substituting \eqref{Odef} into \eqref{gluonO}
\begin{equation}
\begin{split}\label{Osoft}
    \CO_{+ }^a(p)  &=  - \lim_{\o\to0}\frac{4\sqrt{2}\pi}{g}  \frac{1}{\o}\big( N_z^{+a} - N_z^{-a} \big) + O(\o^0)  .
\end{split}
\end{equation}
Repeating this procedure, we could similarly get for an outgoing negative helicity gluon
\begin{align}\label{Osoftneg}
    \CO_{-}^a(p)  &=  - \lim_{\o\to0}\frac{4\sqrt{2}\pi}{g}  \frac{1}{\o}\big( N_\bz^{+a} - N_\bz^{-a} \big) + O(\o^0)  .
\end{align}

\section{Soft Factorization of the \texorpdfstring{$S$}{}-matrix}\label{sec:softfactor}

The $S$-matrix defined in \eqref{Andef} is a complicated quantity that, in general, depends on all the details of the theory. However, as we will show in this section, its dependence on the vacuum state is completely fixed, and the result is given by \eqref{mainresult-intro} in the introduction. We will derive this result in Section~\ref{sec:mainresult} (see \eqref{mainresult}). In the subsequent subsections, we show how the leading single and consecutive double soft theorems follow from \eqref{mainresult}.

\subsection{Ward Identity}\label{sec:mainresult}

Consider the insertion of the operator $C^{ab}(z,\bz)$ into a scattering amplitude involving vacua $|U,+\ra$ and $|U',-\ra$, two eigenstates of $C^{ab}$. Since $C^{ab}(z,\bz)$ commutes with all operators with energies not strictly zero (including the soft operator $N_z^{+a} - N_z^{-a}$, which was shown in \eqref{Osoft} to arise as a soft limit), we have
\begin{equation}
\begin{split}
\!\!\!\! \bra{ U,+  } C^{ab}(z,\bz) T \{ \CO_1(p_1) \cdots \CO_n(p_n) \} \ket{ U' ,-}  = \bra{ U ,+ } T \{ \CO_1(p_1) \cdots \CO_n(p_n) \} C^{ab}(z,\bz)  \ket{ U' ,- } . 
\end{split}
\end{equation}
The vacua are $C^{ab}$ eigenstates, so it follows by \eqref{Ceigenstatedef} and the fact that $C^{ab}$ is Hermitian that
\begin{equation}
\begin{split}
	\left[ U^{ab}(z,\bz) - U'^{ab}(z,\bz) \right] \bra{ U,+ } T \{ \CO_1(p_1) \cdots \CO_n(p_n) \} \ket{ U',- }   = 0 ,
\end{split}
\end{equation}
which in turn implies
\begin{equation}
\begin{split}\label{prop1}
 \bra{ U ,+ } T \{ \CO_1(p_1) \cdots \CO_n(p_n) \} \ket{ U',- }   = \d(U-U') \avg{  \CO_1(p_1) \cdots \CO_n(p_n) }_{U} ,
\end{split}
\end{equation}
where the time-ordering operator is henceforth implicitly included in the $\la \cdots \ra_{U}$ correlator. To evaluate the right-hand-side, we first note that obviously
\begin{equation}\label{mainint}
\begin{split}
	\avg{ \O_\g^{-1} \CO_1(p_1) \cdots \CO_n(p_n) \O_\g }_{U} = \big\langle [ \O_\g^{-1} \CO_1(p_1) \O_\g ] \cdots  [ \O_\g^{-1} \CO_n(p_n) \O_\g ] \big\rangle_{U} .
\end{split}
\end{equation}
Recalling from \eqref{finitelgt} that 
\begin{equation}
\begin{split}
  \O_\g^{-1} \CO_k(p_k) \O_\g = [ \CO_k ]_\g (p_k) = R_k ( \g(z_k,\bz_k) ) \CO_k(p_k), 
\end{split}
\end{equation}
substituting this and \eqref{finitelgt1} into \eqref{mainint} yields
\begin{equation}
\begin{split}
\avg{  \CO_1(p_1) \cdots \CO_n(p_n)  }_{U} = R_1(\g(z_1,\bz_1))  \cdots R_n(\g(z_n,\bz_n))\avg{  \CO_1(p_1) \cdots \CO_n(p_n)  }_{\g^{-1} U} . 
\end{split}
\end{equation}
As this is true for any $\g \in \CG$, we can set $\g = U$ so that
\begin{equation}
\begin{split}\label{prop2}
\avg{  \CO_1(p_1) \cdots \CO_n(p_n)  }_{U} = R_1(U(z_1,\bz_1))  \cdots R_n(U(z_n,\bz_n))\avg{  \CO_1(p_1) \cdots \CO_n(p_n)  }_{U=1} . 
\end{split}
\end{equation}
Substituting this into \eqref{prop1} and reinstating the explicit color/flavor indices $i_k$ and $j_k$, we immediately obtain
\begin{align}\label{mainresult}
\begin{split}
&\bra{U,+} T \{ \CO_1^{i_1}(p_1) \cdots \CO_n^{i_n}(p_n) \} \ket{ U',- } \\
&\qquad \qquad = \d(U-U') R_1(U(z_1,\bz_1))^{i_1}{}_{j_1} \cdots  R_n(U(z_n,\bz_n))^{i_n}{}_{j_n} \avg{  \CO_1^{j_1}(p_1) \cdots \CO_n^{j_n}(p_n) }_{U=1} , 
\end{split}
\end{align}
which is exactly \eqref{mainresult-intro} as promised. Note that the scattering amplitude on the right-hand-side is simply the standard $U=1$ perturbative QFT $S$-matrix element, which we can evaluate using Feynman diagrams and soft theorems. Therefore, the above equation allows us to determine the scattering amplitude between any $in$-vacuum $|U',-\ra$ with any $out$-vacuum $|U,+\ra$. Since the $|U,\pm\ra$ vacua form a complete basis of the vacuum sector, this means we can now compute the scattering amplitude involving any hard operators as well as the soft limit of such operators between any two arbitrary $in$- and $out$-vacua. 

We conclude this subsection with the following observation. Recall that the operator $C$ is only defined up to the identification \eqref{gidentification}, which means if $\g \in \CG$ is a constant in spacetime, we must have
\begin{equation}
\begin{split}
\avg{  \CO_1(p_1) \cdots \CO_n(p_n)  }_{U=\g}  = \avg{  \CO_1(p_1) \cdots \CO_n(p_n)  }_{U=1}.  
\end{split}
\end{equation}
Applying \eqref{prop2} to this case involving a constant $\g$, we get 
\begin{equation}
\begin{split}
\label{colorcons}
\avg{  \CO_1(p_1) \cdots \CO_n(p_n)  }_{U=1} = R_1(\g) \cdots R_n(\g) \avg{  \CO_1(p_1) \cdots \CO_n (p_n)  }_{U=1} . 
\end{split}
\end{equation}
This is simply the statement of global color charge conservation of the $S$-matrix, since if $\g = 1 + \ve^a X^a$ is an infinitesimal global gauge transformations, then \eqref{colorcons} becomes 
\begin{equation}
\begin{split}\label{colorconsprop}
\sum_{k=1}^n T^a_k \avg{  \CO_1(p_1) \cdots \CO_n (p_n)  }_{U=1}  = 0 ,
\end{split}
\end{equation}
which is the standard global color conservation equation.
More generally, we can similarly derive that in a non-trivial $U$ vacuum, global gauge conservation is given by
\begin{equation}
\begin{split}
	\sum_k U^{ba}(z_k,\bz_k) T^b_k \avg{  \CO_1(p_1) \cdots \CO_n(p_n)  }_U &= 0,
\end{split}
\end{equation}
where we used \eqref{usefulprop} to write it in the above form.

\subsection{Single Soft Gluon Limit}
\label{sec:singlesoft1}

The soft gluon theorem describes the factorization of a scattering amplitude in which one or many gluons are soft, i.e. they have energies much smaller than the typical energy scale of the scattering amplitude. If $m$ gluons have soft momenta $q_i$ in an $(n+m)$-point scattering amplitude, the factorization is of the form
\begin{equation}
\begin{split}
\CA_{n+m} \xrightarrow{q_i \to 0} \SS_m \CA_{n} ,
\end{split}
\end{equation}
where $\CA_{n}$ is the scattering amplitude involving the remaining $n$ hard particles, and the soft factor $\SS_m$  (which may be an operator) depends on the quantum numbers (e.g. momentum, color, flavor) of the external states but does not depend on other details of the theory.\footnote{The subleading soft gluon theorem, which we will not discuss here, depends \emph{very} loosely on the interaction terms in the Lagrangian. In particular, the kinematical structure of the subleading term in $\SS_m$ is universal, but the overall normalization is not.} Expanding $\SS_m$ as a Laurent series in terms of the energies of the soft gluons, it has at leading order the structure
\begin{equation}
\begin{split}
	\SS_m = \frac{1}{q^0_1 \cdots q^0_m } \left[ {\hat \SS}_m  + \CO(q_1,\cdots,q_m) \right] ,
\end{split}
\end{equation}
where ${\hat \SS}_m$ depends on (1) the momentum, color, and flavors of the hard particles, (2) the directions of the soft gluons and (3) the ratio of energies between various individual soft gluons. In the rest of this subsection, we will explore the soft factor ${\hat \SS}_m$ for $m=1$, i.e. the single soft gluon limit, and show that the factorization of the $S$-matrix follows from the Ward identity \eqref{mainresult}. We will then generalize this to include multiple consecutive soft gluon limits in the next subsection.

For the case $m=1$, we can derive using Feynman diagrams in perturbative QFT (which means the fields live in the $U=1$ vacuum) that the amplitude in our conventions factorizes in the soft limit as
\begin{equation}
\begin{split}
\label{softgthm}
	\avg{\CO^a_h(q) \CO_1(p_1) \cdots \CO_n(p_n)  }_{U=1} \xrightarrow{q\to0} i  g \sum_{k=1}^n  \frac{p_k \cdot \ve_{(h)}(q) }{ p_k \cdot q - i \e  } T^a_k  \avg{  \CO_1(p_1) \cdots \CO_n(p_n)  }_{U=1} ,
\end{split}
\end{equation}
where the correlators $\la \cdots \ra$ implicitly include a time-ordering operator. Using the momentum parameterization \eqref{mompar} to write $q \equiv (\o,z,\bz)$ and $p_k \equiv (\o_k,z_k,\bz_k)$, as well as \eqref{polpar}, the soft theorem \eqref{softgthm} becomes, depending on the helicity $h$,
\begin{equation}
\begin{split}
\label{softgthm1}
 \lim_{\o \to 0} \o \avg{\CO^a_{+}(q) \CO_1(p_1) \cdots \CO_n(p_n)  }_{U=1} = \sqrt{2}  i  g \sum_{k=1}^n \frac{T^a_k}{z-z_k} \avg{  \CO_1(p_1) \cdots \CO_n(p_n)  }_{U=1}  \\
  \lim_{\o \to 0} \o \avg{\CO^a_{- }(q) \CO_1(p_1) \cdots \CO_n(p_n)  }_{U=1} = \sqrt{2}  i  g \sum_{k=1}^n \frac{T^a_k}{\bz-\bz_k} \avg{  \CO_1(p_1) \cdots \CO_n(p_n)  }_{U=1} .
\end{split}
\end{equation}
We will now show that \eqref{softgthm1} follows directly from the Ward identity \eqref{mainresult}. First, we recall from \eqref{Osoft} that inserting an outgoing positive (or incoming negative) helicity soft gluon inside the $S$-matrix corresponds to
\begin{equation}\label{singleward}
\begin{split}
&\lim_{\o \to 0} \o \avg{ \CO_{+}^a(q) \CO_1(p_1) \cdots \CO_n(p_n)  }_U= - \frac{4\sqrt{2}\pi}{g} \avg{  \big( N_z^{+a} - N_z^{-a}  \big) \CO_1(p_1) \cdots \CO_n(p_n)  }_U  .
\end{split}
\end{equation}
Because of the implicit time-ordering operator in the correlator, $N_z^+$ and $N_z^-$ are moved all the way to the left and right, respectively, to act on the vacuum. Using \eqref{Nzsol} and  \eqref{Naction}, we know that
\begin{equation}
\begin{split}
N_z^{\pm a} \ket{U,\pm} &= - \frac{ig^2}{4\pi}  \int d^2 y \,\frac{ U^{ab}(z,\bz) U^{cb}(y,\by) }{ z - y }  \mfd_{U(y,\by)}^c \ket{ U,\pm } . 
\end{split}
\end{equation}
It follows after some algebra that
\begin{align}\label{Nz-insertion}
\begin{split}
	&\avg{ U,+ | T\big\{ \big(N_z^{+a} - N_z^{-a} \big) \CO_1\cdots\CO_n \big\} |U',- } \\
	 &\qquad\qquad =  \frac{ig^2}{4\pi}\delta(U-U') \int d^2y\, \frac{ U^{ab}(z,\bz)U^{cb}(y,\by)}{z-y} \mfD^c_{U(y,\by)} \la \CO_1\cdots\CO_n \ra_U,
\end{split}
\end{align}
where we had to use the second and third lines of \eqref{Dproperties}. Using \eqref{prop1} and substituting this into the right-hand-side of \eqref{singleward}, we get
\begin{equation}
\begin{split}\label{softlimitres}
&\lim_{\o \to 0} \o \avg{ \CO_{+}^a(p) \CO_1(p_1) \cdots \CO_n(p_n)  }_U \\
&\qquad\qquad = - \sqrt{2} i g   \int d^2 y \,\frac{ U^{ab}(z,\bz) U^{cb}(y,\by) }{ z - y }  \mfd_{U(y,\by)}^c  \avg{ \CO_1(p_1) \cdots \CO_n(p_n)  }_U  .
\end{split}
\end{equation}
Finally, expanding out $\la \CO_1 \cdots \CO_n \ra_U$ via \eqref{prop2} and acting on all the factors of $U$ with $\mfd^c_{U(y,\by)}$ using \eqref{Uaction}, we obtain the single soft gluon theorem in a general $U$ vacuum state:
\begin{equation}
\begin{split}\label{softres1}
&\lim_{\o \to 0} \o \avg{ \CO_{+}^a(p) \CO_1(p_1) \cdots \CO_n(p_n)  }_U =  \sqrt{2} i g \sum_{k=1}^n  \frac{ U^{ab}(z,\bz) U^{cb}(z_k,\bz_k) }{ z - z_k }  T_k^c  \avg{ \CO_1(p_1) \cdots \CO_n(p_n)  }_U.
\end{split}
\end{equation}
Setting $U=1$ results in the first line of \eqref{softgthm1}. The second line of \eqref{softgthm1} is shown similarly, except we would start by computing the $S$-matrix element involving an outgoing negative (or incoming positive) helicity soft gluon $\CO^a_{-}$ instead of \eqref{singleward}. This completes our derivation of the single soft gluon theorem from the Ward identity \eqref{mainresult}.

\subsection{Multiple Consecutive Soft Gluon Limits}
\label{sec:multisoft1}

In the previous subsection, we considered the single soft gluon limit and computed the soft factor $\SS_{m=1}$ from \eqref{mainresult}. This result can be easily extended to the case where $m>1$ gluons are taken to be soft in a consecutive manner. Since the gluons are taken to be soft one at a time, we can determine the soft gluon factor $\SS_m$ by repeatedly applying the single soft gluon theorem $m$ times. We can then derive the multiple consecutive soft gluon theorem from the Ward identity \eqref{mainresult} by applying the argument given in the previous subsection $m$ times.

Nevertheless, it is interesting to compute the commutator of consecutive soft limits
\begin{equation}\label{conscom}
\begin{split}
\left[ \lim_{\o \to 0} , \lim_{\o' \to 0} \right] \o\o' \avg{ \CO_h^{a}(q)  \CO_{h'}^{a'}(q')  \CO_1(p_1) \cdots \CO_n(p_n) }_U ,
\end{split}
\end{equation}
since there are two methods of doing the computation, and they should certainly agree! The first method is to simply take the approach mentioned in the above paragraph and evaluate \eqref{conscom} by taking the two single soft limits one at a time and then taking their difference. The second method is to demonstrate that the commutator of limits in \eqref{conscom} can be related to the commutator of soft modes, which will allow us to use the commutators \eqref{softcomm1} (or equivalently \eqref{conscomm1}) and reduce \eqref{conscom} to a single soft limit. We will now demonstrate that these two methods yield the same answer, thereby serving as a verification of the commutators \eqref{softcomm1} on the Hilbert space. For simplicity, though, we will only work in the $U=1$ vacuum.

We begin by evaluating \eqref{conscom} using the first method. For conciseness, we will denote the set of hard operators collectively as
\begin{align}
    \mxx \equiv \CO_1(p_1)\cdots \CO_n(p_n).
\end{align}
Now, taking $q'$ soft first (i.e. $q'\ll q , p_k$) and using the single soft theorem \eqref{softgthm}, we find
\begin{equation}
\begin{split}
	\avg{  \CO^a_{h}(q)  \CO^{a'}_{h'}(q') \mxx }_{U=1}  \xrightarrow{q'\to0} i g \left[  f^{aa'c} \frac{ q \cdot \ve_{(h')}(q')  }{ q \cdot q' - i \e } + \d^{ac} \sum_{{k'}=1}^n \frac{ p_{k'} \cdot \ve_{(h')}(q') }{ p_{k'} \cdot q' - i \e } T^{a'}_{k'} \right] \avg{  \CO^c_{h}(q)   \mxx }_{U=1} ,
\end{split}
\end{equation}
where we used the fact $\CO_h^a(q)$ transform in the adjoint representation. Next, taking $q$ soft and repeating the procedure, we find that the consecutive double soft gluon theorem is
\begin{equation}\label{cons2soft}
\begin{split}
&\avg{  \CO^a_h(q)  \CO^{a'}_{h'}(q') \mxx }_{U=1}  \\
&\qquad \xrightarrow{q'\to0~\text{then}~q\to0} (i g)^2 \left[  f^{aa'c} \frac{ q \cdot \ve_{(h')}(q')  }{ q \cdot q' - i \e } + \d^{ac} \sum_{{k'}=1}^n \frac{ p_{k'} \cdot \ve_{(h')}(q') }{ p_{k'} \cdot q' - i \e } T^{a'}_{k'} \right] \sum_{k=1}^n \frac{ p_k \cdot \ve_{(h)}(q) }{ p_k \cdot q - i \e } T^c_k \avg{  \mxx }_{U=1} .
\end{split}
\end{equation}
Because we want to compute the commutator of soft limits in \eqref{conscom}, we also need to compute the two soft limits in the opposite order. This is easily determined by exchanging the primed and unprimed quantities in \eqref{cons2soft}, and the answer is
\begin{equation}
\begin{split}
&\avg{  \CO^a_h(q)  \CO^{a'}_{h'}(q') \mxx }_{U=1}  \\
&\qquad \xrightarrow{q\to0~\text{then}~q'\to0}  (i g)^2 \left[ f^{a'ac} \frac{ q' \cdot \ve_{(h)}(q)  }{ q' \cdot q - i \e } + \d^{a'c} \sum_{k=1}^n \frac{ p_k\cdot \ve_{(h)}(q) }{ p_k \cdot q - i \e } T^a_k \right] \sum_{k'=1}^n \frac{ p_{k'} \cdot \ve_{(h')}(q') }{ p_{k'} \cdot q' - i \e } T^c_{k'} \avg{  \mxx }_{U=1} .
\end{split}
\end{equation}
This means the commutator of soft limits is
\begin{equation}\label{commsoftlim}
\begin{split}
\left[ \lim_{\o \to 0} , \lim_{\o' \to 0} \right] \o\o'  \avg{  \CO^a_h(q)  \CO^{a'}_{h'}(q') \mxx }_{U=1} = - (i g)^2 f^{aa'c} \o\o' \sum_{k=1}^n \Delta_{(h,h')}(p_k)  T^c_k \avg{  \mxx }_{U=1}  ,
\end{split}
\end{equation}
where
\begin{equation}
\begin{split}\label{Deltahdef}
\Delta_{(h,h')}(p_k)  &\equiv  \frac{ p_k\cdot \ve_{(h)}(q) }{ p_k \cdot q - i \e } \frac{ p_k \cdot \ve_{(h')}(q') }{ p_k \cdot q' - i \e }  + \frac{ \ve_{(h)}(q) \cdot \ve_{(h')}(q')}{ q \cdot q' - i \e } \\
&\qquad \qquad \qquad \qquad - \frac{ q \cdot \ve_{(h')}(q')  }{ q \cdot q' - i \e } \frac{ p_k \cdot \ve_{(h)}(q) }{ p_k \cdot q - i \e } - \frac{ q' \cdot \ve_{(h)}(q)  }{ q' \cdot q - i \e }    \frac{ p_k \cdot \ve_{(h')}(q') }{ p_k \cdot q' - i \e } . 
\end{split}
\end{equation}
Note that in deriving \eqref{commsoftlim} we used color conservation \eqref{colorconsprop}. Evaluating $\Delta_{(h,h')}(p_k)$ in flat null coordinates, we get
\begin{equation}
\begin{split}
\Delta_{(\pm,\pm)}(p_k) = 0 , \qquad \Delta_{(+,-)}(p_k) = - \frac{1}{\o\o'} \frac{2}{|z-z'|^2} \frac{ \bz - \bz_k }{ z - z_k } \frac{ z' - z_k }{ \bz' - \bz_k } .
\end{split}
\end{equation}
Substituting this back into \eqref{commsoftlim}, we find that the commutator of two consecutive soft limits is
\begin{equation}
\begin{split}\label{softcommutatorresults}
\left[ \lim_{\o \to 0} , \lim_{\o' \to 0} \right] \o\o'  \avg{  \CO^a_{\pm}(q)  \CO^{a'}_{\pm}(q') \mxx }_{U=1} &=  0   \\
\left[ \lim_{\o \to 0} , \lim_{\o' \to 0} \right] \o\o'  \avg{  \CO^a_{+}(q)  \CO^{a'}_{-}(q') \mxx }_{U=1} &=   (\sqrt{2} i g)^2  \frac{f^{aa'c}  }{|z-z'|^2}  \sum_{k=1}^n\frac{ \bz - \bz_k }{ z - z_k } \frac{ z' - z_k }{ \bz' - \bz_k }  T^c_k \avg{  \mxx }_{U=1}  .
\end{split}
\end{equation}

We now want to show that we obtain the same answer using the second method of evaluating \eqref{conscom}. Begin by considering the first line of \eqref{softcommutatorresults}, where we are inserting the soft operators $\CO^a_+(q)\CO^{a'}_+(q')$. Taking $q'$ soft and using \eqref{Osoft}, we get
\begin{equation}
\begin{split}
	&\lim_{\o' \to 0} \o' \avg{ \CO_{+}^{a}(q)  \CO^{a'}_{+}(q')  \mxx }_{U=1}  = - \frac{4\sqrt{2}\pi}{g}  \avg{  \CO^{a}_{+}(q) \big( N_{z'}^{+a'} - N_{z'}^{-a'} \big) \mxx }_{U=1} . 
\end{split}
\end{equation}
Because there is an implicit time-ordering operator in the correlator $\la \cdots \ra_{U=1}$, this means $N_z^+$ is moved all the way to the left and $N_z^-$ is moved all the way to the right. Next, taking $q$ soft, we get
\begin{equation}\label{1order}
\begin{split}
	&\lim_{\o \to 0}\lim_{\o' \to 0} \o \o' \avg{  \CO_{+}^{a}(q)  \CO^{a'}_{+}(q')   \mxx }_{U=1} \\
	&\qquad\qquad = \frac{32\pi^2}{g^2}   \avg{ \big( N_{z'}^{+a'}  N_z^{+a} - N_{z'}^{+a'}  N_z^{-a} - N_z^{+a} N_{z'}^{-a'} + N_z^{-a} N_{z'}^{-a'} \big) \mxx }_{U=1}  .
\end{split}
\end{equation}
Note that the ordering of the operators is determined by the fact $N_{z'}^{+a'}$ is on the left of $N_{z}^{+a}$ since the $q'\to 0$ limit is taken first, so $N_{z'}^{+a'}$ is moved to the left first. Similarly, $N^{-a'}_{z'}$ is on the right of $N_z^{-a}$ (the ordering of the two remaining terms is just due to the implicit time-ordering operator). Since we want to compute a commutator of two soft limits, we also need to compute the correlator when the soft limits are taken in reverse. Repeating the above procedure yields
\begin{equation}\label{2order}
\begin{split}
	&\lim_{\o' \to 0} \lim_{\o \to 0}\o \o' \avg{ \CO_{+}^{a}(q)  \CO^{a'}_{+}(q')  \mxx }_{U=1} \\
	&\qquad\qquad = \frac{32\pi^2}{g^2}   \avg{ \big( N_z^{+a}  N_{z'}^{+a'} - N_{z'}^{+a'}  N_z^{-a} - N_z^{+a} N_{z'}^{-a'} +  N_{z'}^{-a'} N_z^{-a} \big) \mxx }_{U=1}  .
\end{split}
\end{equation}
Subtracting \eqref{2order} from \eqref{1order}, we find
\begin{equation}\label{commline1}
\begin{split}
	&\left[ \lim_{\o \to 0} , \lim_{\o' \to 0} \right] \o \o' \avg{   \CO_{+}^{a}(q)  \CO^{a'}_{+}(q')  \mxx }_{U=1} = - \frac{32\pi^2}{g^2}   \avg{ \big( \big[    N_z^{+a} , N_{z'}^{+a'}  \big]  - \big[ N_z^{-a} , N_{z'}^{-a'}   \big] \big)\mxx }_{U=1}  .
\end{split}
\end{equation}
We can now evaluate the commutators between the constrained soft modes using \eqref{conscomm1}, and since $\big[ N_z^{\pm a} , N_{z'}^{\pm a'} \big] = 0$, this immediately implies the first line of \eqref{softcommutatorresults}, where both operators have positive helicity. 

Similarly, we want to verify the first line of \eqref{softcommutatorresults} for the case where both soft operators $\CO^a_-(q)\CO^{a'}_{-}(q')$ have negative helicity. The procedure is almost exactly the same as that used to derive \eqref{commline1}, except because we are inserting $\CO_{-}^a\CO_{-}^{a'}$ instead of $\CO_{+}^a\CO_{+}^{a'}$ on the left-hand-side of \eqref{softcommutatorresults}, according to \eqref{Osoftneg} we simply need to replace $N_z^{\pm a}$ with $N_\bz^{\pm a}$ and $N_{z'}^{\pm a'}$ with $N_{\bz'}^{\pm a'}$. Making the replacements in \eqref{commline1} and using the fact that $\big[N_\bz^{\pm a}, N_{\bz'}^{\pm a'}\big] = 0$, the commutator of soft limits vanish as well.

Finally, we want to verify the second line of \eqref{softcommutatorresults} using the second method. As we mentioned in the previous paragraph, because we are inserting $\CO_{-}^{a'}$ instead of $\CO_{+}^{a'}$ on the left-hand-side of \eqref{softcommutatorresults}, we just need to replace $N_{z'}^{\pm a'}$ with $N_{\bz'}^{\pm a'}$ in \eqref{commline1}, resulting in
\begin{equation}
\begin{split}
	&\left[ \lim_{\o \to 0} , \lim_{\o' \to 0} \right] \o \o' \avg{ \CO_{+}^{a}(q)  \CO^{a'}_{-}(q')  \mxx }_{U=1} = - \frac{32\pi^2}{g^2}   \avg{ \big( \big[    N_z^{+a} , N_{\bz'}^{+a'}  \big]  - \big[ N_z^{-a} , N_{\bz'}^{-a'}   \big] \big)\mxx }_{U=1}  .
\end{split}
\end{equation}
Evaluating the commutators using \eqref{conscomm1}, we get
\begin{equation}\label{intstep}
\begin{split}
	&\left[ \lim_{\o \to 0} , \lim_{\o' \to 0} \right] \o \o' \avg{  \CO_{+}^{a}(q)  \CO^{a'}_{-}(q') \mxx }_{U=1} \\
&\quad = 4 i f^{aa'c} \int \!\! d^2 y \, \left[ \frac{ 1 }{ ( z - y )^2 ( {\bz'} - \by )^2   } - 4\pi^2 \d^2(z-y) \delta^2 ( z - {z'} )   \right] \avg{ \big( N^{+ c}(y,\by)  - N^{-c}(y,\by)  \big)\mxx }_{U=1}  .
\end{split}
\end{equation}
Now, we want to derive the insertion of $N^{+a} - N^{-a}$ between two $U=1$ vacua. Using \eqref{Naction} and following similar steps used in deriving \eqref{Nz-insertion}, we obtain
\begin{align}
\begin{split}
	& \avg{ \big( N^{+a}(y,\by) - N^{-a}(y,\by) \big) \mxx }_{U} =  - \frac{ig^2}{4\pi} \sum_{k=1}^n  \ln |y-z_k|^2 U^{ba}(z_k,\bz_k) T_k^b \avg{  \mxx }_{U},
\end{split}
\end{align}
which means for the special case where $U=1$, we have
\begin{equation}
\begin{split}
& \avg{ \big( N^{+a}(y,\by) - N^{-a}(y,\by) \big) \mxx }_{U=1} =  - \frac{ig^2}{4\pi} \sum_{k=1}^n  \ln |y-z_k|^2 T^{a}_k \avg{  \mxx }_{U=1} .
\end{split}
\end{equation}
Substituting this back into \eqref{intstep} we obtain
\begin{equation}
\begin{split}
&\left[ \lim_{\o \to 0} , \lim_{\o' \to 0} \right] \o \o' \avg{ \CO_{+}^{a}(q)  \CO^{a'}_{-}(q')  \mxx }_{U=1} \\
&\qquad \qquad  =  \frac{g^2}{\pi}  f^{aa'c} \sum_{k=1}^n    \left[ \int d^2 y  \frac{\ln |y - z_k|^2}{ ( z - y )^2 ( {\bz'} - \by )^2   } - 4\pi^2 \delta^2 ( z - {z'} )\ln |z - z_k|^2   \right]  T^a_k \avg{  \mxx }_{U=1} \\
&\qquad \qquad = -\frac{2g^2}{|z-z'|^2}   f^{aa'c} \sum_{k=1}^n  \frac{ \bz - \bz_k }{ z - z_k } \frac{ z' - z_k }{ \bz' - \bz_k }    T^c_k \avg{  \mxx }_{U=1},
\end{split}
\end{equation}
where in the last step we used global color conservation \eqref{colorconsprop}. Comparing this with the second line of \eqref{softcommutatorresults}, we see that they match exactly.

\appendix

\section*{Acknowledgements}

We would like to thank Daniel Kapec, Alok Laddha, Sebastian Mizera, Sabrina Pasterski, and Shu-Heng Shao for useful discussions. TH is grateful to be supported by U.S. Department of Energy grant DE-SC0009999 and by funds from the University of California. PM gratefully acknowledges support from U.S. Department of Energy grant DE-SC0009988 and from the Infosys Fellowship.

\section{Explicit Derivation of Select Equations}
\label{app:explicitcalculations}

\subsection{Derivation of (\ref{id1})}
\label{app:id12proof}

In this section, we prove the identities
\begin{equation}
\begin{split}\label{A1}
\nabla_\mu \CA^{\mu\nu} + \CR^\nu{}_{\mu\rho\sigma} \CB^{\mu\rho\sigma} = 0 , \qquad \CA^{[\mu\nu]}  + \nabla_\rho \CB^{\rho\mu\nu} = 0 ,
\end{split}
\end{equation}
where $\CA^{\mu\nu}$ and $\CB^{\mu\nu\rho}$ are given in \eqref{ABdef} to be
\begin{equation}
\begin{split}\label{ABdef1}
\CA^{\mu\nu} &\equiv  g^{\mu\nu} \CL + \sum_i \sum_{n=1}^\infty \sum_{k=1}^n (-1)^k \nabla_{\mu_2 \cdots \mu_k} \Pi_i^{\mu \mu_2 \cdots \mu_n}  \nabla^\nu \nabla_{\mu_{k+1} \cdots \mu_n} \varphi^i  \\
\CB^{\mu\nu\rho} &\equiv \frac{1}{2} \sum_i \sum_{n=1}^\infty \sum_{k=1}^n (-1)^k \nabla_{\mu_2 \cdots \mu_k} \Pi_i^{\mu \mu_2 \cdots \mu_n} \big( \S_i^{\nu\rho} \big)_{\mu_{k+1} \cdots \mu_n}{}^{\nu_{k+1} \cdots \nu_n}\nabla_{\nu_{k+1} \cdots \nu_n} \varphi^i .
\end{split}
\end{equation}
Starting with the first equation in \eqref{A1}, we use \eqref{ABdef1} to write
\begin{equation}
\begin{split}\label{id1test1}
	&\nabla_\mu \CA^{\mu\nu} + \CR^\nu{}_{\mu\rho\sigma} \CB^{\mu\rho\sigma} \\
	&\qquad = \nabla^\nu \CL + \sum_i\sum_{n=1}^\infty \sum_{k=1}^n (-1)^k \nabla_\mu  \nabla_{\mu_2 \cdots \mu_k} \Pi_i^{\mu \mu_2 \cdots \mu_n}  \nabla^\nu \nabla_{\mu_{k+1} \cdots \mu_n} \varphi^i \\
	&\qquad\qquad + \sum_i\sum_{n=1}^\infty \sum_{k=1}^n (-1)^k \nabla_{\mu_2 \cdots \mu_k} \Pi_i^{\mu \mu_2 \cdots \mu_n}  \nabla_\mu \nabla^\nu \nabla_{\mu_{k+1} \cdots \mu_n} \varphi^i \\
	&\qquad\qquad + \frac{1}{2} \CR^\nu{}_{\mu\rho\s} \sum_i\sum_{n=1}^\infty \sum_{k=1}^n (-1)^k \nabla_{\mu_2 \cdots \mu_k} \Pi_i^{\mu \mu_2 \cdots \mu_n} \big( \S_i^{\rho\s} \big)_{\mu_{k+1} \cdots \mu_n}{}^{\nu_{k+1} \cdots \nu_n}\nabla_{\nu_{k+1} \cdots \nu_n} \varphi^i \\
	&\qquad = \sum_i \sum_{n=0}^\infty \Pi_i^{\mu_1\cdots\mu_n} \nabla^\nu \nabla_{\mu_1 \cdots \mu_n} \varphi^i + \sum_i\sum_{n=1}^\infty \sum_{k=1}^n (-1)^k \nabla_{\mu_1 \cdots \mu_k} \Pi_i^{\mu_1 \cdots \mu_n}  \nabla^\nu \nabla_{\mu_{k+1} \cdots \mu_n} \varphi^i \\
	&\qquad \qquad  -  \sum_i\sum_{n=1}^\infty \sum_{k=0}^{n-1} (-1)^k \nabla_{\mu_1 \cdots \mu_{k} } \Pi_i^{\mu_1 \cdots \mu_n}   \nabla^\nu  \nabla_{\mu_{k+1} \cdots \mu_n} \varphi^i ,
\end{split}
\end{equation}
where in obtaining the second equality we used \eqref{covdevantisym} to rewrite the last term, used the fact $\Pi_i^{\mu\mu_2\cdots\mu_n}$ is completely symmetric to symmetrize covariant derivatives, and then relabeled the summation index.

It is clear that the second and third terms cancel when $1\leq k\leq n-1$. Therefore, we have
\begin{align}
\begin{split}
	\nabla_\mu \CA^{\mu\nu} + \CR^\nu{}_{\mu\rho\sigma} \CB^{\mu\rho\sigma} &= \sum_i \sum_{n=0}^\infty \Pi_i^{\mu_1\cdots\mu_n} \nabla^\nu \nabla_{\mu_1\cdots\mu_n} \varphi^i  + \sum_i \sum_{n=1}^\infty (-1)^n \nabla_{\mu_1\cdots\mu_n} \Pi_i^{\mu_1\cdots\mu_n}\nabla^\nu \varphi^i  \\
	&\qquad - \sum_i \sum_{n=1}^\infty   \Pi_i^{\mu_1\cdots\mu_n}\nabla^\nu\nabla_{\mu_1 \cdots \mu_n} \varphi^i \\
	&= \sum_i \sum_{n=0}^\infty (-1)^n \nabla_{\mu_1\cdots\mu_n} \Pi_i^{\mu_1\cdots\mu_n}\nabla^\nu \varphi^i \\
	&= 0,
\end{split}
\end{align}
where we used in the second equality that the first and third term cancel except for the $n=0$ case, and in the last equality we noted that the term vanishes by \eqref{eom1} with the fact $*\CE_i = 0$. This proves the first equality in \eqref{A1}.

Next, we turn to the second equation in \eqref{A1}. Again, we find using \eqref{ABdef1}
\begin{equation}
\begin{split}\label{A8}
\CA^{[\mu\nu]} + \nabla_\rho \CB^{\rho\mu\nu} &= \frac{1}{2} \sum_{n=1}^\infty \sum_{k=1}^n (-1)^k \nabla_{\mu_2 \cdots \mu_k} \Pi_i^{\mu\mu_2 \cdots \mu_n}  \nabla^{\nu} \nabla_{\mu_{k+1} \cdots \mu_n} \varphi^i  \\
&\quad - \frac{1}{2} \sum_{n=1}^\infty \sum_{k=1}^n (-1)^k \nabla_{\mu_2 \cdots \mu_k} \Pi_i^{\nu\mu_2 \cdots \mu_n}  \nabla^{\mu} \nabla_{\mu_{k+1} \cdots \mu_n} \varphi^i   \\
&\quad + \frac{1}{2} \sum_{n=1}^\infty \sum_{k=1}^n (-1)^k \nabla_{\mu_1 \cdots \mu_k} \Pi_i^{\mu_1 \cdots \mu_n} \big( \S_i^{\mu\nu} \big)_{\mu_{k+1} \cdots \mu_n}{}^{\nu_{k+1} \cdots \nu_n}\nabla_{\nu_{k+1} \cdots \nu_n} \varphi^i \\
&\quad + \frac{1}{2} \sum_{n=1}^\infty \sum_{k=1}^n (-1)^k \nabla_{\mu_2 \cdots \mu_k} \Pi_i^{\rho \mu_2\cdots \mu_n} \big( \S_i^{\mu\nu} \big)_{\mu_{k+1} \cdots \mu_n}{}^{\nu_{k+1} \cdots \nu_n} \nabla_{\rho}  \nabla_{\nu_{k+1} \cdots \nu_n} \varphi^i ,
\end{split}
\end{equation}
where we have used the fact that the Lorentz generators $\S_i$ are covariantly constant and used $\Pi_i^{\mu_1\cdots\mu_n}$ to symmetrize the covariant derivatives. To simplify this expression, we note the identity
\begin{equation}
\begin{split}\label{A10}
	\big( \S_i^{\mu\nu} \big)_{\s ( \mu_{k+1} \cdots \mu_n)}{}^{\rho (\nu_{k+1} \cdots \nu_n)} = \d^\rho_\s \big( \S_i^{\mu\nu} \big)_{(\mu_{k+1} \cdots \mu_n)}{}^{(\nu_{k+1} \cdots \nu_n)} + \big( \S_\text{vec}^{\mu\nu} \big)_\s{}^\rho \delta_{(\mu_{k+1}}^{\nu_{k+1}} \cdots \delta_{\mu_n)}^{\nu_n}, 
\end{split}
\end{equation}
which means using the explicit form of $\S^{\mu\nu}_\text{vec}$ from Footnote~\ref{vector-rep-lorentz}, we have
\begin{equation}
\begin{split}\label{A11}
\d^\rho_\s \big( \S_i^{\mu\nu} \big)_{(\mu_{k+1} \cdots \mu_n)}{}^{(\nu_{k+1} \cdots \nu_n)} = \big( \S_i^{\mu\nu} \big)_{\s ( \mu_{k+1} \cdots \mu_n)}{}^{\rho (\nu_{k+1} \cdots \nu_n)} - \big(\delta^\mu_\sigma g^{\nu\rho} - \delta^\nu_\sigma g^{\mu\rho}\big)\d_{(\mu_{k+1}}^{\nu_{k+1}} \cdots \delta_{\mu_n)}^{\nu_n}. 
\end{split}
\end{equation}
Using this, the last term in \eqref{A8} becomes
\begin{align}
\begin{split}
	&\frac{1}{2} \sum_{n=1}^\infty \sum_{k=1}^n (-1)^k \nabla_{\mu_2 \cdots \mu_k} \Pi_i^{\rho\mu_2 \cdots \mu_n} \big( \S_i^{\mu\nu} \big)_{\mu_{k+1} \cdots \mu_n}{}^{\nu_{k+1} \cdots \nu_n} \nabla_{\rho}  \nabla_{\nu_{k+1} \cdots \nu_n} \varphi^i \\
	&\qquad = \frac{1}{2}\sum_i \sum_{n=1}^\infty \sum_{k=1}^n (-1)^k \nabla_{\mu_2\cdots\mu_k} \Pi_i^{\sigma\mu_2\cdots\mu_n} \big(\S_i^{\mu\nu}\big)_{\sigma(\mu_{k+1}\cdots\mu_n)}{}^{\rho(\nu_{k+1}\cdots\nu_n)} \nabla_{\rho}\nabla_{\nu_{k+1} \cdots \nu_n} \varphi^i \\
	&\qquad\qquad - \frac{1}{2}\sum_i \sum_{n=1}^\infty \sum_{k=1}^n (-1)^k \nabla_{\mu_2\cdots\mu_k} \Pi_i^{\mu\mu_2\cdots\mu_n}   \nabla^{\nu}\nabla_{\mu_{k+1} \cdots \mu_n} \varphi^i \\
	&\qquad\qquad + \frac{1}{2}\sum_i \sum_{n=1}^\infty \sum_{k=1}^n (-1)^k \nabla_{\mu_2\cdots\mu_k} \Pi_i^{\nu\mu_2\cdots\mu_n}  \nabla^{\mu}\nabla_{\mu_{k+1} \cdots \mu_n} \varphi^i
\end{split}
\end{align}
Substituting this back into \eqref{A8} and cancelling terms, we get
\begin{equation}
\begin{split}\label{A12}
	&\CA^{[\mu\nu]} + \nabla_\rho \CB^{\rho\mu\nu} \\
	&\qquad = \frac{1}{2} \sum_i\sum_{n=1}^\infty \sum_{k=1}^n (-1)^k \nabla_{\mu_1 \cdots \mu_k} \Pi_i^{\mu_1 \cdots \mu_n} \big( \S_i^{\mu\nu} \big)_{\mu_{k+1} \cdots \mu_n}{}^{\nu_{k+1} \cdots \nu_n}\nabla_{\nu_{k+1} \cdots \nu_n} \varphi^i \\
	&\qquad\qquad + \frac{1}{2} \sum_i\sum_{n=1}^\infty \sum_{k=1}^n (-1)^k \nabla_{\mu_2 \cdots \mu_k} \Pi_i^{\s \mu_2 \cdots \mu_n}\big( \S_i^{\mu\nu} \big)_{\s ( \mu_{k+1} \cdots \mu_n)}{}^{\rho (\nu_{k+1} \cdots \nu_n)} \nabla_{\rho}  \nabla_{\nu_{k+1} \cdots \nu_n} \varphi^i .
\end{split}
\end{equation}
Finally, we note that since $\Pi_i^{\sigma\mu_2\cdots\mu_n}$ is completely symmetric, all the lower indices of $\S_i^{\mu\nu}$ in the second sum are naturally symmetrized. Since Lorentz transformations do not modify the symmetry properties of a tensor and $\S_i^{\mu\nu}$ lies in the tensor representation, symmetrization of the lower indices implies the symmetrization of the upper indices as well. Thus, \eqref{A12} becomes
\begin{align}
\begin{split}
	\CA^{[\mu\nu]} + \nabla_\rho \CB^{\rho\mu\nu}  &= \frac{1}{2}\sum_i \sum_{n=1}^\infty \sum_{k=1}^n (-1)^k \nabla_{\mu_1\cdots\mu_k} \Pi_i^{\mu_1\cdots\mu_n} \big(\S_i^{\mu\nu}\big)_{\mu_{k+1}\cdots\mu_n}{}^{\nu_{k+1}\cdots\nu_n} \nabla_{\nu_{k+1} \cdots \nu_n} \varphi^i \\
	&\qquad - \frac{1}{2}\sum_i \sum_{n=1}^\infty \sum_{k=0}^{n-1} (-1)^k \nabla_{\mu_1\cdots\mu_{k}} \Pi_i^{\mu_1\cdots\mu_{n}} \big(\S_i^{\mu\nu}\big)_{\mu_{k+1}\cdots\mu_n}{}^{\nu_{k+1}\cdots\nu_n} \nabla_{\nu_{k+1}\cdots \nu_n} \varphi^i,
\end{split}
\end{align}
where we have relabeled the indices $\mu_2 \to \mu_1,\ldots,\mu_n \to \mu_{n-1}, \sigma \to \mu_n$, and $\rho \to \nu_k$ and then changed the summation index in the second term from $k \to k+1$. Noting that for $1 \leq k \leq n-1$ the terms cancel, we obtain
\begin{align}
\begin{split}
	\CA^{[\mu\nu]} + \nabla_\rho \CB^{\rho\mu\nu} &=  \frac{1}{2}\sum_i \sum_{n=1}^\infty (-1)^n \nabla_{\mu_1\cdots\mu_n} \Pi_i^{\mu_1\cdots\mu_n}  \varphi^i  \\
	&\qquad \qquad - \frac{1}{2}\sum_i \sum_{n=1}^\infty  \Pi_i^{\mu_1\cdots\mu_{n}} \big(\S_i^{\mu\nu}\big)_{\mu_{1}\cdots\mu_n}{}^{\nu_{1}\cdots\nu_n} \nabla_{\nu_1\cdots \nu_n} \varphi^i \\
	&= 0,
\end{split}
\end{align}
where we noted in the second equality that we can include the $n=0$ term in each sum as they cancel out, and that the resulting terms vanish by \eqref{eom1} with the fact $*\CE_i =0$ and by local Lorentz invariance of the Lagrangian \eqref{local-Lorentz-inv}.

\subsection{Derivation of (\ref{eom-explicit}) and (\ref{spcd-explicit})}
\label{app:explicitcalculations1}

We provide here a detailed derivation of the equations of motion and the symplectic potential current density for a gauge theory with Lagrangian
\begin{align}\label{AgenformLag}
\begin{split}
L = \e \, \CL \left(  D_{\a_1 \cdots \a_n} F_{\mu\nu} , D_{\a_1 \cdots \a_n} \Phi^i , D_{\a_1 \cdots \a_n} (\Phi^i)^{\ct} \right),
\end{split}
\end{align}
where $D_{\a_1\cdots\a_n} \equiv D_{(\a_1} \cdots D_{\a_n)}$ is the symmetrized gauge covariant derivative. The procedure implemented here is very similar to the one described in Section~\ref{sec:solspace}, but due to the fact that the derivatives here are gauge covariant derivatives, a few additional complications arise and we discuss these here.

We recall that the variation of $L$ with respect to a generic vector $\X$ takes the general form
\begin{align}\label{varaction1}
\begin{split}
\X(L) = \tr{ \X(A) \wedge \CE^A } + \sum_{i=1}^N \left( (\CE_i^{\Phi})^\ct \X(\Phi^i) + \hc \right) + \dt \bt(\X) .
\end{split}
\end{align}
Note that $\CE^A$ and $\CE^{\Phi}_i$ are the equations of motion and are $(d-1)$- and $d$-forms on $\CM$ respectively. Setting $L = \e\CL$ and taking the spacetime Hodge dual on both sides, we have
\begin{equation}
\begin{split}
\label{genXL}
\X(\CL) &= - \tr{\ast \big( \X(A) \wedge \CE^A \big) } - \sum_{i=1}^N \left( \ast (\CE_i^{\Phi})^\ct\X(\Phi^i)  + \hc \right) - \ast \dt \bt(\X) \\
&=  -\tr{ \X(A_{\mu})(\ast\CE^A)^{\mu}} - \sum_{i=1}^N \left( \ast (\CE_i^{\Phi})^\ct\X(\Phi^i)  + \hc \right) -  \nabla_{\mu} [ \ast \bt(\X) ]^\mu.
\end{split}
\end{equation}
However, we can also write the variation of the Lagrangian density with respect to $\X$ as
\begin{equation}
\begin{split}\label{B1}
\X(\CL) = \sum_{n=0}^\infty \tr{ \Pi^{\a_1 \cdots \a_n ; \mu\nu}  \X(\D_{\a_1 \cdots \a_n} F_{\mu\nu}) } + \sum_{i=1}^N \sum_{n=0}^\infty \big( \Pi^{\a_1 \cdots \a_n}_i \X(\D_{\a_1 \cdots \a_n}\Phi^i) + \hc \big) ,
\end{split}
\end{equation}
where we have defined for convenience
\begin{align}
\label{Pidef}
\Pi^{\a_1\cdots\a_n;\mu\nu} \equiv \frac{\p\CL}{\p\big( D_{\a_1 \cdots \a_n} F_{\mu\nu}\big)}, \qquad \Pi_i^{\a_1\cdots\a_n} \equiv \frac{\p\CL}{\p\big( D_{\a_1\cdots\a_n}\Phi^i\big)},
\end{align}
By recasting \eqref{B1} into the form \eqref{genXL}, we can thus obtain explicitly the equations of motion and the symplectic potential current density $\bt(\X)$. We proceed to analyze the terms in \eqref{B1} one at a time. Denote the terms depending on $\Pi^{\a_1\cdots\a_n;\mu\nu}$ collectively as $\X(\CL)|_{F}$ (i.e. the first term in \eqref{B1}), and the terms depending on $\Pi_i^{\a_1\cdots\a_n}$ collectively as $\X(\CL)|_{\Phi^i}$. We first compute $\X(\CL)|_{\Phi^i}$ as it is the easiest. Recalling the action of the gauge covariant derivative,
\begin{align}\label{covD}
\D_\mu \Phi^i = \nabla_\mu \Phi^i + R_i(A_\mu) \Phi^i  ,
\end{align}
we get
\begin{equation}
\begin{split}\label{A19}
\X(\CL) \big|_{\Phi^i} &= \sum_{n=0}^\infty \Pi^{\a_1 \cdots \a_n}_i \X(\D_{\a_1 \cdots \a_n}\Phi^i)\\
&= \Pi_i \X ( \Phi^i ) + \sum_{n=0}^\infty \Pi^{\a_1 \cdots \a_n}_i \D_{\a_1} \X( \D_{\a_2 \cdots \a_n}\Phi^i)  + \sum_{n=0}^\infty \Pi^{\a_1 \cdots \a_n}_i R_i(\X(A_{\a_1})) \D_{\a_2 \cdots \a_n}\Phi^i .
\end{split}
\end{equation}
We now perform ``integration by parts''-style manipulations (IBP) in the second term to get 
\begin{align}\label{XLPhi}
\begin{split}
\X(\CL) \big|_{\Phi^i} &= \Pi_i\X(\Phi^i) + \nabla_{\a_1}\sum_{n=1}^\infty \Pi_i^{\a_1\cdots\a_n}  \X\big( D_{\a_2 \cdots \a_n}  \Phi^i \big)  \\
&\qquad  - \sum_{n=1}^\infty  D_{\a_1}\Pi_i^{\a_1\cdots\a_n}  \X\big( D_{\a_2 \cdots \a_n} \Phi^i \big) + \sum_{n=1}^\infty \Pi_i^{\a_1\cdots\a_n}    R_i( \X(A_{\a_1})) D_{\a_2 \cdots \a_n} \Phi^i  .
\end{split}	
\end{align}
Noting that we can write any of the symmetrized covariant derivatives $D_{\a_2 \cdots\a_n}$ above as $D_{\a_2}D_{\a_3\cdots\a_n}$ since $\Pi_i^{\a_1\cdots\a_n}$ is completely symmetric, and that for any $k=1,\ldots,n$ we have
\begin{align}
\begin{split}
\X\big( D_{\a_k} D_{\a_{k+1} \cdots \a_n} \Phi^i \big) &= D_{\a_k}\X\big( D_{\a_{k+1} \cdots \a_n} \Phi^i \big)  + R_i( \X(A_{\a_k})) D_{\a_{k+1} \cdots \a_n} \Phi^i ,
\end{split}
\end{align}
we can substitute this for $k=2$ into the third term of the last equality of \eqref{XLPhi} to get
\begin{align}
\begin{split}
\X(\CL)\big|_{\Phi^i}  &=  \Pi_i\X(\Phi^i) \\
&~~+  \nabla_{\a_1}\left( \sum_{n=1}^\infty \Pi_i^{\a_1\cdots\a_n}  \X\big( D_{\a_2 \cdots \a_n}  \Phi^i \big)  - \sum_{n=1}^\infty  D_{\a_2}\Pi_i^{\a_1\cdots\a_n}  \X\big( D_{\a_3 \cdots \a_n} \Phi^i \big) \right)  \\
&~~ + \left( \sum_{n=1}^\infty \Pi_i^{\a_1\cdots\a_n}    R_i( \X(A_{\a_1})) D_{\a_2 \cdots \a_n} \Phi^i  - \sum_{n=1}^\infty  D_{\a_2}\Pi_i^{\a_1\cdots\a_n}  R_i( \X(A_{\a_1})) D_{\a_{3} \cdots \a_n} \Phi^i \right) \\
&~~  +  \sum_{n=1}^\infty  D_{\a_1\a_2} \Pi_i^{\a_1\cdots\a_n}  \X\big( D_{\a_3 \cdots \a_n} \Phi^i \big). 
\end{split}
\end{align}
Repeating this process of using IBP until all the derivatives have been removed from $\X(\Phi^i)$ in the last term, we obtain
\begin{align}\label{XLPhisol}
\begin{split}
	\X(\CL)\big|_{\Phi^i} &=  \nabla_{\a_1}\sum_{n=1}^\infty\sum_{k=1}^n (-1)^{k-1}  D_{\a_2 \cdots \a_k} \Pi_i^{\a_1\cdots\a_n}  \X\big( D_{\a_{k+1} \cdots\a_n} \Phi^i \big)    \\
	&\qquad + \sum_{n=1}^\infty \sum_{k=1}^n (-1)^{k-1}  D_{\a_2 \cdots \a_k} \Pi_i^{\a_1\cdots\a_n}    R_i( \X(A_{\a_1})) D_{\a_{k+1} \cdots \a_n} \Phi^i  \\
	&\qquad  +  \sum_{n=0}^\infty (-1)^n  D_{\a_1 \cdots \a_n} \Pi_i^{\a_1\cdots\a_n}  \X\big( \Phi^i \big) .
\end{split}
\end{align}

Next, we now want to similarly evaluate $\X(\CL)|_{F}$. Following nearly identical reasoning and noting that $F_{\mu\nu}$ transforms in the adjoint representation, which means that \eqref{covD} reduces to
\begin{align}
\begin{split}
D_\a F_{\mu\nu} = \nabla_\a F_{\mu\nu} + \big[ A_{\a} , F_{\mu\nu}\big]  ,
\end{split}
\end{align}
we have
\begin{align}\label{XLF}
\begin{split}
	\X(\CL)\big|_{F} &=  \sum_{n=0}^\infty \tr{ \Pi^{\a_1\cdots\a_n;\mu\nu} \X\big(  D_{\a_1 \cdots \a_n} F_{\mu\nu} \big) } \\
	&= \Pi^{\mu\nu}\X\big( F_{\mu\nu}\big) +  \text{tr}\Bigg[ \nabla_{\a_1}\sum_{n=1}^\infty \Pi^{\a_1\cdots\a_n;\mu\nu}  \X\big( D_{\a_2 \cdots \a_n} F_{\mu\nu} \big)   \\
	&\qquad - \sum_{n=1}^\infty  D_{\a_1}\Pi^{\a_1\cdots\a_n;\mu\nu}  \X\big( D_{\a_2 \cdots \a_n} F_{\mu\nu} \big) + \sum_{n=1}^\infty \Pi^{\a_1\cdots\a_n;\mu\nu} \big[ \X( A_{\a_1}), D_{\a_2 \cdots \a_n} F_{\mu\nu}\big]   \Bigg].   
\end{split}
\end{align}
This has exactly the same structure as \eqref{XLPhi}, except $ D_\a$ is acting in the adjoint representation instead of representation $R_i$, and there is an overall trace over the adjoint representation. This allows us to repeat exactly the same steps as above to obtain
\begin{align}\label{XLFsol'}
\begin{split}
	\X(\CL)\big|_{F} &= \nabla_{\a_1}\sum_{n=1}^\infty\sum_{k=1}^n (-1)^{k-1} \text{tr}\bigg[  D_{\a_2 \cdots \a_k} \Pi^{\a_1\cdots\a_n;\mu\nu}  \X\big( D_{\a_{k+1} \cdots \a_n} F_{\mu\nu} \big)  \bigg]  \\
	&\qquad + \sum_{n=1}^\infty \sum_{k=1}^n (-1)^{k-1} \text{tr}\bigg[  D_{\a_2 \cdots \a_k}  \Pi^{\a_1\cdots\a_n;\mu\nu}    \Big[ \X(A_{\a_1}) ,  D_{\a_{k+1} \cdots \a_n} F_{\mu\nu}\Big] \bigg] \\
	&\qquad  +  \sum_{n=0}^\infty (-1)^n \text{tr}\bigg[  D_{\a_1 \cdots \a_n} \Pi^{\a_1\cdots\a_n;\mu\nu}  \X\big( F_{\mu\nu} \big) \bigg] .
\end{split}
\end{align}
To write the last term using $\X(A_{\mu})$ instead of $\X(F_{\mu\nu})$, we observe
\begin{align}\label{XF}
\begin{split}
	\X(F_{\mu\nu}) &= \p_\mu \X(A_\nu) -\p_\nu\X(A_\mu) + \big[ \X(A_\mu), A_\nu \big] + \big[ A_\mu, \X( A_\nu ) \big] \\
	&=  D_\mu \X(A_\nu) -  D_\nu \X(A_\mu) .
\end{split}
\end{align}
Substituting this into \eqref{XLFsol'} yields
\begin{align}\label{XLFsol}
\begin{split}
	\X(\CL)\big|_{F} &= \nabla_{\a_1}\sum_{n=1}^\infty\sum_{k=1}^n (-1)^{k-1} \text{tr}\bigg[  D_{\a_2 \cdots \a_k} \Pi^{\a_1\cdots\a_n;\mu\nu}  \X\big( D_{\a_{k+1} \cdots \a_n} F_{\mu\nu} \big)  \bigg]  \\
	&\qquad +  2\nabla_\mu \sum_{n=1}^\infty (-1)^n \text{tr}\bigg[  D_{\a_1 \cdots \a_n} \Pi^{\a_1\cdots\a_n;\mu\nu}  \X(A_{\nu} ) \bigg] \\
	&\qquad + \sum_{n=1}^\infty \sum_{k=1}^n (-1)^{k-1} \text{tr}\bigg[  D_{\a_2 \cdots \a_k} \Pi^{\a_1\cdots\a_n;\mu\nu}    \Big[ \X(A_{\a_1}) ,  D_{\a_{k+1} \cdots \a_n} F_{\mu\nu}\Big] \bigg] \\
	&\qquad  -  2\sum_{n=0}^\infty (-1)^n \text{tr}\bigg[  D_\mu  D_{\a_1 \cdots \a_n} \Pi^{\a_1\cdots\a_n;\mu\nu}  \X(A_{\nu} ) \bigg].
\end{split}
\end{align}
To summarize, we have
\begin{align}
\begin{split}
	\X(\CL) &= \X(\CL)\big|_{F} + \sum_{i=1}^N \Big( \X(\CL)\big|_{\Phi^i} + \hc \Big) ,
\end{split}
\end{align}
where the right-hand-side is explicitly given by \eqref{XLPhisol} and \eqref{XLFsol}.

We can now match this with \eqref{genXL} to obtain $\ast\CE^A$, $\ast\CE_i^\Phi$, and $\ast\bt$. Starting with $\ast\bt$, this is just the negative of the sum of terms that are inside a total derivative from \eqref{XLPhisol} and \eqref{XLFsol}:
\begin{align}\label{starTh}
\begin{split}
[\ast\bt(\X)]^\mu &=  \sum_{n=1}^\infty\sum_{k=1}^n (-1)^{k} \text{tr}\bigg[  D_{\a_2 \cdots \a_k} \Pi^{\mu\a_2\cdots\a_n;\a\b}  \X\big( D_{\a_{k+1} \cdots \a_n} F_{\a\b} \big)  \bigg] \\
	&\qquad  - 2 \sum_{n=0}^\infty (-1)^n \text{tr}\bigg[  D_{\a_1 \cdots \a_n} \Pi^{\a_1\cdots\a_n;\mu\nu}  \X(A_{\nu} ) \bigg] \\
	&\qquad + \bigg( \sum_{i=1}^N \sum_{n=1}^\infty\sum_{k=1}^n (-1)^{k}  D_{\a_2 \cdots \a_k} \Pi_i^{\mu\a_2\cdots\a_n}  \X\big( D_{\a_{k+1} \cdots \a_n} \Phi^i \big)   + \hc \bigg).
\end{split}
\end{align}
Next, $\ast\CE^A$ can be determined by noting that it depends on $\X(A^\mu)$, and hence the second term of \eqref{XLPhisol} (and its conjugate transpose) and last two terms of \eqref{XLFsol}. The second term of \eqref{XLPhisol} can be written as
\begin{align}\label{trick}
\begin{split}
	&\sum_{n=1}^\infty \sum_{k=1}^n (-1)^{k-1}  D_{\a_2 \cdots \a_k}  \Pi_i^{\a_1\cdots\a_n}  \X(A_{\a_1}^a)  T_i^a  D_{\a_{k+1} \cdots \a_n} \Phi^i \\
	&\quad = \text{tr}\big[ X^a  \X(A_{\a_1}) \big] \sum_{n=1}^\infty \sum_{k=1}^n (-1)^{k-1}  D_{\a_2 \cdots \a_k} \Pi_i^{\a_1\cdots\a_n}   T_i^a  D_{\a_{k+1} \cdots \a_n} \Phi^i \\
	&\quad = \text{tr}\bigg[ \X(A_{\mu}) \sum_{n=1}^\infty \sum_{k=1}^n (-1)^{k-1}  D_{\a_2 \cdots \a_k} \Pi_i^{\mu\a_2\cdots\a_n}   T_i^a  D_{\a_{k+1} \cdots \a_n} \Phi^i X^a\bigg] ,
\end{split}
\end{align}
where in the first equality we wrote $A_{\a_1}^a$ as a trace, and in the last equality we pulled the rest of the expression (which is a $c$-number) into the trace. The last two terms of \eqref{XLFsol} can be written as
\begin{align}
\begin{split}\label{trick2}
	& \sum_{n=1}^\infty \sum_{k=1}^n (-1)^{k-1} \text{tr}\Big[  D_{\a_2 \cdots \a_k} \Pi^{\a_1\cdots\a_n;\mu\nu}    \big[ \X(A_{\a_1}) ,  D_{\a_{k+1} \cdots \a_n}  F_{\mu\nu}\big] \Big]   \\
	&\qquad -  2\sum_{n=0}^\infty (-1)^n \text{tr}\big[  D_\mu  D_{\a_1 \cdots \a_n}\Pi^{\a_1\cdots\a_n;\mu\nu}  \X(A_{\nu} ) \big] \\
	= \quad &  -\sum_{n=1}^\infty \sum_{k=1}^n (-1)^{k-1} \text{tr}\Big[ \X(A_{\mu})     \big[ D_{\a_2 \cdots \a_k} \Pi^{\mu\a_2\cdots\a_n;\a\b} ,  D_{\a_{k+1} \cdots \a_n} F_{\a\b}  \big] \Big]   \\
	&\qquad + 2\sum_{n=0}^\infty (-1)^n \text{tr}\big[ \X(A_{\mu} )  D_\nu  D_{\a_1 \cdots \a_n} \Pi^{\a_1\cdots\a_n;\mu\nu}  ) \big].
\end{split}
\end{align}
It follows $(\ast\CE^A)^\mu$, i.e. the negative of the term multiplying $\X(A_\mu)$, is
\begin{align}\label{starEA}
\begin{split}
(\ast\CE^A)^\mu &= -  2\sum_{n=0}^\infty (-1)^n   D_\nu  D_{\a_1 \cdots \a_n} \Pi^{\a_1\cdots\a_n,\mu\nu}   \\
&\qquad \qquad  - \sum_{n=1}^\infty \sum_{k=1}^n (-1)^{k} \big[ D_{\a_2 \cdots \a_k} \Pi^{\mu\a_2\cdots\a_n;\a\b}  ,  D_{\a_{k+1} \cdots \a_n} F_{\a\b}  \big]   \\
&\qquad \qquad + \sum_{i=1}^N\sum_{n=1}^\infty \sum_{k=1}^n (-1)^{k} \big(  D_{\a_2 \cdots \a_k} \Pi_i^{\mu\a_2\cdots\a_n}   T_i^a  D_{\a_{k+1} \cdots \a_n} \Phi^i + \hc \big) X^a .
\end{split}
\end{align}
Finally, $(\ast\CE_i^{\Phi})^\ct$ is negative of the remaining term, i.e. last term in \eqref{XLPhisol}: 
\begin{align}\label{starEi}
\begin{split}
	(\ast\CE_i^{\Phi})^\ct &= -\sum_{n=0}^\infty (-1)^{n}  D^{\a_1 \cdots \a_n} (\Pi_i)_{\a_1\cdots\a_n}.
\end{split}
\end{align}
Collecting our results, we have
\begin{align}\label{B21}
\begin{split}
(\ast \CE^A)^\mu &= - 2 \sum_{n=0}^\infty (-1)^n \D_\nu \D_{\a_1\cdots \a_n} \Pi^{\a_1 \cdots \a_n ; \mu\nu}  \\
&\qquad \qquad - \sum_{n=0}^\infty  \sum_{k=1}^n (-1)^k  \left[  \D_{\a_2\cdots \a_k}  \Pi^{\mu \a_2 \cdots \a_n ; \a\b},   \D_{\a_{k+1} \cdots \a_n} F_{\a\b} \right] \\
& \qquad \qquad + \sum_{i=1}^N \sum_{n=0}^\infty \sum_{k=1}^n (-1)^k  \left( \D_{\a_n \cdots \a_k}  \Pi^{\mu \a_2 \cdots \a_n}_i T^a_i  \D_{\a_{k+1} \cdots \a_n}\Phi^i + \hc \right) X^a  \\
\ast ( \CE^{\Phi}_i )^\ct &= - \sum_{n=0}^\infty (-1)^n  \D_{\a_1\cdots\a_n}  \Pi^{\a_1 \cdots \a_n}_i  \\
[\ast \bt (\X)]^\mu &= - 2 \sum_{n=0}^\infty (-1)^n \tr{  \D_{\a_1\cdots \a_n} \Pi^{\a_1 \cdots \a_n ; \mu\nu} \X(A_\nu)  }  \\
&\qquad \qquad +  \sum_{n=0}^\infty \sum_{k=1}^n (-1)^k  \tr{  \D_{\a_2 \cdots \a_k }  \Pi^{\mu \a_2 \cdots \a_n ; \a\b} \X ( \D_{\a_{k+1} \cdots \a_n} F_{\a\b} ) } \\
&\qquad \qquad + \sum_{i=1}^N \sum_{n=0}^\infty \sum_{k=1}^n (-1)^k \left( \D_{\a_2 \cdots \a_k}  \Pi^{\mu \a_2 \cdots \a_n}_i  \X(\D_{\a_{k+1} \cdots \a_n}\Phi^i) + \hc \right) ,
\end{split}
\end{align}
which is precisely \eqref{eom-explicit} and \eqref{spcd-explicit}.

\subsection{Derivation of (\ref{gauge-transform-symp})}
\label{app:explicitcalculations2}

We give a detailed derivation of the pre-symplectic potential and form acting on the vector $\X_\ve$ generating gauge transformations. First, we need to compute $\bt(\X_\ve)$ and $\bo(\Y,\X_\ve)$. Recall that the vector generating gauge transformations is given by \eqref{Xepdef} to be
\begin{equation}
\begin{split}\label{AXepdef}
\X_\ve = \int_\CM \e \left[ - \D_\mu \ve^a \frac{\d}{\d A_\mu^a} + \sum_{i=1}^N \left(  - \ve^a (\Phi^i)^\ct T^a_i  \frac{\d}{\d (\Phi^i)^\ct}  + \hc  \right) \right] \in T \bs{\mathfrak{F}},
\end{split}
\end{equation}
which implies
\begin{equation}\label{infgaugeac}
\begin{split}
\X_\ve ( A_\nu ) &= - D_\nu \ve  \\
\X_\ve ( \D_{\a_{k+1} \cdots \a_n} F_{\a\b} )  &= - \left[  \D_{\a_{k+1} \cdots \a_n} F_{\a\b}  , \ve \right]  \\
\X_\ve ( \D_{\a_{k+1} \cdots \a_n}\Phi^i ) &= \ve^a T^a_i \D_{\a_{k+1} \cdots \a_n}\Phi^i .
\end{split}
\end{equation}
Substituting this into \eqref{B21} with $\X = \X_\ve$, we get
\begin{equation}\label{Atheta-calc}
\begin{split}
[\ast \bt(\X_\ve)]^\mu &=  2 \sum_{n=0}^\infty (-1)^n \tr{  \D_{\a_1\cdots \a_n} \Pi^{\a_1 \cdots \a_n ; \mu\nu} D_\nu \ve  }  \\
&\qquad \qquad - \sum_{n=1}^\infty \sum_{k=1}^n (-1)^k  \tr{  \D_{\a_2 \cdots \a_k }  \Pi^{\mu \a_2 \cdots \a_n ; \a\b} [  \D_{\a_{k+1} \cdots \a_n} F_{\a\b} , \ve ] } \\
&\qquad \qquad + \sum_{i=1}^N \sum_{n=1}^\infty \sum_{k=1}^n (-1)^k \tr{ \ve  \big( \D_{\a_2 \cdots \a_k}  \Pi^{\mu \a_2 \cdots \a_n}_i  T^a_i \D_{\a_{k+1} \cdots \a_n}\Phi^i + \hc  \big) X^a }  \\
&= 2 \nabla_\nu \tr{ \ve   \sum_{n=0}^\infty (-1)^n  \D_{\a_1\cdots \a_n} \Pi^{\a_1 \cdots \a_n ; \mu\nu} } + \tr{  \ve \left( \ast \CE^A \right)^\mu } ,
\end{split}
\end{equation}
where we used the trick introduced in \eqref{trick} to write the last term in the first equality as a trace, and then used IBP on the first term to get the second equality. The second term in the last line vanishes on-shell, so recalling \eqref{Qdef}, which defines
\begin{equation}\label{AQ-def}
\begin{split}
	\left( \ast \CQ \right)^{\mu\nu} \equiv 2 \sum_{n=0}^\infty  (-1)^n \D_{\a_1 \cdots \a_n} \Pi^{\a_1 \cdots \a_n ; \mu\nu} ,
\end{split}
\end{equation}
we then have on-shell
\begin{equation}
\begin{split}
[\ast \bt(\X_\ve)]^\mu &= \nabla_\nu \tr{ \ve (\ast \CQ)^{\mu\nu}  }  \quad \implies \quad \bt (\X_\ve) = \dt\big( \tr{ \ve \CQ }\big)  . 
\end{split}
\end{equation}
Integrating over $\S$, it then follows by Stokes' theorem that
\begin{equation}
\begin{split}
\wt \bT_\S (\X_\ve) = \oint_{\p\S} \tr{ \ve \CQ } ,
\end{split}
\end{equation}
proving the first equation of \eqref{gauge-transform-symp}.

To prove the second equation of \eqref{gauge-transform-symp}, recall that the symplectic potential current density between two arbitrary vectors $\X$ and $\Y$ is given by \eqref{omega-explicit} to be
\begin{align}\label{Aomega-explicit}
\begin{split}
[\ast \bo(\X,\Y)]^\mu  &=  - 2 \sum_{n=0}^\infty (-1)^n \tr{ \X  ( \D_{\a_1 \cdots \a_n} \Pi^{\a_1 \cdots \a_n ; \mu\nu}   )  \Y (A_\nu ) }  \\
&\qquad \qquad + \sum_{n=1}^\infty \sum_{k=1}^n (-1)^k \tr{ \X  ( \D_{\a_2\cdots \a_k} \Pi^{\mu \a_2 \cdots \a_n ; \a\b}    )  \Y ( \D_{\a_{k+1} \cdots \a_n } F_{\a\b} ) }  \\
&\qquad \qquad + \sum_{i=1}^N  \sum_{n=1}^\infty  \sum_{k=1}^n (-1)^k \left[ \X  ( \D_{\a_2 \cdots \a_k } \Pi_i^{\mu \a_2 \cdots \a_n}  )  \Y (  \D_{\a_{k+1} \cdots \a_n } \Phi^i ) + \hc \right] \\
&\qquad \qquad - (\X \leftrightarrow \Y) . 
\end{split}
\end{align}
Setting $\X = \X_\ve$ and using \eqref{infgaugeac}, we obtain upon substituting into \eqref{Aomega-explicit}
\begin{equation}
\begin{split}
[\ast \bo(\Y,\X_\ve)]^\mu &=   \nabla_\nu \tr{  \ve  \Y \left( ( \ast \CQ )^{\mu\nu} \right)  } +  \tr{ \ve \Y \big( (\ast \CE^A)^\mu  \big) } ,
\end{split}
\end{equation}
where we used exactly the same methods as those employed in \eqref{Atheta-calc}, and $\ast\CQ$ is given in \eqref{AQ-def}. Since the second term vanishes on-shell, we have
\begin{equation}
\begin{split}
[\ast \bo(\Y,\X_\ve)]^\mu  =  \nabla_\nu \tr{  \ve  \Y \big( ( \ast \CQ )^{\mu\nu} \big)  }  \quad \implies \quad \bo(\Y,\X_\ve) = \dt \big(\tr{ \ve \Y(\CQ) }\big) . 
\end{split}
\end{equation}
Integrating over $\S$ and using Stokes' theorem, we find
\begin{equation}
\begin{split}
\wt \bO_\S(\Y,\X_\ve) = \oint_{\p\S} \tr{ \ve \Y(\CQ) } ,
\end{split}
\end{equation}
which is the second equation of \eqref{gauge-transform-symp}.

\subsection{Derivation of (\ref{stress-tensor-explicit})}
\label{app:explicitcalculations3}

In this section, we derive the isometry charge, $H_\xi[\S]$. The procedure employed here is very similar to the one in Section~\ref{sec:isometries}. However, given the special form of the gauge theory Lagrangian, we will discover that our boundary term contains a particularly interesting piece.

Recall from \eqref{isom-ch-1} that the isometry charge is
\begin{equation}\label{A-iso}
\begin{split}
H_\xi[\S] = - \int_\S d\S_\mu \,( \ast \bt(\X_\xi) - \ast  i_\xi L )^\mu . 
\end{split}
\end{equation}
Starting with \eqref{B21} and \eqref{isometrydef}, we have
\begin{equation}
\begin{split}\label{B33}
(\ast \bt (\X_\xi) - \ast i_\xi L)^\mu &= \xi^\mu \CL - 2 \sum_{n=0}^\infty (-1)^n \tr{  \D_{\a_1\cdots \a_n} \Pi^{\a_1 \cdots \a_n ; \mu\nu} \pounds_\xi A_\nu  }  \\
&\qquad \qquad +  \sum_{n=1}^\infty \sum_{k=1}^n (-1)^k  \tr{  \D_{\a_2 \cdots \a_k }  \Pi^{\mu \a_2 \cdots \a_n ; \a\b} \pounds_\xi \D_{\a_{k+1} \cdots \a_n} F_{\a\b}  } \\
&\qquad \qquad + \sum_{i=1}^N \sum_{n=1}^\infty \sum_{k=1}^n (-1)^k \Big( \D_{\a_2 \cdots \a_k}  \Pi^{\mu \a_2 \cdots \a_n}_i  \pounds_\xi\D_{\a_{k+1} \cdots \a_n}\Phi^i + \hc \Big) . 
\end{split}
\end{equation}
We begin by simplifying the second term above. Using the fact that
\begin{equation}
\begin{split}
\pounds_\xi A_\nu &= \xi^\rho \nabla_\rho A_\nu + A_\rho \nabla_\nu \xi^\rho = \xi^\rho F_{\rho\nu}  + \nabla_\nu ( \xi^\rho A_\rho ) + [A_\nu , \xi^\rho A_\rho ] ,
\end{split}
\end{equation}
we get upon substituting this into the second term of \eqref{B33} and using IBP
\begin{equation}\label{2ndterm}
\begin{split}
&- 2 \sum_{n=0}^\infty (-1)^n \tr{  \D_{\a_1\cdots \a_n} \Pi^{\a_1 \cdots \a_n ; \mu\nu} \pounds_\xi A_\nu  } \\
&\qquad =  -  \nabla_\nu \tr{  \xi^\rho A_\rho (\ast \CQ)^{\mu\nu} }  - 2 \xi^\rho \sum_{n=0}^\infty (-1)^n \tr{  \D_{\a_1\cdots \a_n} \Pi^{\a_1 \cdots \a_n ; \mu\nu}  F_{\rho\nu}    } \\
&\quad\qquad\qquad  + 2 \sum_{n=0}^\infty (-1)^n \tr{ \xi^\rho A_\rho \D_\nu  \D_{\a_1\cdots \a_n} \Pi^{\a_1 \cdots \a_n ; \mu\nu}   } ,
\end{split}
\end{equation}
where $*\CQ$ is given in \eqref{AQ-def}. Next, to simplify the last two terms of \eqref{B33}, recall that the Lie derivative is defined in \eqref{Liederdef} to act via
\begin{equation}
\begin{split}
\pounds_\xi \varphi^i = \xi^\mu \nabla_\mu \varphi^i + \frac{1}{2} \nabla_{[\mu} \xi_{\nu]} \S_i^{\mu\nu} \varphi^i = \xi^\mu \D_\mu \varphi^i   + \frac{1}{2} \nabla_{[\mu} \xi_{\nu]} \S_i^{\mu\nu} \varphi^i - \xi^\mu R_i(A_\mu) \varphi^i  ,
\end{split}
\end{equation}
where we replaced the (spacetime) covariant derivative with a gauge covariant derivative at the cost of a term involving the gauge field. It follows that the third term in \eqref{B33} becomes 
\begin{align}\label{3rdterm}
\begin{split}
&\sum_{n=1}^\infty \sum_{k=1}^n (-1)^k  \text{tr}\big[ D_{\a_2 \cdots \a_k }  \Pi^{\mu \a_2 \cdots \a_n ; \a\b} \pounds_\xi D_{\a_{k+1} \cdots \a_n} F_{\a\b}  \big] \\
&= \xi_\nu\sum_{n=1}^\infty \sum_{k=1}^n (-1)^k  \text{tr}\big[ D_{\a_2 \cdots \a_k }  \Pi^{\mu \a_2 \cdots \a_n ; \a\b}  D^\nu D_{\a_{k+1} \cdots \a_n} F_{\a\b}  \big] \\
& + \frac{1}{2}\nabla_{[\nu}\xi_{\rho]} \sum_{n=1}^\infty \sum_{k=1}^n (-1)^k  \tr{ D_{\a_2 \cdots \a_k }  \Pi^{\mu \a_2 \cdots \a_n ; \a\b} \big(\S^{\nu\rho}\big)_{\a_{k+1}\cdots\a_n \a\b}{}^{\a'_{k+1}\cdots\a'_n;\a'\b'} D_{\a'_{k+1} \cdots \a'_n} F_{\a'\b'}  } \\
& - \sum_{n=1}^\infty \sum_{k=1}^n (-1)^k  \tr{ D_{\a_2 \cdots \a_k }  \Pi^{\mu \a_2 \cdots \a_n ; \a\b} \xi^\rho \big[A_\rho, D_{\a_{k+1} \cdots \a_n} F_{\a\b} \big]  },
\end{split}
\end{align}
where we used the fact $D_{\a_{k+1}\cdots\a_n} F_{\a\b}$ lives in the adjoint representation. Likewise, the fourth term in \eqref{B33} becomes
\begin{align}\label{4thterm}
\begin{split}
&\sum_{i=1}^N \sum_{n=1}^\infty \sum_{k=1}^n (-1)^k \left( D_{\a_2 \cdots \a_k}  \Pi^{\mu \a_2 \cdots \a_n}_i  \pounds_\xi D_{\a_{k+1} \cdots \a_n}\Phi_i + \hc \right) \\
&=  \xi_\nu \sum_{i=1}^N \sum_{n=1}^\infty \sum_{k=1}^n (-1)^k \left( D_{\a_2 \cdots \a_k}  \Pi^{\mu \a_2 \cdots \a_n}_i  D^\nu D_{\a_{k+1} \cdots \a_n}\Phi_i + \hc \right) \\
&\quad + \frac{1}{2}\nabla_{[\nu}\xi_{\rho]} \sum_{i=1}^N \sum_{n=1}^\infty \sum_{k=1}^n (-1)^k \left( D_{\a_2 \cdots \a_k}  \Pi^{\mu \a_2 \cdots \a_n}_i  \big(\S^{\nu\rho}\big)_{\a_{k+1}\cdots\a_n}{}^{\a'_{k+1} \cdots \a'_{n}}  D_{\a'_{k+1} \cdots \a'_n}\Phi_i + \hc \right) \\
&\quad - \sum_{i=1}^N \sum_{n=1}^\infty \sum_{k=1}^n (-1)^k \text{tr}\Big[ \big( \xi^\rho A_\rho D_{\a_2 \cdots \a_k}  \Pi^{\mu \a_2 \cdots \a_n}_i  T_i^a D_{\a_{k+1} \cdots \a_n}\Phi_i + \hc \big)X^a \Big],
\end{split}
\end{align}
where we used the trick introduced in \eqref{trick} to write the last term as a trace. Substituting \eqref{2ndterm}, \eqref{3rdterm}, and \eqref{4thterm} into \eqref{B33} and noting that on-shell $*\CE^A$ from \eqref{B21} vanishes, we obtain 
\begin{equation}
\begin{split}
(\ast \bt (\X_\xi) - \ast i_\xi L)^\mu &= \CA^{\mu\nu} \xi_\nu + \CB^{\mu\nu\rho} \nabla_{[\nu} \xi_{\rho]}  -  \nabla_\nu \tr{  \xi^\rho A_\rho (\ast \CQ)^{\mu\nu} }  ,
\end{split}
\end{equation}
where 
\begin{equation}
\begin{split}
\CA^{\mu\nu} &= g^{\mu\nu} \CL - 2  \sum_{n=0}^\infty (-1)^n \tr{  \D_{\a_1\cdots \a_n} \Pi^{\a_1 \cdots \a_n ; \mu\rho}  F^\nu{}_{\rho}    } \\
&\quad +   \sum_{n=1}^\infty \sum_{k=1}^n (-1)^k  \tr{  \D_{\a_2 \cdots \a_k }  \Pi^{\mu \a_2 \cdots \a_n ; \a\b} \D^\nu \D_{\a_{k+1} \cdots \a_n} F_{\a\b}  } \\
&\quad +  \sum_{i=1}^N \sum_{n=1}^\infty \sum_{k=1}^n (-1)^k \Big( \D_{\a_2 \cdots \a_k}  \Pi^{\mu \a_2 \cdots \a_n}_i  \D^\nu \D_{\a_{k+1} \cdots \a_n}\Phi^i + \hc \Big) \\
\CB^{\mu\nu\rho} &= \frac{1}{2}  \sum_{n=1}^\infty \sum_{k=1}^n (-1)^k  \tr{  \D_{\a_2 \cdots \a_k }  \Pi^{\mu \a_2 \cdots \a_n ; \a\b} \big(\S^{\nu\rho}\big)_{\a_{k+1} \cdots \a_n;\a\b}{}^{\a'_{k+1} \cdots \a'_n;\a'\b'} \D_{\a'_{k+1} \cdots \a'_n} F_{\a'\b'}  } \\
&\quad + \frac{1}{2}   \sum_{i=1}^N \sum_{n=1}^\infty \sum_{k=1}^n (-1)^k \left( \D_{\a_2 \cdots \a_k}  \Pi^{\mu \a_2 \cdots \a_n}_i  \big(\S_i^{\nu\rho}\big)_{\a_{k+1} \cdots \a_n}{}^{\a'_{k+1} \cdots \a'_n}  \D_{\a'_{k+1} \cdots \a'_n}\Phi^i + \hc \right).
\end{split}
\end{equation}

We can now follow precisely the steps described in Appendix \ref{app:id12proof} to prove the identities \eqref{A1}. Making the same definitions as in \eqref{st-def}, we can then write
\begin{equation}
\begin{split}
(\ast \bt (\X_\xi) - \ast i_\xi L)^\mu &= T^{\mu\nu} \xi_\nu  + \nabla_\nu (\ast \SH_\xi)^{\mu\nu} - \nabla_\nu \tr{ \xi^\rho A_\rho (\ast \CQ)^{\mu\nu} }  .
\end{split}
\end{equation}
Thus, by \eqref{A-iso} the isometry charge is
\begin{equation}
\begin{split}
H_\xi[\S] = - \int_\S d\S_\mu\, T^{\mu\nu} \xi_\nu  + \oint_{\p\S} \SH_\xi - Q_{i_\xi A}[\S] ,
\end{split}
\end{equation}
which is \eqref{stress-tensor-explicit}.

\section{The Derivative Operator \texorpdfstring{$\mfd_{U(z,\bz)}^a$}{}} \label{app:deriv}

\subsection{The Explicit Form of \texorpdfstring{$\mfd_{U(z,\bz)}^a$}{}}

Recall that the operator $\mfd^a_{U(z,\bz)}$ was defined to act on $U(w,\bw)$ via
\begin{equation}\label{ADaction}
\begin{split}
\mfd^a_{U(z,\bz)} U(w,\bw) \equiv - X^a U(z,\bz) \d^2(z-w) .  
\end{split}
\end{equation}
To see that this is indeed a derivative operator, we would like to work out what $\mfd^a_{U(z,\bz)}$ is explicitly. Begin by writing $U = \exp\phi$, so that 
\begin{equation}
\begin{split}
U(z,\bz) = e^{\phi(z,\bz)} = \sum_{n=0}^\infty \frac{1}{n!} \phi(z,\bz)^n .
\end{split}
\end{equation}
We will now show that if $\mfd^a_{U(z,\bz)}$ takes the form
\begin{equation}\label{Aassumed-form}
\begin{split}
\mfd^a_{U(z,\bz)}  = M^{ab}(\phi(z,\bz)) \frac{\delta}{\delta\phi^b(z,\bz)} , \qquad\text{where}\quad M^{ab}(\phi) X^b = \sum_{m=0}^\infty a_m [ \overbrace{\phi , [ \phi , \cdots , [ \phi}^{\text{$m$ times}} , X^a ] \cdots ] ]  ,
\end{split}
\end{equation}
then there exists a set of $a_m$ such that \eqref{ADaction} is satisfied. This would prove that $\mfd^a_{U(z,\bz)}$ is a derivative operator.

To determine $a_m$, we must first prove that
\begin{equation}\label{Pidentity}
\begin{split}
	P(m) \equiv [ \overbrace{\phi , [ \phi , \cdots , [ \phi}^{\text{$m$ times}} , X^a ] \cdots ] ]  = \sum_{p=0}^m  (-1)^{m+p} {m\choose p} \phi^p X^a \phi^{m-p} . 
\end{split}
\end{equation}
We prove this via induction. For $m=0$ we have $P(0) = X^a$, which is trivially true. Assuming that \eqref{Pidentity} holds for $m$, we want to prove that it also holds for $m+1$. We compute
\begin{equation}
\begin{split}
P(m+1) = [ \phi , P(m) ] &= \sum_{p=0}^m  (-1)^{m+p} {m\choose p} \phi^{p+1} X^a \phi^{m-p} - \sum_{p=0}^m  (-1)^{m+p} {m\choose p} \phi^p X^a \phi^{m+1-p} \\
&= \sum_{p=0}^{m+1}  (-1)^{m+1+p} \left[ {m\choose p-1} + {m\choose p}  \right]  \phi^{p} X^a \phi^{m+1-p} \\
&= \sum_{p=0}^{m+1}  (-1)^{m+1+p} {m+1\choose p}  \phi^{p} X^a \phi^{m+1-p} ,
\end{split}
\end{equation}
thus completing the proof. Noting that we can always interchange two sums via
\begin{align}\label{sum-inter}
\begin{split}
	\sum_{n=0}^\infty \sum_{m=0}^n F_{m,n-m} = \sum_{m=0}^\infty \sum_{n=m}^\infty F_{m,n-m} = \sum_{m=0}^\infty \sum_{n=0}^\infty F_{m,n},
\end{split}
\end{align}
by substituting \eqref{Pidentity} into \eqref{Aassumed-form}, we get
\begin{equation}\label{newMX}
\begin{split}
M^{ab}(\phi) X^b &= \sum_{m=0}^\infty \sum_{p=0}^m  a_m (-1)^{m+p} {m\choose p} \phi^p X^a \phi^{m-p} \\
&= \sum_{p=0}^\infty \sum_{m=0}^\infty   a_{m+p} (-1)^{m} {m + p \choose p} \phi^p X^a \phi^{m} .
\end{split}
\end{equation}
It follows upon using \eqref{Aassumed-form} and ignoring the overall $\delta^2(z-w)$ factor that (we keep the $(z,\bz)$ dependence for $\phi$ implicit for notational simplicity)
\begin{align}\label{mfd-inter}
\begin{split}
	\mfd^a_{U(z,\bz)} U(w,\bw) &= M^{ab}(\phi) \frac{\delta}{\delta \phi^b} \sum_{n=0}^\infty \frac{1}{n!} \phi^n \\
&=  -M^{ab}(\phi) \sum_{n=0}^\infty \sum_{k=0}^n \frac{1}{(n+1)!} \phi^k X^b \phi^{n-k} \\
&=  M^{ab}(\phi)  \sum_{k=0}^\infty  \sum_{n=0}^\infty\frac{(-1)^{k+1}}{(n+k+1)!} \phi^k X^b \phi^{n} \\
&=  \sum_{k=0}^\infty  \sum_{n=0}^\infty  \sum_{m=0}^\infty \sum_{p=0}^\infty  \frac{(-1)^{m+k+1}}{(n+k+1)!}  {m + p \choose p} a_{m+p} \phi^{p+k} X^a \phi^{m+n} \\
&= -\sum_{n=0}^\infty X^a \phi^{n}   \sum_{m=0}^n \frac{a_m}{(n-m+1)!}   \\
&\qquad\qquad +  \sum_{k=1}^\infty   \sum_{n=0}^\infty \phi^{k} X^a \phi^{n} \sum_{m=0}^n \sum_{p=0}^k  \frac{(-1)^{k-p+1} }{(n-m+k-p+1)!}{m + p \choose p}  a_{m+p},
\end{split}
\end{align}
where we repeatedly used \eqref{sum-inter}. We know from \eqref{ADaction} that this must equal $-\sum_{n=0}^\infty \frac{1}{n!} X^a \phi^n$ (again ignoring the overall delta function $\delta^2(z-w)$). Therefore, by comparing this with \eqref{mfd-inter}, we see that in order for $\mfd^a_{U(z,\bz)}$ to have the form assumed in \eqref{Aassumed-form}, we require the coefficients $a_m$ to satisfy
\begin{equation}\label{2req}
\begin{split}
	n \geq 0 &: \qquad \sum_{m=0}^n \frac{n!}{(n-m+1)!} a_{m} = 1  \\
	n \geq 0 , ~ k \geq 1 &: \qquad \sum_{m=0}^n \sum_{p=0}^k  \frac{(-1)^{k-p+1}}{(n-m+k-p+1)!}  {m + p \choose p} a_{m+p}  = 0 . 
\end{split}
\end{equation}

We can determine $a_m$ from the first equation as follows. Multiplying both sides of the first equation with $\frac{x^n}{n!}$ and then summing over $n$, we find
\begin{equation}
\begin{split}
	e^x &= \sum_{n=0}^\infty \sum_{m=0}^n \frac{x^n}{(n-m+1)!} a_{m} = \sum_{m=0}^\infty   a_{m}  x^m \sum_{n=0}^\infty \frac{x^{n} }{(n+1)!} = \frac{e^x - 1}{x}\sum_{m=0}^\infty   a_{m}  x^m    ,
\end{split}
\end{equation}
where we used \eqref{sum-inter}. It follows
\begin{equation}\label{am-sum}
\begin{split}
	\sum_{m=0}^\infty   a_{m}  x^m = \frac{ x e^x }{ e^x - 1 } =  \sum_{m=0}^\infty \frac{B_m(1)x^m}{m!} , 
\end{split}
\end{equation}
where we used the definition of the Bernoulli polynomials $B_m(x)$ \cite{abramowitz+stegun}. Since $B_m(1) = B_m^+$ are the Bernoulli numbers,\footnote{There are two conventions for Bernoulli numbers, where $B_m^+ = B_m(1)$ and $B_m^-= B_m(0)$, and the only difference is $B_1^\pm = \pm \frac{1}{2}$. Mathematica uses the convention $B_m^-$.} comparing the coefficients on both sides results in
\begin{equation}\label{am-def}
\begin{split}
	a_m =  \frac{ B_m^+ }{m!}.
\end{split}
\end{equation}

We now need to check that the second equation in \eqref{2req} is satisfied given \eqref{am-def}. Repeating a similar procedure as above, we multiply both sides of the second equation by $x^ny^k$ and then sum over $n$ and $k$ to get
\begin{align}\label{mid-calc}
\begin{split}
	0 &= \sum_{k=1}^\infty \sum_{n=0}^\infty \sum_{m=0}^n \sum_{p=0}^k  \frac{(-1)^{k-p+1}}{(n-m+k-p+1)!}  \binom{m + p}{p} x^ny^k a_{m+p} \\
	&= \sum_{m=0}^\infty \sum_{n=0}^\infty \sum_{p=0}^\infty\sum_{k=0}^\infty x^{n+m}y^{k+p}  \frac{(-1)^{k+1}}{(n+k+1)!}  {m + p \choose p} a_{m+p} - \sum_{m=0}^\infty \sum_{n=0}^\infty \frac{(-1)}{(n+1)!}   x^{n+m} a_{m} \\
	&= -\sum_{n=0}^\infty \sum_{k=0}^\infty x^ny^k \frac{(-1)^{k}}{(n+k+1)!}\sum_{p=0}^\infty \sum_{m=0}^\infty x^my^p\binom{m+p}{p}a_{m+p} + \frac{1}{x}\sum_{n=0}^\infty \frac{x^{n+1}}{(n+1)!}\sum_{m=0}^\infty a_mx^m,
\end{split}
\end{align}
where we applied \eqref{sum-inter}. The sum over $n$ in the second term just yields $e^x-1$, while the sum over $n$ and $k$ in the first term yields
\begin{align}
\begin{split}
	\sum_{n=0}^\infty \sum_{k=0}^\infty x^ny^k \frac{(-1)^{k}}{(n+k+1)!} &= \sum_{n=0}^\infty \sum_{k=0}^n x^{n-k}y^k \frac{(-1)^k}{(n+1)!} \\
	&= \frac{1}{x+y}\sum_{n=0}^\infty \frac{x^{n+1} - (-y)^{n+1} }{(n+1)!} \\
	&= \frac{e^{x} - e^{-y}}{x+y}.
\end{split}
\end{align}
Substituting these all back into \eqref{mid-calc} and using \eqref{sum-inter} one last time yields
\begin{align}
\begin{split}
	0 &= -\frac{e^{x} - e^{-y}}{x+y} \sum_{p=0}^\infty \sum_{m=0}^\infty x^my^p\binom{m+p}{p}a_{m+p} + \frac{e^x-1}{x}\sum_{m=0}^\infty a_mx^m \\
	&= -\frac{e^{x} - e^{-y}}{x+y}  \sum_{m=0}^\infty\sum_{p=0}^m x^{m-p} y^p\binom{m}{p}a_{m} + \frac{e^x-1}{x}\sum_{m=0}^\infty a_mx^m \\
	&= -\frac{e^{x} - e^{-y}}{x+y}\sum_{m=0}^\infty a_m (x+y)^m + \frac{e^x-1}{x}\sum_{m=0}^\infty a_mx^m.
\end{split}
\end{align}
To prove that the $a_m$ from \eqref{am-def} satisfies this above equation, it suffices to substitute \eqref{am-sum} into the right-hand-side of the above equation and show that the equation holds. Indeed, we have
\begin{align}
\begin{split}
	&-\frac{e^{x} - e^{-y}}{x+y}  \sum_{m=0}^\infty a_m (x+y)^m + \frac{e^x-1}{x}\sum_{m=0}^\infty a_mx^m \\
	&\qquad = -\left(\frac{e^{x} - e^{-y}}{x+y}\right)\left( \frac{(x+y)e^{x+y}}{e^{x+y}-1}\right) + \left(\frac{e^x-1}{x} \right) \left(\frac{xe^x}{e^x -1} \right) \\
	&\qquad = 0,
\end{split}
\end{align}
proving that \eqref{am-def} also satisfies the second equation of \eqref{2req}. Thus, substituting \eqref{am-def} into \eqref{Aassumed-form}, we see that $\mfd^a_{U(z,\bz)}$ is the derivative operator
\begin{align}
\begin{split}
	\mfd^a_{U(z,\bz)}  = M^{ab}(\phi(z,\bz)) \frac{\delta}{\delta \phi^b(z,\bz)} , \qquad\text{where}\quad M^{ab}(\phi) X^b = \sum_{m=0}^\infty \frac{ B_m^+ }{m!}  [ \overbrace{\phi , [ \phi , \cdots , [ \phi}^{\text{$m$ times}} , X^a ] \cdots ] ] .
\end{split}
\end{align}

\subsection{Properties of \texorpdfstring{$\mfd_{U(z,\bz)}^a$}{}}

We now show that $\mfD^a_{U(z,\bz)}$ satisfies some useful properties. First, it is straightforward to check via induction that for any power $k$, we have
\begin{align}
\begin{split}
	\Big[ \mfD^a_{U(z,\bz)}, \mfD^b_{U(w,\bw)} \Big]U(y,\by)^k &=   f^{abc} \delta^2(z-w)\mfD_{U(z,\bz)}^c U(y,\by)^k.
\end{split}
\end{align}
Assuming an arbitrary function $f(U)$ can be written as a Taylor series in $U$, this means
\begin{align}\label{Dprop1}
	\Big[ \mfD^a_{U(z,\bz)}, \mfD^b_{U(w,\bw)} \Big]f(U(y,\by)) = f^{abc} \delta^2(z-w)\mfD_{U(z,\bz)}^c f(U(y,\by)).
\end{align}

Next, we want to define integration over the measure $[dU]$. Note that this is not trivial, since the functions $U(z,\bz)$ are constrained, e.g. they satisfy non-trivial identities like \eqref{Cprop}. However, we can define the measure to be the left-invariant Haar measure, so that $[dU] = [d(gU)]$. It follows
\begin{align}
\begin{split}\label{prop3deriv}
	\int [dU]\, f(U) &= \int [d(gU)]\, f(gU) = \int [dU]\, f(e^\ve U),
\end{split}
\end{align}
where in the last step we wrote $g = \exp\ve$ for some Lie algebra element $\ve \in \mfg$. This then implies
\begin{align}\label{deltaprop}
	\delta(U-U') = \delta(gU -gU').
\end{align}
Now, we can write 
\begin{align}\label{U-taylor}
\begin{split}
	e^{\ve(z,\bz) }U(z,\bz) &= U(z,\bz) + \ve(z,\bz) U(z,\bz) + O(\ve^2) \\
	&= U(z,\bz) - \int d^2w\, \ve^a(w,\bw)\mfD^a_{U(w,\bw)}U(z,\bz) + O(\ve^2).
\end{split}
\end{align}
More generally, we can also prove via induction that the above equation holds if we replace $U(z,\bz)$ with $U(z,\bz)^k$. Therefore, assuming $f(U)$ admits a Taylor series, we have 
\begin{align}\label{fgaugeexp}
	f\big(e^{\ve(z,\bz)} U(z,\bz) \big) = f\big(U(z,\bz)\big) - \int d^2w\, \ve^a(w,\bw)\mfd^a_{U(w,\bw)}f\big(U(z,\bz)\big) + O(\ve^2). 
\end{align}
Using this equation along with \eqref{deltaprop} with $g = \exp\ve$, we have
\begin{align}
\begin{split}
\delta(U-U') &= \delta(e^\ve U - e^\ve U') \\
&= \delta(U-U') - \int d^2z\,\ve^a(z,\bz)\left( \mfd^a_{U(z,\bz)} + \mfd^a_{U'(z,\bz)} \right) \delta(U-U'),
\end{split}
\end{align}
which implies
\begin{align}\label{Dprop2}
\begin{split}
	\Big( \mfD^a_{U(z,\bz)} + \mfD^a_{U'(z,\bz)} \Big) \delta(U-U') &=  0.
\end{split}
\end{align}
Finally, note that applying \eqref{fgaugeexp} to \eqref{prop3deriv}, we get
\begin{align}
	\int [dU] \, f(U) &= \int [dU] \, f(U) - \int d^2w \, \ve^a(w,\bw)\int [dU]\, \mfd^a_{U(z,\bz)} f(U) + O(\ve^2).
\end{align}
Requiring that this equation be satisfied to linear order in $\ve$, we obtain
\begin{align}\label{Dprop3}
	\int [dU]\,\mfd^a_{U(z,\bz)}f(U) = 0.
\end{align}
Collecting \eqref{Dprop1}, \eqref{Dprop2}, and \eqref{Dprop3}, we have
\begin{align}\label{Dproperties}
\begin{split}
	\Big[ \mfD^a_{U(z,\bz)}, \mfD^b_{U(w,\bw)} \Big]  &=    f^{abc} \delta^2(z-w)\mfD_{U(z,\bz)}^c  \\
	\Big( \mfD^a_{U(z,\bz)} + \mfD^a_{U'(z,\bz)} \Big) \delta(U-U') &=  0 \\
	\int [dU]\, \mfD^a_{U(z,\bz)} &= 0.
\end{split}
\end{align}

\bibliography{YMbib}{}
\bibliographystyle{utphys}

\end{document}